\PassOptionsToClass{balance=false}{acmart}

\documentclass[sigconf, nonacm]{acmart}

\usepackage{pvldb}

\usepackage{amsfonts}
\usepackage{bbm}
\usepackage{tabularx}
\usepackage{xspace}
\usepackage{placeins}
\usepackage{tikz}
\usepackage{colortbl}
\usepackage{listings}

\usetikzlibrary{arrows.meta,positioning}

\definecolor{cmCodeBlue}{RGB}{34,82,145}
\definecolor{cmCodeTeal}{RGB}{20,112,112}
\definecolor{cmCodeString}{RGB}{154,83,20}

\lstdefinestyle{codeminer}{
  language=Python,
  basicstyle=\ttfamily\scriptsize,
  keywordstyle=\color{cmCodeBlue}\bfseries,
  keywordstyle=[2]\color{cmCodeTeal}\bfseries,
  commentstyle=\color{black!55},
  stringstyle=\color{cmCodeString},
  morekeywords=[2]{clock,build,derive_capabilities,save,load_skill_metadata,requirements,load_views,load_skills,run},
  backgroundcolor=\color{black!2},
  columns=fullflexible,
  keepspaces=true,
  showstringspaces=false,
  numbers=left,
  numberstyle=\tiny\color{black!55},
  numbersep=6pt,
  frame=tb,
  framerule=0.45pt,
  rulecolor=\color{black!35},
  xleftmargin=1.4em,
  framexleftmargin=1.2em,
  framexrightmargin=0.3em,
  framesep=4pt,
  aboveskip=0.6\baselineskip,
  belowskip=0.4\baselineskip,
  captionpos=b
}

\newcommand{\ind}{\mathbf{1}}

\newif\ifvldbsubmission
\ifdefined\PVLDBSUBMISSION
  \vldbsubmissiontrue
\else
  \vldbsubmissionfalse
\fi
\newcommand{\detailsref}[1]{%
  \ifvldbsubmission supplemental material\else Appendix~\ref{#1}\fi}
\newcommand{\Detailsref}[1]{%
  \ifvldbsubmission Supplemental material\else Appendix~\ref{#1}\fi}

\newcommand{\qref}[2]{\hyperref[#2]{Q#1}}

\ifvldbsubmission
\else
\fi

\AtEndEnvironment{figure}{%
  \ifvldbsubmission\else\vspace{-7pt}\fi}
\AtEndEnvironment{figure*}{%
  \ifvldbsubmission\else\vspace{-7pt}\fi}

\renewcommand\vldbdoi{XX.XX/XXX.XX}
\renewcommand\vldbpages{XXX-XXX}
\renewcommand\vldbavailabilityurl{https://github.com/sysevol-ai/CodeNib}

\newcommand {\nickname}{\textsc{CodeNib}\xspace}

\makeatletter
\newif\if@qendtitle
\newcommand\qendtitle{?\global\@qendtitletrue}
\def\showarticletitle#1{\global\@qendtitlefalse#1%
  \if@qendtitle\expandafter\@qendtitle@gobble\fi}
\def\@qendtitle@gobble#1{\ifx.#1\else#1\fi}
\makeatother

\begin{document}
\ifvldbsubmission
  \title{CodeNib: A Multi-View Data System for Serving Repository Context to
  Coding Agents}
\else
  \title{\texorpdfstring{%
    \raisebox{-0.30em}{%
      \includegraphics[height=1.35em]{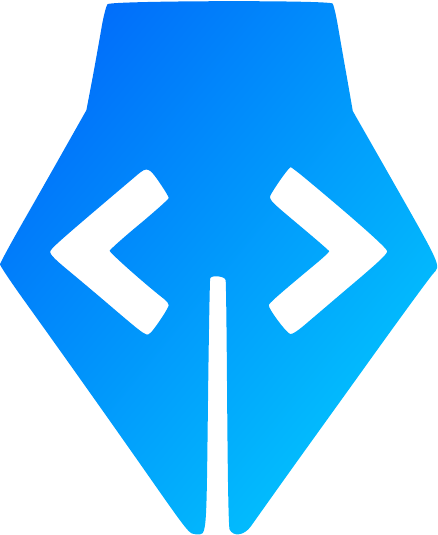}}%
    \hspace{0.32em}}{}%
    CodeNib: A Multi-View Data System for Serving Repository Context to
    Coding Agents}
\fi
\ifvldbsubmission
\else
  \makeatletter
  \g@addto@macro\@titlenotes{%
    \begingroup
      \renewcommand{\thefootnote}{}
      \footnotetext[0]{Code and artifacts:
      \url{https://github.com/sysevol-ai/CodeNib}.}
    \endgroup}
  \makeatother
\fi
\author{Zhongming Yu}
\affiliation{%
  \institution{UC San Diego}
  \city{La Jolla}
  \state{California}
  \country{USA}}
\email{zhy025@ucsd.edu}

\author{Hengjia Yu}
\affiliation{%
  \institution{UC San Diego}
  \city{La Jolla}
  \state{California}
  \country{USA}}
\email{hey015@ucsd.edu}

\author{Boqin Yuan}
\affiliation{%
  \institution{UC San Diego}
  \city{La Jolla}
  \state{California}
  \country{USA}}
\email{b4yuan@ucsd.edu}

\author{Shuting Zhao}
\affiliation{%
  \institution{UC San Diego}
  \city{La Jolla}
  \state{California}
  \country{USA}}
\email{shz106@ucsd.edu}

\author{Yizhao Chen}
\affiliation{%
  \institution{UC San Diego}
  \city{La Jolla}
  \state{California}
  \country{USA}}
\email{yic138@ucsd.edu}

\author{Aryan Dokania}
\affiliation{%
  \institution{UC San Diego}
  \city{La Jolla}
  \state{California}
  \country{USA}}
\email{adokania@ucsd.edu}

\author{Mihir Jagtap}
\affiliation{%
  \institution{UC San Diego}
  \city{La Jolla}
  \state{California}
  \country{USA}}
\email{mjagtap@ucsd.edu}

\author{Jiayu Chang}
\affiliation{%
  \institution{Stanford University}
  \city{Stanford}
  \state{California}
  \country{USA}}
\email{cjy1125@stanford.edu}

\author{Yitong Ma}
\affiliation{%
  \institution{UC San Diego}
  \city{La Jolla}
  \state{California}
  \country{USA}}
\email{yim030@ucsd.edu}

\author{Yash Jayswal}
\affiliation{%
  \institution{UC San Diego}
  \city{La Jolla}
  \state{California}
  \country{USA}}
\email{yjayswal@ucsd.edu}

\author{Wentao Ni}
\affiliation{%
  \institution{UC San Diego}
  \city{La Jolla}
  \state{California}
  \country{USA}}
\email{w2ni@ucsd.edu}

\author{Hejia Zhang}
\affiliation{%
  \institution{UC San Diego}
  \city{La Jolla}
  \state{California}
  \country{USA}}
\email{hez024@ucsd.edu}

\author{Zhaoling Chen}
\affiliation{%
  \institution{UC Riverside}
  \city{Riverside}
  \state{California}
  \country{USA}}
\email{zhaoling.chen@email.ucr.edu}

\author{Gangda Deng}
\affiliation{%
  \institution{University of Southern California}
  \city{Los Angeles}
  \state{California}
  \country{USA}}
\email{gangdade@usc.edu}

\author{Jishen Zhao}
\affiliation{%
  \institution{UC San Diego}
  \city{La Jolla}
  \state{California}
  \country{USA}}
\email{jzhao@ucsd.edu}
\renewcommand{\shortauthors}{Zhongming Yu et al.}
\renewcommand{\vldbauthors}{Zhongming Yu, Hengjia Yu, Boqin Yuan, Shuting Zhao,
Yizhao Chen, Aryan Dokania, Mihir Jagtap, Jiayu Chang, Yitong Ma, Yash Jayswal,
Wentao Ni, Hejia Zhang, Zhaoling Chen, Gangda Deng, and Jishen Zhao}

\begin{abstract}
An incremental code index can complete every update without error and still
return the wrong graph half the time. Coding agents depend on such reused
views for search and navigation, yet the systems that supply them never
measure where a precomputed answer may stand
in for the trusted route: localization agents rebuild a graph per task, tool
servers query a live language server per request, and code-intelligence
databases serve developers, not agents.

\nickname is a multi-view data system that makes that measurement possible.
It is the first to compile lexical, dense, and structural views of one commit
behind a single manifest and source-address contract, so that a lexical hit,
a graph occurrence, and a dense block can be checked against each other and
against a rebuild or a live server, and it serves search, navigation, and
bounded context to agents through one cost-visible runtime.

The comparisons overturn three assumptions. Execution status is not
correctness: symbol-level repair completes on every held-out change but
reproduces the rebuild on 13 of 24, so the runtime admits it per repository,
while vector reuse is exact on 27 of 30 held-out changes at 40$\times$ and
static definitions match the live server on 86\% of requests at a 1.2\,s
ready point. Shorter answers finish more tasks:
behind LocAgent, \nickname's views let the agent finish 78 of 90 runs against
65 at lower cost, and selected context policies use 50--87\% fewer tokens
than grep/read across five models. And context pays most when it is taken
back: on 200 SWE-bench Verified tasks, handing the agent \nickname's
candidates and retiring them after its first source read resolves 79 issues,
the most of five context policies, against 66 for on-demand exploration and
56 for Aider's repository map.
\end{abstract}

\maketitle
\ifvldbsubmission
\else
  \enlargethispage{18pt}
  \raggedbottom
\fi

\ifvldbsubmission
  \vldbtopmatter
\fi

\section{Introduction}\label{sec:intro}

Coding agents access evolving repositories through lexical and semantic
search, symbol and dependency navigation, source reads, and bounded history.
These operations share source state while yielding distinct evidence.

\begin{figure}[!t]
  \centering
  \includegraphics[width=0.78\columnwidth]{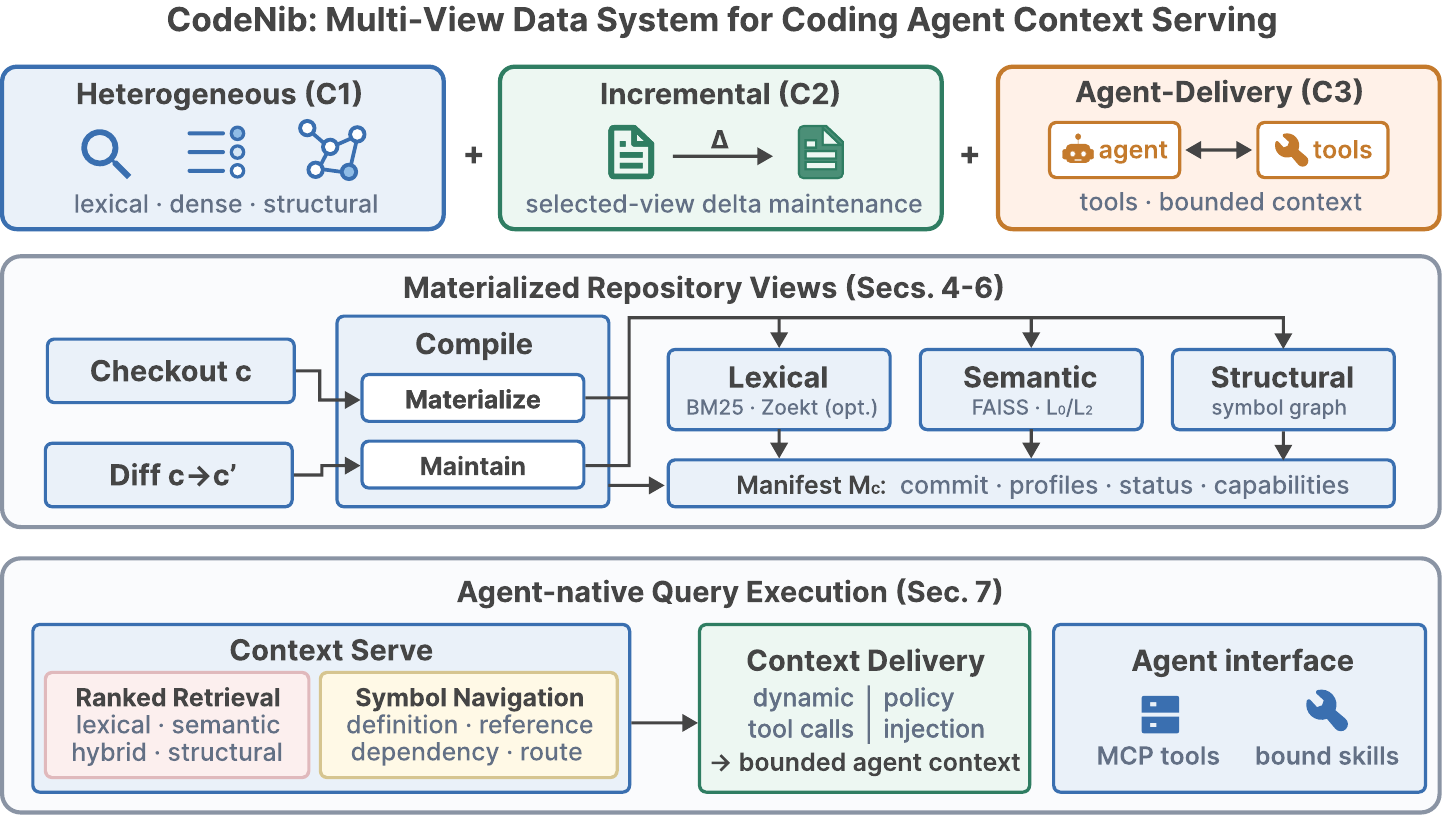}
  \Description{The top row states three coupled challenges: repository changes
  require selected-view maintenance, agent requests need heterogeneous lexical,
  dense, and structural evidence, and retrieved evidence must enter bounded
  context. The middle panel, labeled Sections 4--6, sends a repository checkout
  through the Compile path and changes through the Maintain path; both produce specialized
  views whose commit, profiles, status, and capabilities are recorded in a
  manifest. The bottom panel, labeled Section 7, sends ranked retrieval and
  symbol navigation through dynamic tool calls or policy injection before MCP
  tools or bound skills expose bounded context to an agent.}
  \caption{CodeNib compiles and maintains heterogeneous repository views,
  then serves them as bounded agent context.}
  \label{fig:motivation}
\end{figure}

Repository artifacts span text indexes, embedding stores, semantic graphs,
language servers, and prompt state, each with distinct layouts, costs, update
paths, and output contracts. Retrieval yields ranked candidates; navigation
yields locations. Agent systems obtain this evidence in two ways. Localization agents build a
repository graph per task with syntactic
parsers~\cite{locagent,orcaloca,codexgraph,repograph}, which leaves cross-file
symbol resolution to heuristics and confines their evaluations to Python. Tool
servers instead reach a live language server on every navigation
call~\cite{serena}, which resolves symbols exactly and pays the server's
startup in every session. Code-intelligence infrastructure does precompute
resolved facts~\cite{glean,scip}, but it answers developer queries rather than
an agent's mixed diet of ranked candidates and source locations, and it does
not report where a stored answer may stand in for a live one. That boundary is
what an agent runtime needs before it can reuse anything, and composing these
capabilities creates a data-lifecycle coordination problem for the runtime
(Sec.~\ref{sec:bg:retrieval}).

We take a data-systems view: a commit is immutable base data; chunks, postings,
embeddings, occurrences, and relationships are derived views; agent requests
are view-specific queries; and prompt context is a bounded delivery result.
The key design is to reuse repository-derived state across tasks without
collapsing ranked candidates, source locations, and prompt history into one
abstraction.
Figure~\ref{fig:motivation} summarizes three coupled challenges.
\textbf{C1: Heterogeneous views.} Lexical, dense, and structural data need
different physical layouts, but their results must map back to the same commit
and repository-relative source ranges. \textbf{C2: Incremental freshness.} An
edit changes each view differently, so graph repair, embedding reuse, and
rebuild paths must be selected and evaluated separately. \textbf{C3: Agent
delivery.} Precomputed evidence must reach the model through tools or bounded
context while build, runtime loading, query, and history costs remain visible.
Otherwise, each issue repeats model-directed grep/read discovery and carries
its task-specific observations through later turns.

We present \nickname, a multi-view data system spanning this lifecycle. The
mechanism is simple yet effective: build several repository views once, map
every result back to source ranges, update each view through its own path, and
load the views needed by each agent operation. Its repository view compiler
normalizes files, scopes, and callables; independent
view builders materialize lexical, dense, and structural artifacts, while
view-specific maintainers use Git/LSP-assisted graph repair and content-addressed
vector reuse. A manifest records each artifact's path, status, commit,
configuration, and supported operations; repository-relative addresses align
their outputs. Independent rebuilds are used only after timing to determine
which update outputs match; this comparison does not execute on the maintainer
path.

The runtime loads required views, lowers ranked requests to physical routes,
and composes source-linked ranked code blocks. It exposes static or live symbol
providers through one location interface.
Bound skills and a stdio MCP adapter~\cite{mcp} connect these operations to the
agent loop, where policies govern grep/read, eager, or eager-plus-compact
context delivery.
Traces retain the view, provider, and token usage across construction,
maintenance, queries, and delivery.

The components are familiar; the contract that joins them is not. Existing
systems each occupy one row of Table~\ref{tab:positioning}: a fact database
answers developer queries, a search index answers text, a tool server answers
a live session, a localization agent builds a graph it throws away. None
serves ranked candidates, resolved locations, and bounded context from one
commit-linked catalog, and none can say which of its answers may stand in for
the trusted route, having no such route to compare against. \nickname's
manifest and source-address contract give every view the same commit and the
same coordinates across three index families and five languages, so a
lexical hit, a graph occurrence, and a dense block can be checked against
each other and against a rebuild or a live server. That is what makes the
system usable as its own instrument: the route-level verdicts below require
exactly this catalog, which prior systems lack, and it
is what an agent runtime needs before it can reuse anything across sessions.
\ifvldbsubmission
Sec.~\ref{sec:bg} positions this boundary against prior systems.
\else
Table~\ref{tab:positioning} positions this boundary: read down the table,
each family covers one representation and one delivery surface, and
\nickname is the row that holds all three under one commit and one address
scheme.
\fi

\ifvldbsubmission
\else
\begin{table}[t]
  \caption{Positioning by system family, maintained repository state, and
  agent delivery. Entries indicate primary emphasis.}
  \label{tab:positioning}
  \centering
  \scriptsize
  \setlength{\tabcolsep}{2.5pt}
  \renewcommand{\arraystretch}{1.06}
  \begin{tabularx}{\columnwidth}{@{}>{\raggedright\arraybackslash}p{1.55cm}>{\raggedright\arraybackslash}p{2.35cm}>{\raggedright\arraybackslash}X@{}}
    \toprule
    \textbf{System family} & \textbf{Maintained state} &
    \textbf{Agent delivery} \\
    \midrule
    \rowcolor{black!5}
    Code search and intelligence~\cite{glean,sourcegraph,zoekt,scip} &
    Persistent code facts and lexical or symbol indexes &
    Lexical results and indexed locations \\
    Retrieval \& graph~\cite{repocoder,locagent,codexgraph,repograph,orcaloca,aider} &
    Repository- or task-scoped retrieval indexes and graphs &
    Ranked code blocks or structural traversal \\
    \rowcolor{black!5}
    Live LSP navigation~\cite{lsp,serena,lsprag} &
    Mutable workspace and live language-server state &
    Live definitions, references, or workspace symbols \\
    \textbf{CodeNib} &
    Manifest-cataloged, commit-linked lexical, dense, and structural views &
    Ranked blocks, static/live navigation, or bounded context under
    per-operation validity \\
    \bottomrule
  \end{tabularx}
\end{table}
\fi

This paper makes three contributions corresponding to C1--C3.

\begin{enumerate}
  \item \textbf{A repository view compiler (C1).} \nickname is the first
  serving layer to build lexical, dense, and structural artifacts of one
  commit behind a single manifest and source-address contract, recording
  which views are available for that commit and mapping every output to
  repository-relative source ranges, so that the three views can be served,
  maintained, and checked against each other.

  \item \textbf{View-specific incremental maintenance with measured validity
  boundaries (C2).} Git/LSP-assisted graph repair and content-addressed vector
  reuse update affected state. We compare timed updates with independent
  rebuilds offline, on an original and a repository-disjoint held-out cohort,
  and report update speedups only for matching outputs. The comparison
  yields an admission rule per route and catches what execution status
  cannot: symbol repair completes without error on every mismatch, so the
  runtime admits it per repository rather than per language. The admitted
  routes are fast: vector reuse runs a median 40$\times$ faster than a
  rebuild and graph repair up to 11.6$\times$.

  \item \textbf{A cost-visible agent runtime evaluated inside and outside our
  own harness (C3).} Ranked plans, static/live navigation providers, and
  bounded context policies connect reusable views to agent tools. Our
  stage-separated quality--cost analyses cover retrieval and reranking,
  dense-index construction and search, static/live navigation per backend and
  capability, incremental maintenance, and bounded context delivery across
  five models. Two further studies place the system behind agents we did not
  write: a paired provider substitution inside LocAgent, and five context
  policies on 200 real SWE-bench Verified repairs scored by the official
  tests. Behind LocAgent, \nickname's views let the agent finish 78 of 90
  runs against 65 at lower cost; on the repairs, handing the agent
  \nickname's candidates and retiring them after its first source read
  resolves 79 tasks, 13 more than on-demand exploration and 23 more than
  Aider's repository map.
\end{enumerate}

\ifvldbsubmission
Graph and vector updates match independent rebuilds on 15/33 and 28/31
source-changing transitions and reach 8.67$\times$/25.44$\times$ median
speedups on those cases. Static navigation preserves normalized live-server
locations on 75\% of requests, with a usable rule per backend and capability.
For each of five models, the lowest-token core arm that meets the localization
margin uses 50--87\% fewer trajectory tokens than paired grep/read.
\else
We evaluate repository localization, context serving, and, in two studies,
real issue repair; concurrent production updates remain outside this study.
The measurements expose where each route holds and where it does not. Static
navigation reproduces the live server's normalized path/start-line set on 705
of 940 compared requests, but the rule that survives is per backend and
capability: definitions on Go, Rust, C/C++, and Python and references on Go
alone, at a 1.23\,s ready point against the live server's 6.35\,s. Vector
reuse matches independent rebuilds on 28/31 and 27/30 source-changing
transitions across two cohorts at 25.44$\times$ and 40.43$\times$ median
speedups, and the same comparison admits symbol-level graph repair per
repository, 15/33 and 13/24, where a per-language rule would have let
silent mismatches through.
For each of five agent models, the lowest-token core arm that meets a common
localization margin uses 50--87\% fewer trajectory tokens than paired
grep/read. Substituting \nickname for LocAgent's provider lets the agent
finish 78 of 90 runs against 65 at lower cost, because the views answer in
fewer bytes, and on 200 real repairs with Haiku 4.5, retiring delivered candidates
after the first source read records the most confirmed resolutions of five
policies by reaching a submission more often under one budget; where the two
policies disagree, retirement wins 22 tasks to 9 under a test fixed before
the run.
Together, \nickname turns repository context from repeated per-task
exploration into reusable views that can be built, updated, and served
independently. Shared commit and source addresses connect their outputs, while
per-operation measurements distinguish exact substitutions from quality--cost
tradeoffs and name the boundaries where a substitution should not be made.
\fi

\section{Background and Positioning}\label{sec:bg}

\begin{figure*}[t]
  \centering
  \includegraphics[width=\textwidth]{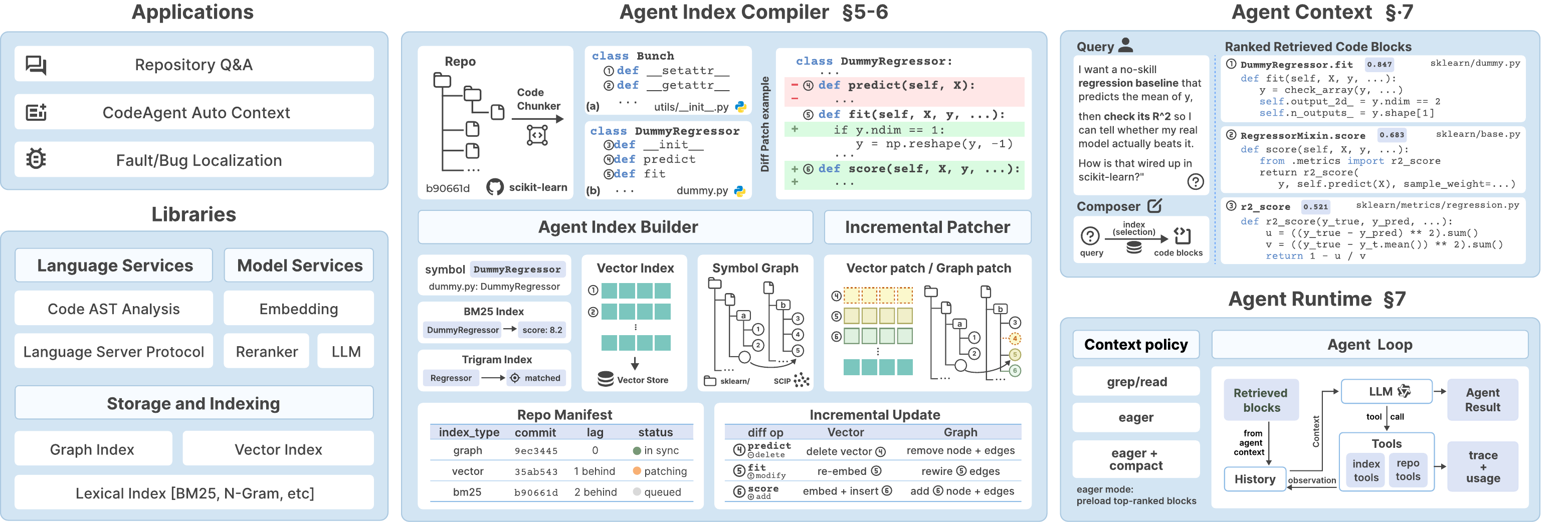}
  \Description{The left column lists repository applications and the language,
  model, storage, and indexing libraries they use. The Repository View Compiler,
  Sections 5--6, contains View Builders for BM25, trigram, vector, and symbol-graph
  views plus View Maintainers for incremental vector and graph updates; a
  repository manifest records per-view commit, lag, and status. In the running
  diff, circled symbol 4 is deleted, 5 is modified, and 6 is added. Query Planning
  and Retrieval, Section 7.2, composes source-linked ranked retrieved code blocks
  for a query. The Agent Runtime, Sections 7.1, 7.3, and 7.4, applies grep/read,
  eager, or eager-plus-compact context policies and exposes index tools around
  one agent loop. Displayed commit hashes, lag values, and scores are
  illustrative, not measurements.}
  \ifvldbsubmission
    \caption{CodeNib's repository-to-agent dataflow. The left column lists
    applications and dependencies, not a pipeline stage. The center compiler's
    \emph{View Builders} and \emph{View Maintainers} create and update
    manifest-linked views (Secs.~\ref{sec:index}--\ref{sec:indexing});
    the upper-right planner composes ranked retrieved code blocks
    (Sec.~\ref{sec:system:plans}); and the lower-right runtime applies context
    policies and index tools around the agent loop
    (Secs.~\ref{sec:system:loop}, \ref{sec:system:lsp},
    and~\ref{sec:system:context}). Displayed fields are illustrative.}
  \else
    \caption{CodeNib's repository-to-agent dataflow. The left column lists
    supported applications and dependencies, not a pipeline stage. The center
    Repository View Compiler's \emph{View Builders} and \emph{View Maintainers}
    create and update manifest-linked views
    (Secs.~\ref{sec:index}--\ref{sec:indexing}). The
    upper-right Query Planning and Retrieval path composes source-linked ranked
    retrieved code blocks (Sec.~\ref{sec:system:plans}); the lower-right Agent
    Runtime applies context policies and exposes index tools around the agent
    loop (Secs.~\ref{sec:system:loop}, \ref{sec:system:lsp},
    and~\ref{sec:system:context}). Hashes, lag, and scores are illustrative, not
    measurements.}
  \fi
  \label{fig:architecture}
\end{figure*}

\subsection{Data-Management Foundations}

Materialized views trade construction and maintenance for query-time
reuse~\cite{incremental-views}; repository indexes make this trade against a
changing commit. Because text, vector, and graph operators use different
records and return different values, \nickname catalogs specialized indexes
behind a manifest. This resembles polystore mediation~\cite{bigdawg}, but uses
curated physical routes rather than searching a costed plan
space~\cite{volcano}.
General agent-data engines such as CocoIndex incrementally maintain declared
source-to-target flows with lineage~\cite{cocoindex}. \nickname instead fixes
the repository commit and build configuration for each view, then measures its
build, update, load, and query costs separately.

\subsection{Materialized Code Intelligence}\label{sec:bg:indexing}

LSP standardizes live requests~\cite{lsp}; LSIF~\cite{lsif} and
SCIP~\cite{scip} serialize locations and relationships. Glean stores code
facts~\cite{glean}; Sourcegraph combines search and code
intelligence~\cite{sourcegraph}; and Zoekt provides trigram
search~\cite{zoekt}. \nickname combines these ideas with dense and sparse
retrieval in a local runtime, without claiming Glean-scale storage or a general
fact language. Glean propagates fact ownership around reindexed
units~\cite{glean-incremental}. Industrial call-graph maintenance invalidates
then patches affected code~\cite{incremental-callgraph}; Stack Graphs provides
file-incremental name resolution~\cite{stack-graphs}; and incremental CodeQL
reuses analysis state~\cite{incremental-codeql}. \nickname instead repairs
source-anchored graph facts. \qref{4}{sec:eval:incremental} qualifies its
speedups against offline independent rebuilds, not on-path comparisons.
Tree-sitter supplies source positions when semantic coverage is absent, and the
manifest records semantic-navigation availability~\cite{treesitter}.

LSPRAG retrieves task-specific definitions and references from live LSP back
ends for real-time unit-test generation~\cite{lsprag}, while TypeScript indexing
can replace per-symbol LSP calls with
compiler-API analysis~\cite{abcoder-ts-index}. We compare static and live normalized locations
across five language groups and retain a live provider because their outputs
can differ.

\subsection{Repository Retrieval and Structural Context}\label{sec:bg:retrieval}

Lexical retrieval matches identifiers. Dense encoders include
CodeBERT~\cite{codebert}, GraphCodeBERT~\cite{graphcodebert},
UniXcoder~\cite{unixcoder}, and Qwen3-Embedding~\cite{qwen3-embed};
Qwen3-Reranker~\cite{qwen3-embed} and RankGPT~\cite{rankgpt} rerank; and
FAISS~\cite{faiss} and HNSW~\cite{hnsw} provide exact, inverted-file, and
graph-based ANN indexes. We evaluate these choices, not a new model or ANN
algorithm.

Repository systems differ in when and how they retrieve: RepoCoder alternates
retrieval and generation~\cite{repocoder}; RepoFormer predicts when retrieval
is useful~\cite{repoformer}; and CoRet trains a structure-aware edit
retriever~\cite{coret}. Direct Corpus Interaction exposes raw grep/read without
a prebuilt index~\cite{dci}; a cross-paradigm study compares similarity,
static-analysis, and navigation context~\cite{one-size-context}. These are
complementary serving choices, not implementations of one operator.

Repository localization ranges from staged narrowing in
Agentless~\cite{agentless}, through iterative exploration in
AutoCodeRover~\cite{autocoderover}, to two-phase graph search in
CoSIL~\cite{cosil}. Graph-guided agents include
CodexGraph~\cite{codexgraph}, RepoGraph~\cite{repograph},
LocAgent~\cite{locagent}, and OrcaLoca~\cite{orcaloca}.
RepoNavigator trains a repository-localization agent with
reinforcement learning around one execution-aware jump-to-definition
tool~\cite{reponavigator}. SWE-Explore
benchmarks ranked relevant code regions under a fixed line budget, measuring
coverage, ranking, and context efficiency~\cite{swe-explore}. RIG serves a
deterministic architectural map~\cite{rig}, whereas Codebase-Memory exposes a
persistent Tree-sitter knowledge graph through MCP~\cite{codebase-memory}. AOCI
combines symbolic and semantic indexing~\cite{aoci}; Code Isn't Memory evaluates
a structural index in a controlled agent harness~\cite{code-isnt-memory};
SpIDER adds graph exploration and reasoning to dense retrieval~\cite{spider};
and AIRCoder fuses text, dependencies, and hierarchy~\cite{aircoder}. \nickname
instead links ranked and navigation views to one commit, preserves their
distinct output contracts, and measures paired one-hop graph effects with the
dense retriever fixed.

\subsection{Agent Interfaces and Context}\label{sec:bg:agents}

Agent interfaces include SWE-agent~\cite{sweagent},
OpenHands~\cite{openhands}, and OpenCode~\cite{opencode}. Devin combines shell, editor, and
browser tools~\cite{devin}; DeepWiki exposes generated documentation and
repository search~\cite{deepwiki}. Serena provides
LSP-backed retrieval and editing~\cite{serena}; Aider supplies a repository
map~\cite{aider}; and Context as a Tool compresses history~\cite{context-as-tool}.
RepoShapley learns completion-context filters~\cite{reposhapley}, while
SpecAgent predicts context during indexing~\cite{specagent}. ContextBench
benchmarks agent context retrieval~\cite{contextbench}; an AGENTS.md study tests
repository context files~\cite{agents-md}; and deterministic anchoring studies
required static structure~\cite{deterministic-anchoring}. CodeStruct provides
AST-scoped actions~\cite{codestruct}; CodeMEM adds session
memory~\cite{codemem}; and other work varies edit-time
representation~\cite{act-context}. AgentDiet removes redundant or expired
trajectory information~\cite{agentdiet}; the Complexity Trap compares raw
history, observation masking, and
summarization~\cite{complexity-trap}; and SWE-Pruner performs task-conditioned
context pruning~\cite{swe-pruner}.
Repository-context compression instead shortens a static repository input
before generation, not observations retained across an interactive
loop~\cite{repo-context-compression}.

We instead hold candidates fixed while varying initial delivery and one-time
compaction during localization. Static/live latency is reported only where
normalized path/start-line sets match, and token savings only under a paired
localization margin; neither claim extends to learned exploration or patch
correctness.

\section{System Overview}\label{sec:overview}

The dataflow in Figure~\ref{fig:architecture} begins in the center panel; the left
column lists supported applications and dependencies rather than another
pipeline stage. The center \emph{Repository View Compiler} is the
commit-indexed data plane (Sec.~\ref{sec:overview:compiler}). The upper-right
\emph{Query Planning and Retrieval} path converts a query into ranked retrieved
code blocks (Sec.~\ref{sec:overview:query}), while the lower-right
\emph{Agent Runtime} loads views, exposes tools, and delivers context around the
agent loop (Sec.~\ref{sec:overview:runtime}). The planes meet at manifest $M_c$
and repository-relative source addresses. We follow this center-to-right flow
below.

\subsection{Repository View Compiler}\label{sec:overview:compiler}

The center panel has two paths. Under \emph{View Builders}, the Code Chunker and
semantic backends derive source-linked units and write graph, vector, BM25, and
optional Zoekt artifacts. Under \emph{View Maintainers}, a Git diff drives
LSP-assisted graph repair or content-addressed vector reuse. The bottom-center
\emph{Repo Manifest} is $M_c$: it catalogs each view's commit, profile, status,
and capabilities. It is the runtime's lookup boundary, not a container for view
payloads. Section~\ref{sec:framework:views} formalizes $M_c$,
Section~\ref{sec:index} defines the views, and Section~\ref{sec:indexing}
details the builder and maintainer paths.

\ifvldbsubmission
\else
The initial-build path runs requested builders sequentially and publishes the
manifest only after every builder reports success or failure. A failed optional
view does not invalidate successful siblings. Initial construction and delta
maintenance have different publication boundaries: builders link a complete
requested artifact set through $M_c$, whereas graph and vector maintainers update
their own stores. This distinction lets the evaluation change one physical path
without silently changing the others.
\fi

\subsection{Query Planning and Retrieval}\label{sec:overview:query}

The request plane exposes two request classes, placed in different regions of
Figure~\ref{fig:architecture}.

\paragraph{Ranked retrieval.}
In the upper-right panel, a query enters the \emph{Composer}, which selects a
dense, lexical, structural, or fused physical plan and returns ordered,
source-linked code blocks. Section~\ref{sec:system:plans} defines this lowering.

\paragraph{Symbol navigation.}
A definition or reference request selects a static occurrence/graph provider
or live JSON-RPC and returns normalized locations rather than ranked code blocks.
The figure represents these providers as the lower-right loop's \emph{index
tools}; Section~\ref{sec:system:lsp} specifies their shared location interface.

\ifvldbsubmission
\else
Provider choice remains trace-visible: a normalized location does not erase
whether it came from persisted occurrences, graph traversal, or a live server.
\fi

\subsection{Agent Runtime}\label{sec:overview:runtime}

The lower-right panel begins with a \emph{Context policy} and ends with an
\emph{Agent Result} and trace. During session setup, the runtime reads $M_c$ and
opens only the required prebuilt views as process-local query state
(Sec.~\ref{sec:system:loop}); it does not build views or insert code into
history. During task execution, the agent may call \emph{index tools} or
\emph{repo tools} dynamically, while the policy may instead preload ranked
$L_2$ code blocks and compact retained observations
(Sec.~\ref{sec:system:context}).
Skills and the stdio MCP adapter expose the same search, dependency, definition,
reference, and route operations.

\ifvldbsubmission
\else
Runtime view loading is capability-driven. MCP loads configured resources once
per process, while the agent path loads only indexes required by selected skills.
\fi

\subsection{Boundaries}\label{sec:overview:boundaries}

\nickname does not choose edits or judge tests. Its manifest is not a
cross-store transaction, and compilation records a commit without locking the
worktree. Because static navigation does not reproduce every live response's
normalized locations, callers requiring workspace semantics must retain a live
LSP path. \qref{4}{sec:eval:incremental} compares updated views with independent
rebuilds only after timing; these comparisons do not run on the maintainer path.

\ifvldbsubmission
\else
These limits follow directly from the implementation. A recorded commit does
not prove that a mutable checkout stayed quiescent during construction; a
manifest does not make separate stores transactional; and source-address
normalization does not make two providers behaviorally equivalent. Each
experiment therefore states the concrete output it compares.
\fi

\paragraph{From dataflow to measurements.}
The remainder follows Figure~\ref{fig:architecture}'s dependency chain.
Section~\ref{sec:framework} first defines the shared commit and source-address
model plus the result and metric for each operation. Section~\ref{sec:index}
instantiates the view state cataloged by $M_c$, and Section~\ref{sec:indexing}
gives the initial and delta paths that create or advance that state.
Section~\ref{sec:system} consumes the same views through ranked plans, symbol
providers, and context policies, while Section~\ref{sec:impl} maps the
abstractions to implementation packages and adapters. Finally,
Section~\ref{sec:eval} preserves this decomposition:
\qref{1}{sec:eval:rag}--\qref{2}{sec:eval:index} measure retrieval and dense
indexes, \qref{3}{sec:eval:lsp} measures symbol navigation,
\qref{4}{sec:eval:incremental} and the lifecycle trace measure maintenance and
cross-stage composition, and \qref{5}{sec:eval:context} measures context delivery.

\section{Repository Views and Request Semantics}\label{sec:framework}

We distinguish four operations by their returned value and measured cost:
ranked retrieval, symbol navigation, structural maintenance, and context
delivery. Their output comparisons define the evaluation; they are not one
shared runtime gate.

\subsection{Views and Source Addresses}\label{sec:framework:views}

Let $c$ identify a repository commit and let $U_c$ be the source units extracted
from its checkout. A source unit records
\[
u=\langle p, r_s, r_e, \ell, \tau, x, s\rangle,
\]
where $p$ is a repository-relative path, $[r_s,r_e]$ is a source range,
$\ell\in\{L_0,L_1,L_2\}$ is its granularity, $\tau$ is its node type, $x$ is
source text, and $s$ is an optional resolved symbol. Files are $L_0$, type-like
scopes are $L_1$, and callable definitions are $L_2$.

\nickname materializes three views over $U_c$:

\begin{description}
  \item[Lexical view $V_c^{\mathrm{lex}}$.] Posting or trigram records for
  identifiers, paths, comments, and source text.
  \item[Dense view $V_c^{\mathrm{dense}}$.] Embeddings of $L_0$ or $L_2$ units
  with a mapping back to their source ranges.
  \item[Structural view $G_c$.] Typed containment and relationship edges among
  source-linked files, scopes, and definitions, plus persisted occurrence
  records where a backend supplies them.
\end{description}

Each view has profile $\theta$ (language, backend, schema, model, and options).
Manifest $M_c=\langle c,V_c^{\mathrm{lex}},V_c^{\mathrm{dense}},G_c,K_c\rangle$
links profiles, artifact status, and capabilities $K_c$ without implying one
storage engine. Results expose address $\langle p,r_s,r_e,\tau\rangle$; range
containment aligns granularities, while backend identifiers remain view-local.
During session setup, the runtime uses $M_c$ to locate required artifacts and
check their recorded status and capabilities.

\subsection{Four Operation Classes}\label{sec:framework:ops}

\paragraph{Ranked retrieval.}
A request maps text query $q$ to ranked source units and lowers to
\[
z=\langle r,k,\rho,h\rangle.
\]
Route $r\in\{A,B,C,D\}$ selects lexical, semantic, hybrid, or structural
retrieval and route-local fusion; only hybrid C owns RRF. Width $k$ contains
retrieval fan-out $k_{\mathrm{ret}}$ and optional pre-rerank cut $k'$; final
$k_{\mathrm{out}}$ remains a caller limit. $\rho$ selects an optional reranker,
and graph expansion $h$ is legal only on D. The tuple records varying decisions,
not a normal form; \qref{1}{sec:eval:rag}'s explicit dense--graph fusion is an ablation outside this
automatic lowering space.

\ifvldbsubmission
\else
The executed result is a ranked list
$[(u_1,\sigma_1),\ldots,(u_{k_{\mathrm{out}}},\sigma_{k_{\mathrm{out}}})]$.
Operationally, $k_{\mathrm{ret}}$ maps to \texttt{retrieve\_top\_k}, while
$k'$ maps to \texttt{rerank\_candidate\_top\_k}; the latter is subordinate to
the width policy rather than a fifth plan coordinate. This separation matters
because increasing retrieval fan-out and increasing reranker work have
different latency and recoverability effects. Named experimental pipelines may
compose operators outside the automatic A--D routes, but their operators and
fusion rules remain explicit.
\fi

\paragraph{Symbol navigation.}
An LSP-shaped request $a$ specifies a capability and source position. Let
$N(\cdot)$ project provider-limited results to deterministic unique
path/start-line pairs. Static and live providers match when
\[
N(P_{\mathrm{static}}(a,M_c))=N(P_{\mathrm{live}}(a,c)).
\]
This normalized location set omits characters, end ranges, and provider
metadata. Equality defines agreement in Figure~\ref{fig:nav-agreement}, but does not
prove interchangeability; agreement and operational cost must be reported
together.

\ifvldbsubmission
\else
The normalized set is intentionally weaker than full location or response
equality. A match says nothing about characters, end ranges, hover text,
workspace diagnostics, or provider metadata, and a latency reduction on the
matched subset says nothing about mismatches. The evaluation therefore reports
coverage and conditional latency as separate quantities rather than treating
the static provider as a universal LSP replacement.
\fi

\paragraph{Structural-view maintenance.}
Let $D$ be the files changed by a transition $c\!\rightarrow\!c'$, let
$\widehat G_{c'}$ be an incrementally repaired graph, and let $G_{c'}$ be a
fresh target graph built independently for offline evaluation. Define
$\mathcal{F}(G)$ as the declared graph-fact projection: the tagged multiset of
vertex identities, types, source/selection lines, and typed edges with source
anchors in $G$. It is not byte-level serialization equality. The offline
graph-output equality check is
\begin{equation}
\mathcal{F}(\widehat G_{c'})=\mathcal{F}(G_{c'}).
\label{eq:graph-fact-equivalence}
\end{equation}
For deterministic definition/reference requests $\mathcal{A}_D$ anchored in
$D$, let $R_G(a)=N(P_{\mathrm{static}}(a;G))$. After persistence and reload, we
apply a second offline serving check:
\begin{equation}
\forall a\in\mathcal{A}_D:\;
R_{\widehat G_{c'}}(a)=R_{G_{c'}}(a).
\label{eq:graph-serving-equivalence}
\end{equation}
Both checks run after timed maintenance against the independently rebuilt
target. The fresh rebuild and comparisons do not execute on the current
maintainer path and are not included in update latency. They qualify the
reported speedup; they do not provide an online correctness oracle. Neither
check covers live-LSP behavior or atomic cross-view publication.

\ifvldbsubmission
\else
The first check compares the declared graph-fact multiset in memory. The replay
check crosses persistence and serving boundaries by reloading both graphs and
comparing deterministic requests. It is a regression suite over the same
transition, not additional statistical evidence. A transition that fails
either check contributes no conditional speedup, even when its patch latency is
low.
\fi

\paragraph{Context delivery.}
A context policy $\pi$ maps ranked candidates and accumulated history to model
prompts. We compare model-directed grep/read, eager $L_2$ injection, and the
same injection followed by a one-time history rewrite. \qref{5}{sec:eval:context}
reports trajectory tokens and $\mathrm{AnswerRecall@5}$ over the first five
deduplicated source spans committed in the final answer. The $@5$ cutoff is a
reporting choice, not an output cap; Figures~\ref{fig:rag-pareto}
and~\ref{fig:retrieval-ablation} retain their separately frozen top-10
retrieval/index contracts. Eager and compact receive identical candidates, so
only their direct contrast isolates retention; comparisons with grep/read also
change candidate delivery. Retained content is charged each time it appears in
a later prompt, independently of prefix-cache reuse.

\subsection{Metrics and Cost Accounting}\label{sec:framework:cost}

Each operation uses its declared output contract: target coverage for retrieval
and final answers, all-target File Success, exact-Flat overlap for ANN, and
output-match predicates for navigation and maintenance before conditional
latency or speedup. \Detailsref{app:metrics} defines the denominators,
deduplication and cutoff rules, and gives worked examples.

\noindent\emph{Maintenance speedup.}
Graph arm $a$ is file replacement or symbol repair. Let
$n_{\mathrm{share}}\geq 1$ count consecutive transitions processed after one
LSP server startup/warmup. This is setup-cost amortization, not reuse of server
results: $n_{\mathrm{share}}=1$ for an isolated update and
$n_{\mathrm{share}}=5$ here because each repository sequence contains five
transitions. The terms $T_f^G$, $T_u^{G,a}$, and $T_s^{G,a}$ are graph rebuild,
update, and server startup/warmup costs;
$T_f^V$ and $T_u^V$ are vector rebuild and update costs after shared model
loading:
\begin{equation}
\Gamma_{G,a}
=\frac{T_f^G}{T_u^{G,a}+T_s^{G,a}/n_{\mathrm{share}}},
\qquad
\Gamma_V=\frac{T_f^V}{T_u^V}.
\label{eq:update-speedups}
\end{equation}
A ratio above one means the update plus its setup share is faster than
rebuilding. Rebuild and post-timing comparison remain outside the update path;
conditional summaries require source-changing output matches, while raw ratios
remain visible.

\noindent\emph{Lifecycle projection.}
Unlike \qref{2}{sec:eval:index}'s post-extraction timing, lifecycle accounting
starts from a prepared checkout. With materialization $B$, fresh-process view
loading $L$, and an $N$-session service trace $S$, the per-session costs at
reuse scale $q$ are $B/q+L+S/N$ for independent loading and $(B+L)/q+S/N$ for
one resident runtime. The measured serve-only term $S/N$ excludes concurrency
and queueing and is not an end-to-end or hardware bound. Plan latency includes
invoked operators; LSP replay measures marginal warm latency.

Agent token usage sums provider-reported prompt and completion tokens over
all model invocations, including answer-format invocations; tool observations
count when serialized into a submitted prompt. \Detailsref{app:metrics} gives
the exact sum and a retained-history example.

For every model, let $\Delta\mathrm{AR@5}(\pi)$ denote the paired mean
$\mathrm{AnswerRecall@5}$ change from grep/read. Policy $\pi$ preserves quality
under our reporting rule when the lower bound of its paired 95\% interval
satisfies the operational margin $\epsilon=0.05$:
\[
\operatorname{LB}_{.95}\!\left[\Delta\mathrm{AR@5}(\pi)\right]
\geq -\epsilon.
\]
Among policies that meet the margin, fewer tokens are better. This is a
reporting threshold, not a pre-registered equivalence test.

\section{Materialized Repository Views}\label{sec:index}

\ifvldbsubmission
\else
One checkout is represented by separate structural, dense, and lexical
artifacts. Their shared fields are a repository-relative source range and the
repository commit recorded in the manifest, not a shared storage engine or a
globally meaningful backend identifier.
\fi

\subsection{Source Units}

Language adapters instantiate the $L_0$/$L_1$/$L_2$ hierarchy using tree-sitter
ranges; a language may omit $L_1$. The hierarchy supports file- or callable-level
indexing, enclosing-file projection, and stable boundaries across backend symbol
IDs. The registry chunks 14 languages and records graph and incremental support
separately. \Detailsref{app:source-units} gives the exact five-language source
unit and repository-filter rules used by the experiments.

\ifvldbsubmission
\else
Adapters preserve zero-based parser coordinates internally and may omit $L_1$
when a language has no applicable named scope. The hierarchy lets retrieval
switch between file and callable units, lets evaluation project a callable back
to its enclosing file, and supplies source boundaries even when a semantic
backend uses a different symbol identity.
\fi

\subsection{Structural View}

Graph $G_c=(V_c,E_c)$ stores directories, files, and typed symbols linked by
\textsf{contain}; semantic backends add \textsf{reference}, \textsf{import},
and \textsf{type-use} edges. Nodes retain path, range, language, name, and an
optional backend symbol ID.

An \texttt{igraph} query layer indexes ranges and edge anchors, supports bounded
neighborhood/dependency queries, and maps positions to enclosing symbols when
no exact occurrence is persisted.

Versioned graph pickles reject mismatched schemas rather than migrate silently.

\ifvldbsubmission
\else
Symbol subtypes include classes, functions, methods, and fields. The query
layer indexes both source ranges and edge anchors, then filters incident
adjacency by containment or semantic family; bounded expansion therefore does
not scan every edge. Position-to-enclosing-symbol lookup is also the fallback
used by static navigation when a backend did not persist an exact occurrence.
Schema rejection makes a stale graph fail with a rebuild requirement instead of
silently interpreting it under a newer node or edge schema.
\fi

\subsection{Dense and Lexical Views}

The dense builder embeds $L_0$ files or $L_2$ callables into FAISS and retains a
source/text side mapping. Flat inner-product search is the default, IVF is
configurable, and \qref{2}{sec:eval:index} rebuilds fixed vectors into HNSW only
for ablation.

BM25 ranks chunk text~\cite{bm25}; optional Zoekt indexes raw-file trigrams for
substring and regex search. MCP normalizes both to paths, ranges, snippets, and
optional scores.

\ifvldbsubmission
\else
The dense side mapping retains the exact source address and text associated
with every vector. \qref{2}{sec:eval:index}'s HNSW path rebuilds the same frozen
vectors solely to compare physical organizations; HNSW is not silently
substituted into the runtime configuration. BM25 and Zoekt also answer
different requests: BM25
ranks source units, whereas Zoekt reports file-level text matches before MCP
normalization.
\fi

\subsection{Manifest Linking}

Each builder writes independently and returns status and metadata. After all
builders finish, $M_c$ records type, path, timestamp, status, configuration, and
duration, then derives capabilities. Thus BM25 can remain available after a
vector failure. The manifest binds requested artifacts to commit $c$ but does
not provide cross-store transactions after edits.

\ifvldbsubmission
\else
This linking protocol provides failure isolation and discovery. Clients inspect
capabilities instead of inferring availability from files on disk, and a query
process can continue serving one successful view after another builder fails.
The entry records what was requested for commit $c$; it does not certify later
worktree state or synchronize independent post-build mutations.
\fi

\section{View Construction and Freshness}\label{sec:indexing}

\ifvldbsubmission
\else
The materialization pipeline converts a checkout into the views of
Sec.~\ref{sec:index}. Its primary correctness responsibility is provenance: every
artifact and exposed capability must identify the repository state, profile,
and backend that produced it. Construction and incremental repair deliberately
retain different freshness and publication boundaries.
\fi

\begin{figure*}[t]
  \centering
  \includegraphics[width=\textwidth]{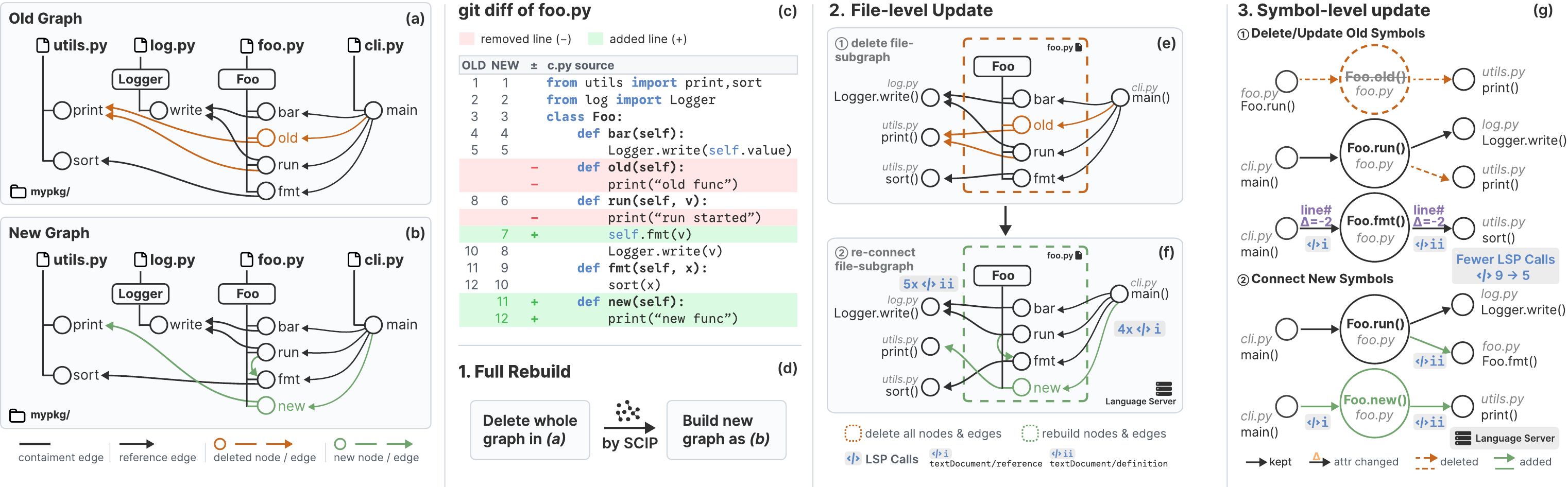}
  \Description{Panels show an old graph, its desired new graph, and a Python
  edit that removes one method, changes one method body, shifts another method,
  and adds a method. Full rebuild replaces the complete graph. File-level update
  deletes and reconnects the changed-file subgraph. Symbol-level update
  classifies deleted, affected, shifted, unchanged, and added symbols, rebases
  locations, and repairs only relationships invalidated by the edit. The example
  reduces displayed reference and definition requests from nine to five.}
  \caption{LSP-assisted incremental graph maintenance. \textbf{(a--b)} Old and
  target graphs for diff \textbf{(c)}. Full rebuild \textbf{(d)} recreates the
  target; file-level update \textbf{(e--f)} deletes and reconnects the entire
  changed-file subgraph. Symbol-level update \textbf{(g)} preserves stable
  graph facts and repairs edit-invalidated state. File-level uses four reference
  plus five definition requests; symbol-level uses one plus four
  ($9\!\rightarrow\!5$, four or 44.4\% fewer). Counts cover displayed LSP
  requests, not all messages or aggregate speedup.}
  \label{fig:incremental-lsp}
\end{figure*}

\subsection{Language Backend Selection}\label{sec:indexing:cold}

A registry independently selects chunking, cold graph, incremental, and live
language-server backends. Mature routes use SCIP; C/C++ uses clangd artifacts;
other languages retain tree-sitter chunks and text retrieval. Capabilities thus
distinguish syntactic hierarchy from resolved cross-file occurrences.

SCIP occurrences associate source ranges with optional symbol identifiers and
role bitsets~\cite{scip}; the decoder
creates graph nodes/edges and can persist character-accurate lookups. Without
exact occurrences, static serving falls back to the coarser graph and records
that result granularity.

Compiled-language preparation may require dependencies, a compilation database,
or a project build. Graph builders therefore fail explicitly rather than
silently substitute weaker data under the same capability.

\ifvldbsubmission
\else
SCIP definitions become source-linked graph vertices and resolved occurrences
become typed relationships. Where available, the decoder also persists an
occurrence table for character-accurate definition and reference lookup; a
graph-only fallback advertises its coarser behavior. This distinction
is especially important for compiled projects, where dependency restoration,
compilation-database generation, or a successful build is part of semantic
coverage. An optional builder failure is recorded instead of being relabeled as
equivalent semantic output. \Detailsref{app:semantic-backends} records the
evaluated Tree-sitter, cold-index, live-server, and static-position routes.
\fi

\subsection{Initial-Materialization Pipeline}

The store derives separate repository/commit and artifact-profile IDs, verifies
the detached worktree, nests profiles below the source identity, and aliases
benchmark IDs.
Requested builders then run sequentially and write $M_c$. BM25 and vector
currently extract units independently, so $U_c$ is logical rather than a shared
materialization; we claim neither extraction sharing nor parallel-build speedup.

\qref{2}{sec:eval:index} build timers start after source extraction and include
embedding plus FAISS, but exclude checkout, dependencies, SCIP, and graph
decoding; they isolate granularity and vector-index choices rather than
repository cold start.

\ifvldbsubmission
\else
\texttt{SourceSnapshot} hashes the canonical repository/commit pair separately
from \texttt{ArtifactProfile}, which hashes language, schema, backend, and
builder options. The store verifies the detached worktree commit, nests profiles
under the commit identity, and treats benchmark instance IDs only as aliases.
These identities prevent measurements from conflating repeated instances with
unique repository/commit pairs. They do not remove the need for a quiescent
checkout during a build.
\fi

\subsection{Static Symbol Serving}\label{sec:indexing:static}

Static and live providers receive the same capability, path, position, and
options. Static serving prefers occurrences, otherwise resolves an enclosing
graph node and follows source-linked records. Both emit normalized locations and
provider, repository commit, and granularity metadata.

Figure~\ref{fig:nav-agreement} measures static/live normalized-location
matches per backend and capability. Because only some backend--capability
pairs pass the agreement rule, the interface retains live JSON-RPC and does
not route unseen requests to static serving automatically.

\ifvldbsubmission
\else
The provider boundary converts coordinates once, applies identical result
limits, and sorts normalized locations before comparison. Persisted occurrences
serve position-based requests directly; the fallback maps a position to a graph
node and follows definition or reference records. Trace metadata makes the path
visible, but cannot predict whether an unseen static response will match a live
workspace server.
\fi

\subsection{Delta Maintenance and Freshness}\label{sec:indexing:warm}

Figure~\ref{fig:incremental-lsp} contrasts full rebuild with file- and
symbol-level repair. A Git hunk need not invalidate every declaration: the
file-level baseline
discards stable symbols and cross-file relationships and treats a line shift as
a semantic edit. The symbol-level path instead combines zero-context hunks with
one new \texttt{documentSymbol} tree to classify symbols as deleted, affected,
shifted, unchanged, or added. In panel (g), \emph{kept} \texttt{bar} and unchanged
\texttt{run} edges are copied; \emph{attribute-changed} \texttt{fmt} facts keep
identity while line/anchor attributes rebase by $-2$; \texttt{old} and
\texttt{run}$\rightarrow$\texttt{print} are \emph{deleted}; and \texttt{new},
\texttt{run}$\rightarrow$\texttt{fmt}, and incident facts are \emph{added} after
all vertices exist. Thus one affected symbol can mix actions.

The file-level path reconnects four symbols and five outgoing relationships with
four \texttt{references} and five \texttt{definition} requests. Symbol-level
reuses preserved facts: one \texttt{references} request discovers incoming edges
for \texttt{new}, while four \texttt{definition} requests validate the two
rebased \texttt{fmt} anchors and the new outgoing relationships from
\texttt{run} and \texttt{new}. The example therefore saves four requests; the
count excludes synchronization, \texttt{documentSymbol}, and other protocol
messages. \qref{4}{sec:eval:incremental} reports this saving only for
transitions matching an independently rebuilt target in offline evaluation.

\ifvldbsubmission
\else
The classifier also retains an old backend-invisible symbol when its declaration
line is unchanged. A file-wide line map treats each edge's
\texttt{anchor\_file} as location authority, preserving anchors on unchanged
lines and removing anchors on edited lines before repair. Creating all new
vertices before edge repair ensures that cross-file targets already exist when
relationships are reconnected.
\fi

The patcher synchronizes changed text, re-resolves relocated call anchors with
position-based definitions, discovers incoming edges with references, and
resolves outgoing edges from changed-range semantic tokens;
\detailsref{app:incremental} gives protocol details. The vector updater reuses
content-addressed embeddings and mutates or rebuilds FAISS according to the
declared delta threshold; BM25 still rebuilds.

\ifvldbsubmission
\else
Live LSP work is confined to this maintenance path: changed text is synchronized
before queries, incoming method references are validated through
source-position definitions, and outgoing relationships are reconstructed from
semantic tokens in changed ranges. \qref{4}{sec:eval:incremental} evaluates the
vector updater under a separate artifact-and-replay protocol; the current BM25
path has no integrated delta update.
\fi

Delta paths do not transactionally advance all views, and finishing a repair
without error says nothing about whether it matches a rebuild.
\qref{4}{sec:eval:incremental} therefore conditions each reported speedup on an
offline comparison with an independently rebuilt target, and finds that the
distinction matters: on a held-out cohort the simpler file-level replacement
is admitted more often than symbol repair, so neither path should be served
without that check. The comparison is not on the maintenance path; burst
throughput and cross-view staleness remain outside scope.

\section{Query and Context Runtime}\label{sec:system}

\subsection{Agent Loop and View Loading}\label{sec:system:loop}

Listing~\ref{lst:agent-runtime} shows the two phases. \textsc{Compile} isolates
view construction and publishes artifact paths, status, and capabilities in
$M$. \textsc{Run-Agent} derives \texttt{needed} from the selected skill
metadata. \texttt{load\_views(M, needed)} performs runtime view loading: it
resolves those views in $M$, validates their entries, and opens or deserializes
their prebuilt artifacts as runtime contexts before binding skills.
It neither builds an index nor inserts retrieved code into model history. A
missing required view aborts setup; the runner then filters unavailable skills
and surfaces staleness warnings before exposing tool schemas. This runtime
preflight checks manifest state, not fresh-target equivalence. Thus no builder
is reachable on the measured request path.

\begin{lstlisting}[style=codeminer, float, floatplacement=t,
  caption={Manifest-mediated offline build and online agent setup.
  \texttt{load\_views} opens existing manifest-linked views as runtime contexts; it
  neither builds views online nor adds retrieved code to model history.},
  label={lst:agent-runtime}]
def COMPILE(checkout, requested):
    M = RepoManifest(commit=checkout.commit)
    for kind in requested:
        try:
            artifact, config = BUILDERS[kind].build(
                checkout)
            M.indexes[kind] = IndexEntry(
                path=artifact, status="fresh",
                config=config)
        except Exception as error:
            M.indexes[kind] = IndexEntry(
                status="failed", error=error)
    M.derive_capabilities()
    M.save("repo_manifest.json")
    return M

def RUN_AGENT(issue, M, skill_ids):
    specs = load_skill_metadata(skill_ids)
    needed = requirements(specs)
    # Open existing views; never build online.
    contexts = load_views(M, needed)
    registry = load_skills(specs, contexts)
    runner = AgentRunner(
        registry=registry,
        manifest=M,  # Runtime preflight
    )
    return runner.run(issue)
\end{lstlisting}

\begin{figure}[b]
  \centering
      \includegraphics[width=\columnwidth]{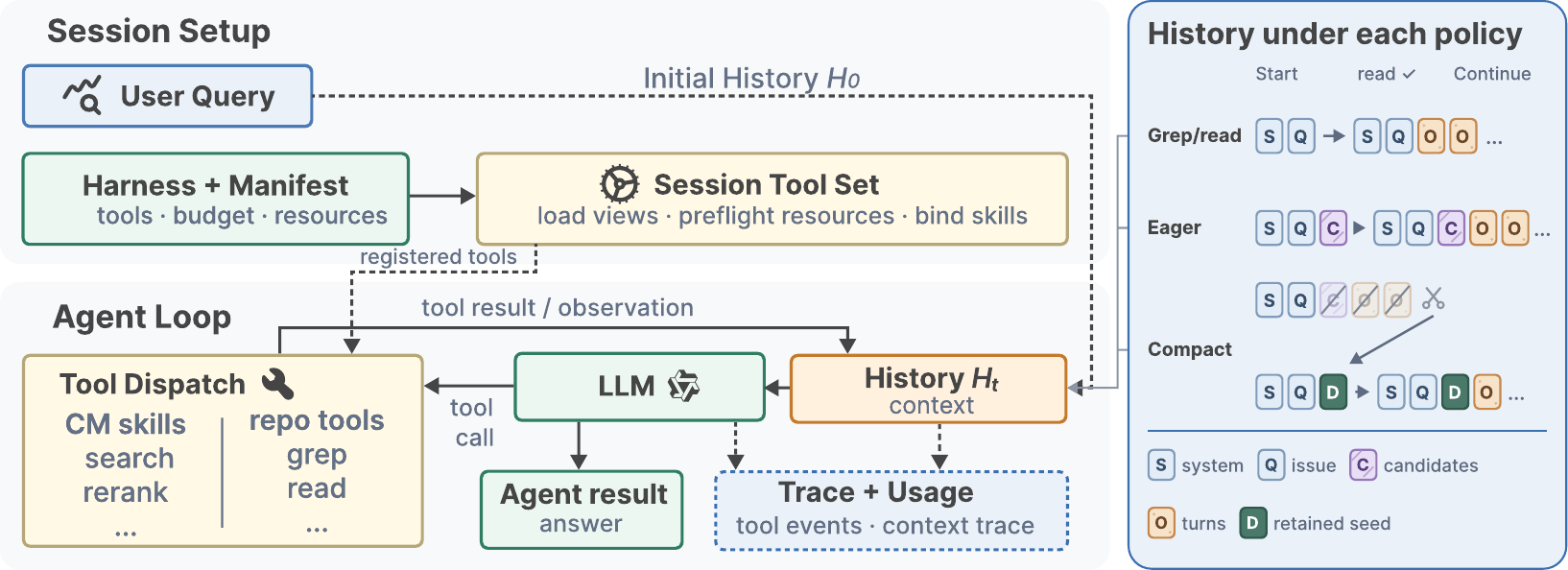}
    \Description{Session setup derives initial history from the user query and
  binds a session tool set after loading and checking manifest resources. Each
  agent turn sends history to the language model, dispatches any tool call, and
  appends the observation while trace and usage accounting records execution.
  The policy panel shows grep/read, eager candidate delivery, and compact's
  one-time rewrite from candidates and observations to a retained seed.}
  \caption{Agent loop, view loading, and evaluated context policies. Manifest
  preflight loads views and binds session tools. S/Q/C/O/D denote system prompt,
  issue, candidates, observations, and retained seed. Arms share issue, tools,
  and budget: Grep/read starts with [S,Q], Eager and Compact with [S,Q,C];
  after the first successful read, Compact alone rewrites once to [S,Q,D] and
  then appends. \qref{5}{sec:eval:context} fixes $k=10$ candidates.}
  \label{fig:agent-runtime-overview}
\end{figure}

The loop follows ReAct's standard reason--action--observation pattern
~\cite{react}. \nickname contributes the view-backed tool surface and the
evaluated delivery policies, not the loop topology.

\ifvldbsubmission
\else
Figure~\ref{fig:agent-runtime-overview} expands the listing's online half:
manifest-driven runtime view loading and preflight construct the session tool
set, while the issue and policy construct $H_0$. Each turn follows
$H_t\rightarrow\textsc{LLM}\rightarrow\textsc{Tool Call}\rightarrow
\textsc{Dispatch}\rightarrow\textsc{Observation}\rightarrow H_{t+1}$; a
terminal answer instead produces the agent result. The trace ledger records
tool and provider provenance alongside usage. The policy changes initial
delivery and, for Compact only, one later history state.
\fi

\subsection{Ranked Query Plans}\label{sec:system:plans}

\begin{figure}[t]
  \centering
      \includegraphics[width=0.86\columnwidth]{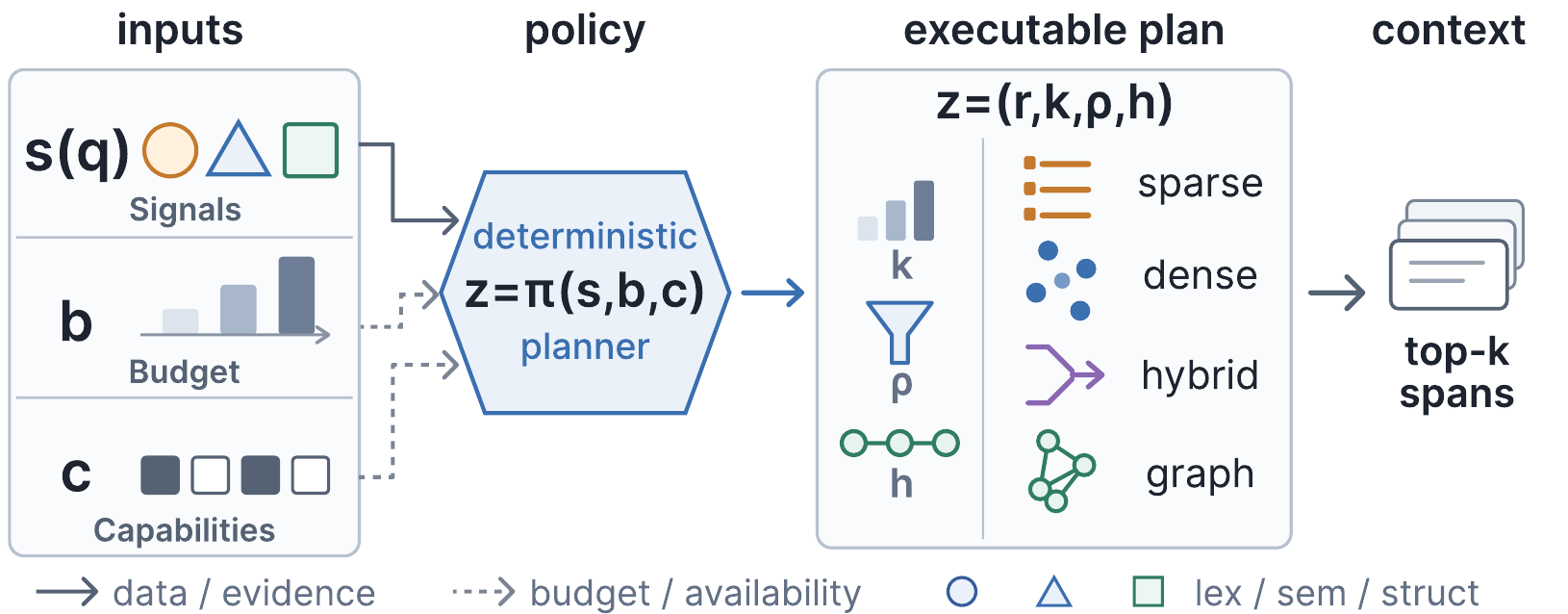}
    \vspace{-0.8ex}

    \small\textbf{(a)} Query-to-plan lowering

    \vspace{0.6ex}
    \includegraphics[width=0.55\columnwidth,trim=16 17 16 16,clip]{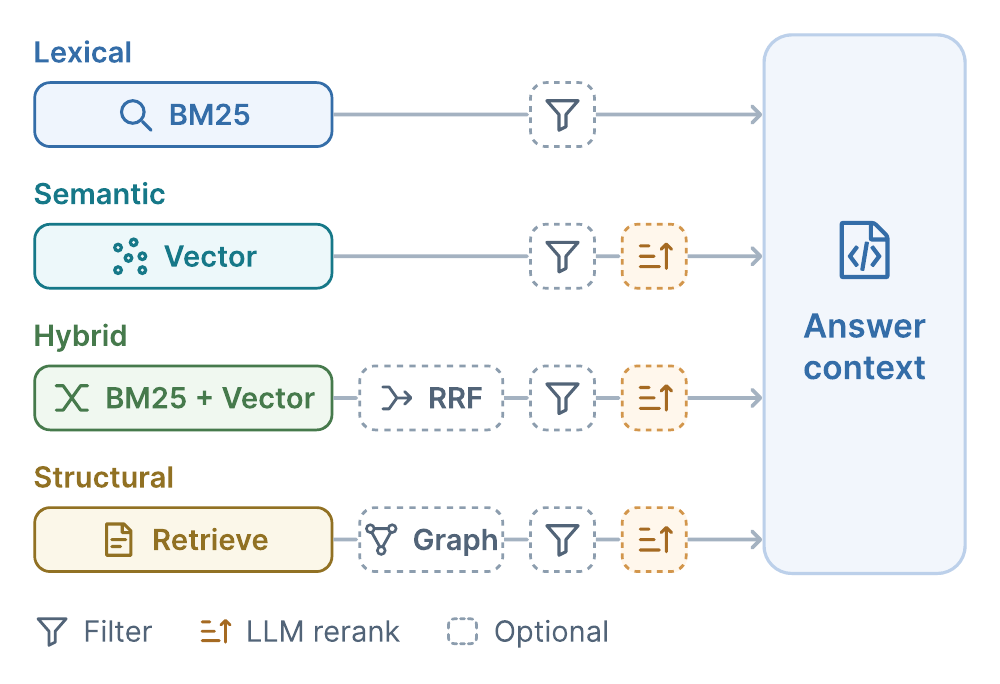}
    \vspace{-0.8ex}

    \small\textbf{(b)} Physical retrieval routes
    \Description{The upper panel lowers query signals, a budget, and available
  capabilities into a route, width policy, optional reranker, and optional
  graph expansion. The lower panel shows lexical, semantic, hybrid, and
  structural routes feeding one answer context. Only the hybrid route fuses
  lists with reciprocal rank fusion; the filter, reranker, and graph expansion
  stages are optional.}
  \caption{Deterministic ranked-query compilation. \textbf{(a)} Signals,
  budget, and capabilities lower to $z=\langle r,k,\rho,h\rangle$.
  \textbf{(b)} Route $r$ selects lexical (A), semantic (B), hybrid (C), or
  structural (D); only hybrid C owns RRF. Width $k$ sets retrieval fan-out and
  the filter cut, $\rho$ is an optional reranker, and graph expansion $h$ is
  legal only on D.}
  \label{fig:rag-lowering}
\end{figure}

Plans may be explicit or selected by a deterministic query/budget heuristic.
\qref{1}{sec:eval:rag} invokes them explicitly and does not evaluate selector
quality. After route execution, the composer projects scored source units to
code blocks; the agent path's top-ten $L_2$ code blocks form
$C_{10}^{\mathrm{ctx}}$ in Sec.~\ref{sec:system:context}.

\ifvldbsubmission
\else
The implemented selector extracts lexical, semantic, and structural signals,
combines them with a latency/quality budget and manifest capabilities, and
lowers them to Figure~\ref{fig:rag-lowering}'s tuple. Because
\qref{1}{sec:eval:rag} invokes plans directly, the paper evaluates physical
operators and not this heuristic's route accuracy.
\fi

\paragraph{Dense search.}
$\mathsf{S}_{\mathrm{dense}}(q,k_{\mathrm{ret}})$ embeds $q$ and returns FAISS
top-$k_{\mathrm{ret}}$ $L_0$ or $L_2$ units. It is
\qref{1}{sec:eval:rag}'s common starting point.

\paragraph{Pointwise reranking.}
$\mathsf{S}_{\mathrm{dense}}\mathsf{R}$ truncates to $k'$ and applies a Qwen3
pointwise reranker before final top-$k_{\mathrm{out}}$.
\qref{1}{sec:eval:rag} sweeps $k'\in\{30,50,100\}$ and model size.

\ifvldbsubmission
\else
Each candidate is paired with the query and scored independently. The pre-rank
cut $k'$ therefore controls both maximum recoverable recall and the number of
model inferences; it is not interchangeable with final output width.
\fi

\paragraph{Graph expansion and experimental fusion.}
Automatic route D expands sparse seeds and optionally reranks without RRF.
\qref{1}{sec:eval:rag} instead instantiates an explicit dense--graph ablation:
$\mathsf{S}_{\mathrm{dense}}\mathsf{X}\mathsf{Fus}$ expands the first $s$ dense
seeds by one semantic-edge hop and combines dense and structural ranks
~\cite{rrf}:
\[
\mathrm{score}(u)=
\frac{w_d}{\kappa+r_d(u)}+\frac{w_g}{\kappa+r_g(u)},
\]
where missing ranks contribute zero. We fix $w_d=1$, $\kappa=60$, select $w_g$
on repository-disjoint tuning repositories, report the ablation on held-out
repositories, and optionally append $\mathsf{R}$. Explicit fusion prevents
graph candidate generation from being mistaken for dense rescoring.

BM25, regex, and Zoekt serve lexical requests but are not compared in
\qref{1}{sec:eval:rag}; we do not infer their relative quality from the
dense-plan experiment.

\subsection{Static and Live Navigation}\label{sec:system:lsp}

Definition/reference requests contain a repository-relative file, position,
options, and limit. MCP converts its one-based line once before the zero-based
provider interface. The server instantiates static graph/occurrence serving;
agent and evaluation paths can instead inject live JSON-RPC. Both normalize to
one location schema, and \qref{3}{sec:eval:lsp} projects it to the
path/start-line set in Sec.~\ref{sec:framework:ops}.

Metadata records backend, capability, commit, and granularity, so a caller
can apply the admitted routes of Figure~\ref{fig:nav-agreement} per request;
\qref{3}{sec:eval:lsp} reports agreement per backend and capability next to
operational cost. The MCP adapter currently answers every capability
statically; applying those routes and the manifest's staleness flag there is
a routing change on metadata the response already carries.

\subsection{Context Policies}\label{sec:system:context}

All arms share issue, tools, budget, turn cap, model parameters, repository
commit, and answer format. Let $C_k^{\mathrm{ctx}}$ denote the frozen top-$k$
embedding-ranked $L_2$ code blocks; \qref{5}{sec:eval:context} fixes $k=10$.
\emph{Grep/read} starts without candidates, while \emph{eager} and
\emph{compact} receive the same $C_{10}^{\mathrm{ctx}}$. Let $j$ index the
model invocation whose completed tool batch first contains a successful read.
After that batch produces $H_j$, compact applies the one-time transition
\begin{equation}
  H_j=[s,q\mathbin\Vert C_{10}^{\mathrm{ctx}},e_{1:j}]
  \;\longrightarrow\;
  \widehat H_j=[s,q\mathbin\Vert d_j],
  \label{eq:compact-transition}
\end{equation}
where $s$ is the system prompt, $q$ the clean issue, $e_{1:j}$ the discarded
exploration transcript, and $d_j$ a direction seed containing deduplicated
read paths, the latest successful read result in full, and a bounded prefix of
the latest nonempty assistant message. Invocation $j+1$ consumes
$\widehat H_j$, and later invocations append normally.
Thus the transition creates one new cache prefix, while every subsequent
invocation extends that fixed prefix and can reuse its KV cache. The rewrite is
deterministic and invokes no summarization model; it changes retention, not
retrieval. \qref{5}{sec:eval:context} sums the full trajectory, including
tokens spent before the rewrite, under the quality threshold in
Sec.~\ref{sec:framework:cost}.

\ifvldbsubmission
\else
Candidate assembly remains experiment-side; bounded history and one-time
compaction are reusable runner mechanisms. At the transition, only the latest
nonempty assistant message is shortened---to its first 600 characters---as a
direction cue; the retained read content and final answer remain intact.
\fi

\paragraph{Real-repair policies.}
The five policies that \qref{7}{sec:eval:repair} runs on real repairs differ
in exactly two things: what the agent sees before its first turn, and what it
keeps afterwards. Everything else is shared, including the issue, the Bash
tool, every limit, the model parameters, the base commit, and the
patch-submission format. \emph{On-demand} starts with the issue alone.
\emph{BM25 eager} and \emph{CodeNib eager} append the top ten lexical or
dense $L_2$ candidates (Qwen3-Embedding-0.6B, exact search), and \emph{Aider
eager} appends the upstream RepoMap~\cite{aider}; each payload is capped at
16\,KiB, and the candidate packs keep at most ten nonoverlapping source
blocks. Delivered bytes therefore differ across these policy-level
comparisons. \emph{CodeNib compact} receives exactly the dense eager
candidate bytes; once the agent has performed its first attested direct source
read, the runtime replaces the candidate block in every later model query with
a content-hash receipt, and the issue and the rest of the exploration, editing,
and testing history stay intact. A read is attested when a literal
\texttt{cat} or \texttt{sed} command succeeds and its output matches the file
bytes checked inside the task container; an ambiguous command leaves the
history unchanged. The transition is deterministic and makes no summarization
call, and all usage before it stays in the record. Initial context describes
the base snapshot; after an edit, the agent reads and tests the live worktree,
and these policies provide neither automatic context refresh nor extra
interactive tools.

\FloatBarrier

\section{Implementation}\label{sec:impl}

\nickname is implemented in Python 3.10+ and comprises compiler, graph, index,
agent-runtime, and serving-adapter packages, including stdio MCP. Evaluation
runners own sampling, arm wiring, timing, and post-run output comparisons; reusable
operators and history policies remain in core packages.

\subsection{Repository View Compiler and Runtime}

\texttt{IndexCompiler} implements the initial-build path and records builder
failures while retaining successful artifacts. \texttt{GraphPatcher} and
\texttt{CodeVectorStore.delta\_update} implement the two evaluated maintainer
paths.
\texttt{ServerContext} loads available resources once for process-resident
serving; agents load only skill-required views. Versioned graph pickles fail with
a rebuild instruction, and a C++/pybind SCIP decoder is tested for schema parity.

LSP-shaped skills call an injected static or live provider and record the chosen
implementation. MCP currently instantiates static serving; live JSON-RPC is an
agent/evaluation path, not an automatic fallback.

\ifvldbsubmission
\else
For MCP, \texttt{ServerContext} loads vector, BM25, graph, and Zoekt resources
once at process startup. The agent runtime instead owns provider selection,
tool schemas, history state, context policy, and per-run traces. LSP-shaped
skills receive an injected provider, so static and live implementations share
the same returned-location schema while traces retain which backend actually
served a request.
\fi

\subsection{Tool and Serving Adapters}

Agent skills and stdio MCP wrap the same operators and normalized results with
separate model-facing schemas. MCP exposes semantic, BM25, regex, and Zoekt
search; dependency, definition, reference, and route tools; and
\texttt{get\_manifest}. Missing optional indexes return explicit errors.
Experiments invoke the operators through benchmark or agent harnesses; MCP is
not an experimental factor.

Search returns snippets, while LSP-shaped tools return compact locations to keep
high-fanout references small. Authentication, multi-tenancy, and network
transport are outside the evaluated stdio implementation.

\ifvldbsubmission
\else
The concrete search surface is \texttt{search\_semantic},
\texttt{search\_bm25}, \texttt{search\_regex}, and
\texttt{search\_zoekt}; structural tools expose dependency subgraphs,
definitions, references, and routes. \texttt{get\_manifest} reports commit,
languages, artifact status, and capabilities. Missing optional indexes surface
an explicit tool error rather than a silent fallback under the same name.
\fi

\subsection{Runtime Entry Points and Observability}

The public \texttt{query} entry point accepts exactly one of a repository path,
caller-opened contexts, or a manifest. Only repository-path mode may compile and cache views
before constructing \texttt{AgentRunner}. Evaluation uses manifest mode, which
opens artifacts named by $M_c$ and cannot build inside the measured loop.

\texttt{AgentRunner} separates index-backed \texttt{SkillRegistry} capabilities
from ordinary \texttt{ToolRegistry} primitives. Loading validates required
artifacts; \texttt{ResourceGuard} filters unavailable skills and warns
on staleness. This setup checks recorded runtime state, not
\qref{4}{sec:eval:incremental}'s independent fresh-build equality.

\texttt{AgentRunTrace} records ordered events and context state, and
\texttt{UsageTracker} records provider tokens. Evaluation consumes these records
without embedding scoring, policy selection, or fresh-build comparisons in the
trace schema.

\section{Evaluation}\label{sec:eval}

We evaluate \nickname from retrieval operators to the agent loop. Five
questions measure what the serving routes deliver against the alternatives
they would replace; two more place the system behind agents we did not write,
on real repairs, and score what those agents produce:

\begin{description}
  \item[\textbf{\qref{1}{sec:eval:rag}}] What quality--latency tradeoffs arise when dense
  retrieval is composed with structural expansion and pointwise reranking?
  \item[\textbf{\qref{2}{sec:eval:index}}] How do dense-index designs trade construction time,
  search latency, and retrieval quality?
  \item[\textbf{\qref{3}{sec:eval:lsp}}] For graph-anchored symbol-navigation requests, how often
  does a static index reproduce the live LSP's normalized path/start-line set,
  and what marginal latency does it save on matching requests?
  \item[\textbf{\qref{4}{sec:eval:incremental}}] How often do incremental graph and vector updates match
  independently rebuilt targets, and what speedup do matching transitions
  achieve?
  \item[\textbf{\qref{5}{sec:eval:context}}] What token--quality tradeoffs arise from context-delivery
  policies across agent models and workload slices?
  \item[\textbf{\qref{6}{sec:eval:provider}}] With agent policy, tools, model,
  tasks, and limits fixed, how does replacing the repository provider behind
  an existing localization agent affect localization, availability, and
  serving cost?
  \item[\textbf{\qref{7}{sec:eval:repair}}] Under a fixed repair harness on
  real issues, how do context selection, initial delivery, and retention
  affect resolution, usage, time, and failures?
\end{description}

The questions follow the system boundary rather than assuming one common
quality metric. \qref{1}{sec:eval:rag} tests ranked plans;
\qref{2}{sec:eval:index} separates dense construction, search, and
approximation; \qref{3}{sec:eval:lsp} tests navigation; and
\qref{4}{sec:eval:incremental} tests maintenance. The mixed trace checks
compiler--runtime composition and stage costs; \qref{5}{sec:eval:context}
measures history delivery on a synthetic localization workload.
\qref{6}{sec:eval:provider} exercises manifest loading, lexical search, and
graph navigation inside LocAgent, and \qref{7}{sec:eval:repair} measures
context policies on real issue repairs scored by the official tests. MCP is
not an experimental factor: no arm varies the serving protocol or attributes
quality or latency to MCP itself.

\begin{figure*}[t]
  \centering
  \includegraphics[width=\textwidth]{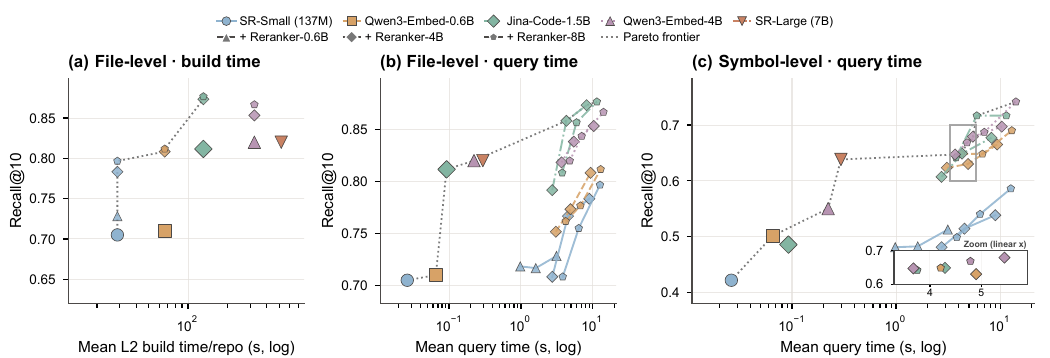}
  \Description{Three scatter plots compare retrieval recall with dense-index
  construction or query time. Points distinguish five embedding models and
  connected candidate-budget sweeps for three pointwise rerankers.}
  \caption{Embedding and pointwise-reranker operating points on the
  100-snapshot corpus; Qwen-4B/8B score means use $n=99/98$ at $k'=30/50$,
  and all other score means use $n=100$.
  \textbf{(a)} File Recall@10 versus mean L2 index-build time for the callable
  index used by retrieval.
  \textbf{(b)} File and \textbf{(c)} symbol Recall@10 versus mean query time.
  The SweRank embeddings SR-Small and SR-Large have approximately 137M and 7B
  stored parameters, respectively.
  Connected rerank points sweep pre-rerank cuts
  $k'\in\{30,50,100\}$; dotted lines show the empirical non-dominated points
  within each panel.}
  \label{fig:rag-pareto}
\end{figure*}

\subsection{Experimental Setup}

\begin{table}[t]
  \caption{Evaluation matrix and frozen record counts.}
  \label{tab:eval-matrix}
  \centering
  \scriptsize
  \setlength{\tabcolsep}{2.6pt}
  \renewcommand{\arraystretch}{0.98}
  \providecommand{\evalstudywidth}{1.15cm}
  \begin{tabularx}{\columnwidth}{@{}>{\raggedright\arraybackslash}p{\evalstudywidth}>{\raggedright\arraybackslash}X>{\raggedleft\arraybackslash}p{2.2cm}@{}}
    \toprule
    \textbf{Study} & \textbf{Compared arms/providers} & \textbf{Frozen records} \\
    \midrule
    \rowcolor{black!5}
    \qref{1}{sec:eval:rag} retrieval & 5 dense embedders; partial 3-reranker matrix & 100 snapshots; rerank $n=98$--100 \\
    \qref{1}{sec:eval:rag} graph & Dense/graph, $\pm$ fixed 4B; tune/freeze $w_g$ & 58 tune/42 held-out; 15/10 repos \\
    \rowcolor{black!5}
    \qref{2}{sec:eval:index} index & $L_0/L_2$ $\times$ 5; Qwen-0.6B Flat/IVF/HNSW & 500 pairs; 100 ANN \\
    \qref{3}{sec:eval:lsp} nav. & Static index versus 5 live LSPs & 1,000 requests; 100 snapshots \\
    \rowcolor{black!5}
    \qref{4}{sec:eval:incremental} maint. & Graph rebuild/file/symbol; vector rebuild/incremental & 40 (8 repos); 33 G/31 V source-changing \\
    Lifecycle & Materialize/load/mixed trace; isolated/resident & 25 snapshots; 1,050 requests (3 reps) \\
    \rowcolor{black!5}
    \qref{5}{sec:eval:context} context & Grep/read, Eager, Eager+Compact; 5 models & 7,500 core trajectories \\
    \qref{6}{sec:eval:provider} provider & LocAgent native / CodeNib; Qwen3.5-27B; 11 repos & 30 tasks; 180 cells \\
    \rowcolor{black!5}
    \qref{7}{sec:eval:repair} repair & Five context policies; Haiku 4.5; 200 of 500 Verified & 200 tasks; 1,100 cells \\
    \midrule
    Embedding & SweRank-Small (137M)/Large (7B)~\cite{swerank}; Qwen3-Embed-0.6B/4B~\cite{qwen3-embed}; Jina-Code-1.5B~\cite{jinacode} & \qref{1}{sec:eval:rag}--\qref{2}{sec:eval:index} \\
    \rowcolor{black!5}
    Reranking & Qwen3-Reranker-0.6B/4B/8B~\cite{qwen3-embed} & \qref{1}{sec:eval:rag} \\
    Agents & Claude Haiku 4.5~\cite{claude-haiku-45}; Qwen3.5-9B/27B~\cite{qwen35}; Gemma 4-12B-IT~\cite{gemma4}; Gemini 2.5 Flash~\cite{gemini25} & \qref{5}{sec:eval:context}--\qref{7}{sec:eval:repair} \\
    \bottomrule
  \end{tabularx}
\end{table}

\paragraph{Workloads.}
Table~\ref{tab:eval-matrix} aligns each study with its compared arms and frozen
record counts. \qref{1}{sec:eval:rag}--\qref{3}{sec:eval:lsp} use frozen
CodeNib Base: Python snapshots come from SWE-bench
Verified~\cite{swebench-verified}; the other four language groups come from
SWE-bench Multilingual~\cite{swebench-multilingual}.
Its five repositories per language group span file-count percentiles, and its
instances span Opus-assigned difficulty strata. The query is the issue body;
targets are files and pre-patch L2 blocks intersected by the developer patch.
Only graph fusion partitions Base for tuning; the other
\qref{1}{sec:eval:rag} operating points use all 100 snapshots.
\Detailsref{app:datasets} records dataset construction.

\begin{figure*}[t]
  \centering
  \includegraphics[width=0.98\textwidth]{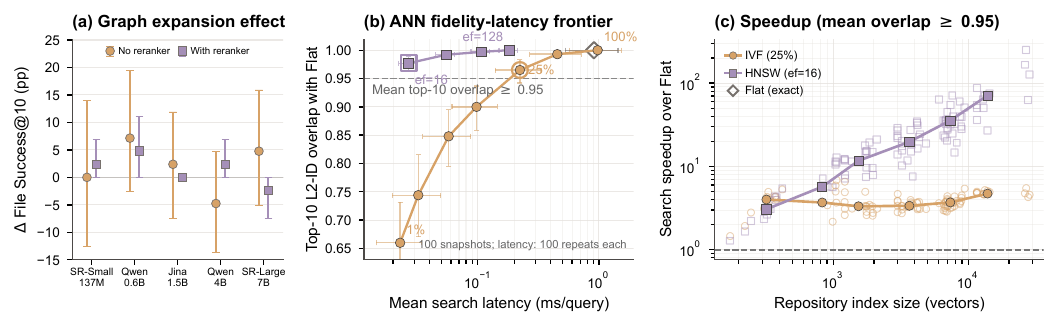}
  \Description{Three panels separate paired graph-minus-dense changes in
  all-target file coverage from ANN top-10 document-identity overlap with exact
  Flat and the per-snapshot speedups of mean-overlap-qualified configurations.}
  \caption{Task-level graph and physical ANN ablations.
  \textbf{(a)} Paired graph-minus-dense $\Delta$File Success@10 (all target
  files in the top-10 distinct-file prefix).
  \textbf{(b--c)} 100 Qwen3-Embedding-0.6B $L_2$ indexes.
  \textbf{(b)} Mean exact-Flat top-10 $L_2$-ID overlap versus search latency;
  outlines mark each family's fastest configuration at mean overlap
  $\geq0.95$.
  \textbf{(c)} Faint marks are snapshot speedups; six large marks per curve are
  log-size-bin medians joined as guides.}
  \label{fig:retrieval-ablation}
\end{figure*}

Agent experiments use CodeNib Synthesis. Anthropic distinguishes pinned model
IDs from convenience aliases~\cite{claude-model-versioning}; this artifact
records only the provider alias \texttt{opus}, not its immutable resolution.
The same observed alias generates candidates and judges them in a separate pass,
then deterministic checks reject duplicate identifiers, empty targets, and
language-inconsistent files. One trajectory per query, policy, and model gives
7,500 trajectories: 2,500 query--model cells with three policies each. Intervals
measure snapshot variation, not repeated model sampling. Haiku's compact arm
was backfilled after its baseline/eager sweep under the same observed model ID
and harness configuration;
both Qwen sizes and Gemma 4-12B use complete frozen three-arm matrices. Gemma is
served by vLLM 0.25.1 from a pinned Hub revision; Gemini 2.5 Flash is accessed
through Vertex AI with thinking disabled. \Detailsref{app:agent-protocol}
records provider controls, observed revisions, and remaining provider-time drift.

The lifecycle trace reuses the 25 synthesis snapshots and fixes 20 source-query
sessions per snapshot. Each query runs once through Qwen3-Embedding-0.6B and
once through BM25; two additional deterministic definition probes include
static navigation. The resulting count is therefore $20(1+1)+2=42$ service
requests, rather than a separately chosen scale. The selected definitions are
nonempty, stable, and return the same normalized path/start-line sets from the
static and live providers in every calibration comparison. References are
excluded because their normalized sets drifted during calibration. This mix is
not an observed agent-tool distribution. Runs use isolated quiescent checkouts
and fresh runtime processes, but warm machine caches;
\Detailsref{app:lifecycle-boundary} specifies the boundary.

Incremental maintenance follows five first-parent commit transitions in each of
eight repositories ($n_{\mathrm{share}}=5$), with two repositories in each of
Go, Python, Rust, and TypeScript/JavaScript. The
graph scope includes language-source tests; the vector builder's declared
scope excludes test-like paths. This yields 33 graph and 31 vector
source-changing transitions. No-source transitions remain in the artifact but
are excluded before computing update ratios. Graph arms start from the same
base artifact and compare a fresh target rebuild, file replacement, and symbol
repair; vector arms compare a fresh target rebuild with content-addressed
embedding reuse and FAISS delta update. The fresh target supplies both the
rebuild baseline and the offline equivalence reference; its construction and
comparison are not added to incremental update latency. Each incremental graph
arm reuses one LSP process per repository sequence, whereas each graph rebuild
constructs an independent target. The study spans eight
repositories but does not estimate production arrival rates or concurrent
update throughput.

\paragraph{Ground truth.}
We project patch line ranges onto pre-patch L2 chunks using the tree-sitter
hierarchy of Sec.~\ref{sec:index}. Retrieval runs against the base commit, so only
pre-patch files and blocks can be targets. Added files and added-only symbols
are excluded because they have no retrievable source unit in that snapshot.
Edited blocks are a localization proxy, not an assertion that every useful
supporting block was changed by the patch.

\paragraph{Metrics and inference.}
We use the operation-specific metrics summarized in
Sec.~\ref{sec:framework:cost}; \detailsref{app:metrics} gives their formal
definitions and examples. Recall and match-rate estimates macro-average
per-instance scores. File Success is an all-target indicator, Neighbor Recall
measures exact-Flat fidelity rather than patch relevance, and
$\mathrm{AnswerRecall@5}$ scores the first five deduplicated source spans
committed in the final answer. An answer with no usable in-scope committed span
receives zero recall.
Agent intervals use 10,000 bootstrap samples clustered by repository snapshot.
Graph effects use 20,000 paired repository-clustered bootstraps:
pointwise in Figure~\ref{fig:retrieval-ablation}(a) and max-deviation
simultaneous over ten cross-embedding contrasts. ANN means use 20,000 clustered
resamples. Latencies are
warm wall-clock measurements unless stated otherwise. Lifecycle medians and
accounting projections use 10,000 snapshot-level percentile bootstraps.
Incremental transitions use the offline output comparisons in
Eqs.~\eqref{eq:graph-fact-equivalence} and
\eqref{eq:graph-serving-equivalence}; graph maintenance also checks the
changed-file projection. Vector maintenance requires exact document identities,
numerical vector equivalence, and exact ordered Flat top-$k$ replay. Transition
ratios follow Eq.~\eqref{eq:update-speedups}. Successful-arm ratios remain visible, but
conditional summaries retain only source-changing transitions that match the
independently rebuilt target.
\qref{6}{sec:eval:provider} scores file and qualified-symbol localization
with the task as its analysis unit and repeats nested inside, under 20,000
repository/task/repeat hierarchical bootstrap draws; a failed execution
receives empty predictions and stays in the denominator.
\qref{7}{sec:eval:repair} scores official repair outcomes, and its inference
was frozen in advance: 20,000 task-paired bootstrap draws, an exact McNemar
test for the primary contrast, and Holm adjustment for the four secondary
resolution tests, with repository-hierarchical intervals as sensitivity
analyses. It keeps every task in its denominator, carrying an unknown outcome
as a range over both values rather than dropping the task, and reports
incomplete usage as an observed lower bound. In both agent studies the runner
sees only the issue ID, repository, base commit, and problem statement; golden
patches and test labels stay inside the independent evaluator, and both sides
use the same immutable per-task container image.

\paragraph{Execution environment and models.}
Unless stated otherwise, local inference uses one NVIDIA H100 PCIe GPU
(80\,GB), and CPU experiments use two Intel Xeon Gold 5416S processors; FAISS
search uses one CPU thread. Agent temperature is zero, and both eager-context
policies share the frozen Qwen3-Embedding-0.6B top-10 $L_2$ result. Historical
retrieval and Haiku/Qwen agent runs record model IDs, not immutable revisions;
the artifact recovers available cache revisions, while Haiku hardware and
revision remain unobserved. The Gemma run pins its Hub commit, serving template,
and vLLM version; the Gemini run records its Vertex configuration and access
date, but not an immutable provider revision. The lifecycle trace separately
pins its embedding snapshot and build configuration at execution time: batch
size 32 and a 2,048-token document cap fixed before the final batch.
\Detailsref{app:retrieval-config} gives model
architectures, prompts, truncation, batching, and operator hyperparameters.

\paragraph{AI-assisted research.}
Codex assisted with protocol design, implementation and analysis code, failure
diagnosis, and table and figure preparation for the held-out maintenance,
provider-substitution, and real-repair studies. Source snapshots, execution
records, targeted tests, and replay document those additions separately from
the dataset-generation models and the evaluated agents.

\subsection{Q1: Retrieval-Plan Tradeoffs}\label{sec:eval:rag}

\paragraph{Dense retrieval and reranking.}
Figure~\ref{fig:rag-pareto} maps empirical operating points and non-dominated
frontiers under three distinct budgets rather than selecting one ``best''
pipeline. Mean dense-query time spans 26--295\,ms. At $k=10$, file recall rises
from 0.705 to 0.820 across embedding families, while symbol recall spans
0.422--0.638. Embedding choice shifts both recall and latency; the measured
higher-recall operating points generally pay at both build and query time.

Pointwise reranking moves the high-recall frontier at a much larger online cost.
For example, Jina plus the 4B reranker at $k'=50$ reaches 0.858 file Recall@10
in 4.29\,s, compared with Jina dense at 0.812 in 92\,ms: a 4.6-point gain at
46.6$\times$ latency. At symbol level, the
highest measured recall is 0.742 for Qwen3-Embedding-4B plus the 8B reranker at
$k'=100$, requiring 14.1\,s. Across the three measured cuts, larger $k'$
generally raises recall and latency. On this host, mean dense-query
latency stays below 300\,ms across embeddings, whereas reranking requires
seconds. The measured file- and symbol-recall maxima occur at different
embedder--reranker pipelines.

\paragraph{Graph expansion.}
To isolate the graph operator, dense and graph arms share the same embedding
index and per-instance candidate budget. We retrieve 300 L2 chunks, expand the
first ten seeds over incoming and outgoing reference edges, and fuse semantic
and structural ranks with weighted RRF~\cite{rrf}. On the 58-snapshot,
15-repository tuning partition, we sweep
$w_g\in\{0.25,0.5,0.75,1,1.25,1.5,2\}$ and select lexicographically by File
Success@10, then @5 and @1. This uniquely selects $w_g=0.5$, which we freeze
before reporting effects on 42 snapshots from 10 disjoint repositories, avoiding
evaluation on the labels that selected the weight.

In Figure~\ref{fig:retrieval-ablation}(a)'s before-rerank series, point estimates
range from
$-4.8$ points (Qwen3-Embedding-4B) to $+7.1$ ($[-2.6,+19.5]$;
Qwen3-Embedding-0.6B). Every model-level interval in both series includes zero;
so do all ten familywise cross-embedding intervals. The study establishes
neither a general gain nor an embedding-specific routing rule. Expansion adds
15--39\,ms to median non-reranked latency, while reranked arms cost
2.7--6.8\,s in total. Panel (a) uses all-target File Success@10 rather than mean
File Recall@10, so its effects are not shifts of
Figure~\ref{fig:rag-pareto}'s frontier. Expansion remains an optional plan
requiring deployment-specific validation.

\ifvldbsubmission\else
\textbf{Answer to \qref{1}{sec:eval:rag}.}
Reranking is a controllable seconds-scale tradeoff; graph expansion remains
unresolved, so $\rho$ and $h$ require separate plan policies.
\fi

\begin{figure*}[t]
  \centering
  \includegraphics[width=0.98\textwidth]{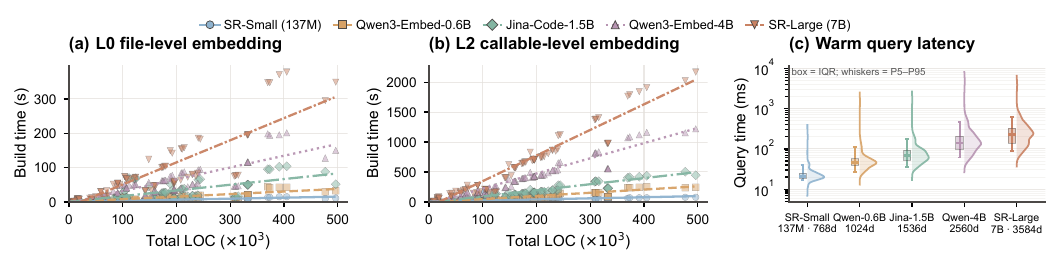}
  \Description{Two construction-time scatter plots show L0 and L2 index build
  time against repository size for five embedding models. A third panel shows
  their warm dense-query latency distributions.}
  \caption{Dense-view construction and query profiles over 100 snapshots.
  \textbf{(a)} L0 file and \textbf{(b)} L2 callable construction time versus
  repository lines of code (LOC); lines are per-model fits. \textbf{(c)} Warm end-to-end $L_2$
  dense-query latency; labels give parameter scale and output dimension $d$.
  Half violins show distributions, boxes show IQR, and whiskers show P5--P95.}
  \label{fig:index-build}
\end{figure*}

\subsection{Q2: Index Construction and Search}\label{sec:eval:index}

\qref{1}{sec:eval:rag} measures composed retrieval plans.
\qref{2}{sec:eval:index} separates offline dense-index construction from online
vector-search cost and approximation error.

\paragraph{Build-time scaling.}
Figure~\ref{fig:index-build} separates repository size, granularity, and model
choice across 500 L0/L2 pairs. Median construction spans 3.8--56.7\,s for L0
and 19.3--285.0\,s for L2; within a model, L2/L0 is 5.0--6.4$\times$.
All ten LOC fits are positive; in parameter-count rank, L0 and L2 slopes rise
from 0.029--0.644 and 0.186--4.282\,s/kLOC. This matches the first-order model
$T_{\mathrm{build}}\approx T_{\mathrm{enc}}(m,W_\ell)+O(N_\ell d_m)$: LOC
proxies encoded-token work $W_\ell$, while Flat writes $N_\ell$ vectors of
dimension $d_m$. The trend is consistent with this decomposition; architecture,
chunking, token lengths, and batching explain
residuals, so the fits are descriptive rather than universal laws. Timers start
after chunking and exclude checkout, compiler/SCIP execution, and graph decoding.

Panel (c) shows median dense-query latency rising from 20.9 to 233.8\,ms.
Output dimension (768--3584) and parameter count have the same rank order, while
the timed query combines encoding, an $O(N_\ell d_m)$ Flat scan, and fixed-$k$
result materialization; separately timed index loading is excluded. The trend
is descriptive, not causal; the FAISS ablation isolates scan organization.

\paragraph{Index-family ablation.}
With vectors and queries fixed, we compare exact
\texttt{IndexFlatIP}~\cite{faiss}, inverted-file \texttt{IndexIVFFlat}, and
\texttt{IndexHNSWFlat}~\cite{hnsw}. IVF sweeps probe fractions
of 1\%, 2\%, 5\%, 10\%, 25\%, 50\%, and 100\%. HNSW fixes $M=32$ and
$ef_{\mathrm{construction}}=200$ while sweeping
$ef_{\mathrm{search}}\in\{16,32,64,128\}$. Each method--snapshot cell uses 10
warmups and 100 timed searches. In this descriptive sweep, each approximate
family contributes its fastest configuration with across-snapshot mean Neighbor
Recall@10 at least 0.95.
Figure~\ref{fig:retrieval-ablation}(b--c) shows that
HNSW with $ef_{\mathrm{search}}=16$ reaches
0.977 neighbor Recall@10 and reduces mean FAISS search from 0.910 to
0.0268\,ms (33.9$\times$). IVF at 25\% probes reaches 0.965 at 0.223\,ms
(4.1$\times$). Both preserve Flat's patch-target-file Recall@10 on every
snapshot (macro mean 0.620).

These ratios are component-local: HNSW reduces mean FAISS search by 0.883\,ms,
whereas the complete dense-query median is 45.1\,ms. It also raises mean
index-only build time from
6.4\,ms (Flat) to 2.00\,s and
mean serialized size from 21.74 to 23.18\,MiB; the IVF means are 0.889\,s and
22.33\,MiB.
At the measured means, extra construction amortizes after about 1,300 searches
for IVF and 2,300 for HNSW; these index-local crossovers exclude loading and updates.
Flat is therefore simpler at this scale; larger indexes and query rates remain
unmeasured, and the FAISS ratios are not end-to-end speedups.

\ifvldbsubmission\else
\textbf{Answer to \qref{2}{sec:eval:index}.}
At this scale, embedding and granularity dominate; ANN's sub-millisecond saving
justifies additional construction only under sufficient query reuse.
\fi

\subsection{Q3: Static versus Live Symbol Compatibility and Latency}\label{sec:eval:lsp}

\qref{1}{sec:eval:rag}--\qref{2}{sec:eval:index} concern ranked retrieval.
\qref{3}{sec:eval:lsp} turns to symbol navigation, which returns a set of source
locations rather than a ranking, and asks for which requests the static
occurrence index returns what the live language server returns.

\begin{figure}[t]
  \centering
  \includegraphics[width=0.95\columnwidth]{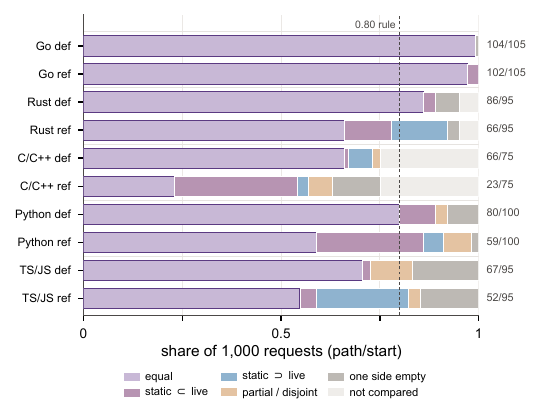}
  \Description{Stacked horizontal bars, one per language and capability, show
  the share of requests on which the static and live answers are equal and how
  the remainder differ: static strict subset, static strict superset, partial
  or disjoint overlap, one side empty, or not compared. A dashed line marks the
  0.80 admission rule and labels give matched over compared counts.}
  \caption{Static/live agreement per language and capability over 1,000
  uncapped, graph-anchored requests (corrected occurrence indexes). Labels
  give matched/compared path/start-line sets; the dashed line is the 0.80
  admission rule. Static answers that are a strict subset of the live answer
  dominate the mismatches for C/C++ and Python references.}
  \label{fig:nav-agreement}
\end{figure}

\begin{figure*}[t]
  \centering
  \includegraphics[width=0.98\textwidth,trim=0 5pt 0 0,clip]
    {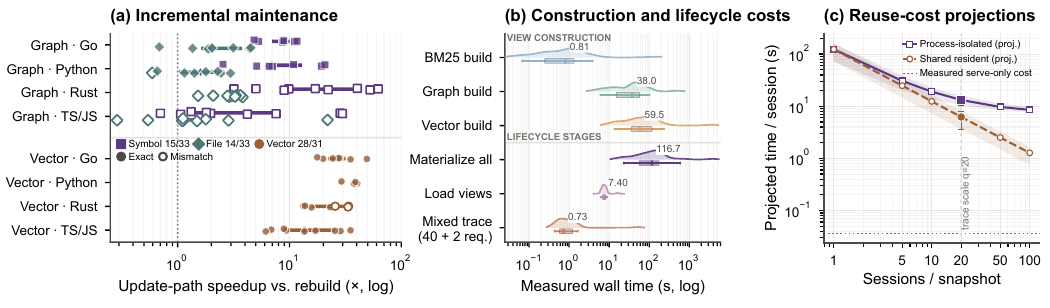}
  \vspace{-4pt}
  \Description{A transition-level dot plot shows graph symbol repair, graph
  file replacement, and vector delta-update speedups with filled marks for
  outputs that exactly match an independent rebuild; legend ratios summarize
  exact matches versus measured source-changing transitions. Two panels show
  component construction costs, complete lifecycle-stage costs, and projected
  per-session cost under process-isolated or shared runtime reuse. In the final
  panel, open markers show six accounting projections connected by lines, while
  filled markers identify the controlled trace scale.}
  \ifvldbsubmission
    \caption{Incremental maintenance and lifecycle accounting.
    \textbf{(a)} Update speedup; fill denotes exact output match, labels give
    exact/measured transitions, and bars show IQR/median. \textbf{(b)}
    View-construction components and complete, non-additive
    materialize--load--serve stages; the trace records $20$ paired dense/BM25
    queries plus two definitions (42 requests). \textbf{(c)} Projections from
    measured $B,L,S$ at $q\in\{1,5,10,20,50,100\}$; open marks are projected
    medians joined as guides, bands are bootstrap 95\% intervals, and filled
    $q=20$ marks identify the controlled trace scale.}
  \else
    \caption{Incremental maintenance and lifecycle accounting.
    \textbf{(a)} Update speedup; fill denotes exact output match, labels give
    exact/measured transitions, and segments/ticks show IQR/median.
    \textbf{(b)} View-construction components and complete, non-additive
    materialize--load--serve stages; the trace records 20 paired dense/BM25
    queries plus two definitions (42 requests). \textbf{(c)} Projections from
    measured $B,L,S$ at $q\in\{1,5,10,20,50,100\}$; open marks are projected
    medians joined as guides, bands are bootstrap 95\% CIs, and filled $q=20$
    marks identify the controlled trace scale.}
  \fi
  \label{fig:maintenance-lifecycle}
\end{figure*}

\begin{figure*}[t]
  \centering
  \includegraphics[width=0.98\textwidth,trim=0 7pt 0 7pt,clip]
    {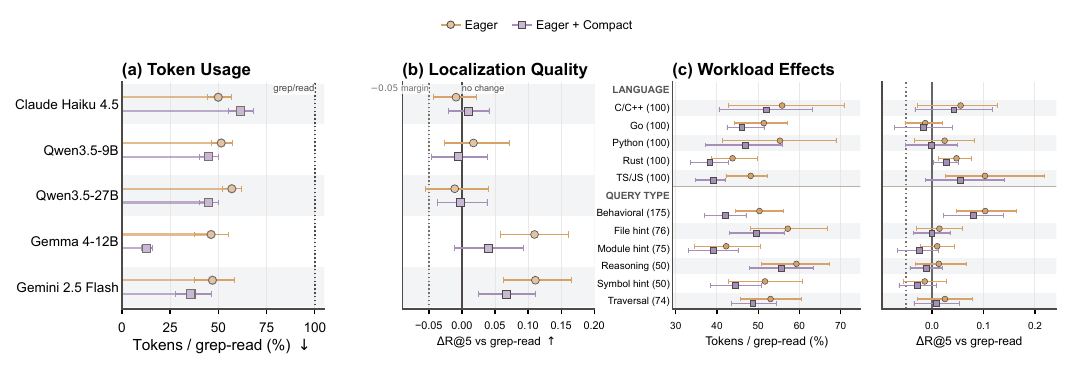}
  \Description{Model-level token and committed-answer Recall-at-5 effects for
  Eager and Eager plus Compact against paired grep/read, followed by pooled
  language and query-type effects.}
  \ifvldbsubmission
  \caption{Agent context-policy effects.
  \textbf{(a--b)} Model-level token and $\mathrm{AnswerRecall@5}$ effects for
  Eager and Eager+Compact against paired grep/read. Both use the same
  candidates, so their direct contrast isolates the one-time history transition.
  \textbf{(c)} Pooled Eager/Compact workload effects; row labels give query
  counts across five models and three policies. Whiskers are
  snapshot-clustered 95\% intervals; quality axes mark zero and $-0.05$.
  Compact starts a new short prefix once; every subsequent turn appends to that
  fixed prefix and can reuse its KV cache. Token axes report provider-reported
  trajectory totals.}
  \else
  \caption{Agent context-policy effects.
  \textbf{(a--b)} Model-level token and $\mathrm{AnswerRecall@5}$ effects for
  Eager/Compact against paired grep/read; candidates are shared.
  \textbf{(c)} Pooled workload effects (labels give query counts across five
  models and three policies). Whiskers are snapshot-clustered 95\% intervals.
  Compact resets the prefix once; tokens are provider-reported trajectory
  totals.}
  \fi
  \label{fig:agent-runtime}
\end{figure*}

This experiment changes only the backend used for an otherwise identical
\texttt{definition} or \texttt{references} request. For each of 100 snapshots
(21 Go, 20 C/C++, 20 Python, 20 Rust, and 19 TS/JS), we sort source-mappable
reference-edge anchors, select five evenly spaced positions, and issue both
capabilities. The replay retains uncapped outputs, compares normalized
path/start-line locations and UTF-32 full ranges, and measures each request
over five repetitions after initialization and stable-response probes. The
live providers are clangd, gopls, basedpyright, rust-analyzer, and
typescript-language-server for C/C++, Go, Python, Rust, and
TypeScript/JavaScript (TS/JS). Because positions originate from the static
graph, the match rate does not estimate arbitrary editor requests. The static
provider serves occurrence indexes decoded directly from SCIP protobufs and
audited on all 80 SCIP snapshots; the superseded capped replay is retained in
\detailsref{app:distributional-views}.

\paragraph{Compatibility by capability and backend.}
Whether the static provider can stand in for the live server depends on what
is asked and of which language. It returns the server's normalized locations
on 705 of 940 compared path/start requests, or 75.0\%, and on 616 of 800 at
full range, or 77.0\% with a repository-clustered 95\% interval of
72.6--81.3. That average hides the split that matters:
Figure~\ref{fig:nav-agreement} admits a capability statically where its
agreement reaches 0.80, and Go qualifies for both capabilities while TS/JS
qualifies for neither, with Rust, C/C++, and Python admitting definitions
alone. The threshold was fixed after the replay, but the admitted set is the
same for any value from 0.71 to 0.80; raising it to 0.85 removes only Python
definitions, which sit exactly at 0.80, and lowering it to 0.70 adds only
TS/JS definitions at 0.705. As with the maintenance gate
(\qref{4}{sec:eval:incremental}), the rule is a measurement to repeat on the
repositories where it will run, not a constant. Six snapshots fail before
measurement, leaving 60 requests unmeasured, and another 140 lack comparable
full ranges; missing observations are counted separately from mismatches.

The residual disagreement lies mostly outside the static route. The protobuf
decoder raised path/start matches on the 790 SCIP-subset comparisons from 303
under the original text decoder to 614, against the same recorded live
responses, and rerunning the live servers moves a further 19 of 940. A
stratified case study of 50 residual reference mismatches (two AI-assisted
passes, $\kappa=0.56$) attributes 37 to the indexers and to the answers
themselves, 16 to indexer coverage and 21 to scope or identity differences,
11 to decoding or normalization, and two to live drift; C/C++ is mostly
coverage and Rust mostly scope
(\detailsref{app:navigation-residual}).

\paragraph{Operational cost.}
The static route is ready almost six times sooner. Over the 94 measured snapshots it
is ready in 1.229\,s at the median, counting graph and occurrence-index
loading and provider initialization, whereas the live server needs 6.347\,s
at the median and 27.8\,s at the 95th percentile including idle wait and
stability probes; the median of the per-snapshot ratios is 5.9$\times$ (95\%
interval 5.5--7.4$\times$), larger than the 5.2$\times$ ratio of the two
medians. Median resident memory is 162\,MB for the static provider against
670\,MB for the live process tree, whose figure can count shared pages more
than once. Frozen paired replay removes agent tool-choice and model API
variance while preserving repository-specific request positions and provider
outputs.

\ifvldbsubmission\else
\textbf{Answer to \qref{3}{sec:eval:lsp}.}
Static navigation reproduces the live server on 75\% of requests, but the
usable rule is per backend and per capability: definitions on Go, Rust,
C/C++, and Python, references on Go alone. Where admitted it is ready five
times sooner; elsewhere callers route to the live server.
\fi

\subsection{Q4: Incremental View Maintenance}
\label{sec:eval:incremental}

\qref{3}{sec:eval:lsp} evaluates requests against a fixed repository commit.
\qref{4}{sec:eval:incremental} evaluates the per-transition speedups in
Eq.~\eqref{eq:update-speedups}, then applies the post-timing output comparisons.
For every graph source change, we execute a fresh LSP rebuild, file-level
replacement, and the symbol-level path in Figure~\ref{fig:incremental-lsp}.
Deterministic definition/reference requests are anchored in changed files.
Vector maintenance compares a fresh Flat index with content-addressed embedding
reuse and deterministic Flat replay.

Figure~\ref{fig:maintenance-lifecycle}(a) reports all 33 graph and 31 vector
source-changing transitions; no-source rows are outside the ratio population,
and mismatches remain open. Symbol repair matches the independent rebuild on
15/33 (45.5\%) transitions; among those transitions, its amortized speedup has
median 8.67$\times$ and IQR 5.95--10.99$\times$. File replacement matches on
14/33, with median 1.95$\times$; on the 14 transitions where both paths match,
symbol repair is
4.25$\times$ faster at the median. All Go (7/7) and Python (8/8) symbol updates
pass. Rust and TS/JS have high median edge F1 (99.12\% and 97.61\%) and serving
agreement (97.40\% and 99.53\%), but none passes both offline whole-graph and
serving checks; their points therefore remain open.

Vector maintenance matches the independent rebuild on 28/31 (90.3\%)
source-changing transitions, with
median speedup 25.44$\times$ and IQR 15.24--35.15$\times$. All 31 persisted
artifacts reproduce target document identities and vectors; three Rust rows
fail exact ordered and set top-10 replay and are excluded from the
matching-transition speedup summary.

A separate four-case same-commit rebuild audit gives median fresh/fresh edge F1
of 99.75\% and serving agreement of 99.35\%, with changed-scope exactness on
3/4 cases. Thus a cold live-LSP rebuild is not a deterministic oracle for every
Rust/TS graph. We report exact-match counts and raw fidelity together rather
than tuning a post hoc tolerance. \Detailsref{app:incremental} gives the protocol,
language breakdown, and failure audit.

\begin{table}[t]
  \caption{Maintenance validity on the original and the repository-disjoint
  held-out cohort. Rebuild-equivalent transitions are counted over each
  view's source changes; a graph transition must match the rebuild on
  declared facts and on every anchored navigation request. Median ratios
  cover admitted transitions only, with graph ratios given for an isolated
  LSP startup and for one shared across five updates. No Rust or TS/JS graph
  source change is admitted in either cohort.}
  \label{tab:maintenance-validity}
  \centering
  \small
  \setlength{\tabcolsep}{5pt}
  \begin{tabular}{@{}llrl@{}}
    \toprule
    \textbf{Arm} & \textbf{Cohort} & \textbf{Equivalent} &
    \textbf{Median ratio} \\
    \midrule
    Symbol repair & original & 15/33 & 3.61/8.67$\times$ \\
    \rowcolor{black!5}
    Symbol repair & held-out & 13/24 & 4.30/11.62$\times$ \\
    File replacement & original & 14/33 & 1.95$\times$ \\
    \rowcolor{black!5}
    File replacement & held-out & 14/24 & not reported \\
    Vector reuse & original & 28/31 & 25.44$\times$ \\
    \rowcolor{black!5}
    Vector reuse & held-out & 27/30 & 40.43$\times$ \\
    \bottomrule
  \end{tabular}
\end{table}

\begin{figure}[t]
  \centering
  \includegraphics[width=0.92\columnwidth]{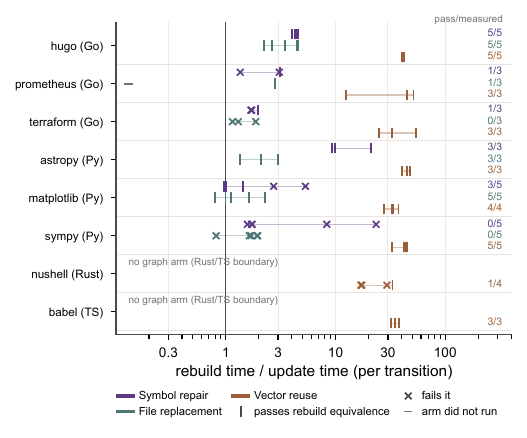}
  \Description{One tick per held-out maintenance transition, grouped by
  repository and arm, at its rebuild-to-update time ratio on a log axis;
  ticks pass the rebuild-equivalence check, crosses fail it, and a label
  gives passed over measured for each arm. Symbol repair fails on every
  sympy transition and two of three terraform transitions while hugo and
  astropy pass throughout; vector reuse fails only on nushell.}
  \caption{Held-out cohort, one tick per transition at its rebuild/update
  time ratio with isolated setup; a cross marks a transition that fails
  rebuild equivalence, and labels count passes over measured transitions.
  Symbol repair is fast on every repository but reproduces the rebuild on no
  sympy transition and one of three terraform transitions; vector reuse
  fails only where nushell's inventory changed.}
  \label{fig:maintenance-holdout}
\end{figure}

\paragraph{The comparison admits graph repair per repository, not per language.}
The original cohort suggests a simple rule, attempt symbol repair on Go and
Python and rebuild elsewhere, and a repository-disjoint follow-up tests it.
The follow-up freezes five first-parent transitions in each of eight new
repositories (hugo, prometheus, terraform, astropy, matplotlib, sympy,
nushell, babel), 40 positions in all, with the rebuild-equivalence checks of
Sec.~\ref{sec:framework:ops} applied to every source change. Symbol repair
completes on all 24 Go/Python source changes and reproduces the rebuild on
13; the rebuild-equivalence check separates them, since seven of the eleven
mismatches occur where independent rebuilds agree. The verdict is
repository-shaped: hugo and astropy pass throughout, every sympy and two of
three terraform transitions miss (Figure~\ref{fig:maintenance-holdout}), so
the runtime admits repair where the check passes; a runtime relying on
execution status alone would have served all eleven mismatched graphs. Where
admitted, repair runs 4.30$\times$ faster than rebuild with isolated setup
and 11.62$\times$ with setup shared across five transitions; file
replacement is admitted on 14 of 24 held-out and 14 of 33 original changes
at 1.95$\times$ (Table~\ref{tab:maintenance-validity}), so the arms are
close in admission and repair wins on speed. A per-language rule would have
saved just 1.21$\times$ of total workload on the original cohort; the
per-repository check is what turns repair into a safe speedup.

\paragraph{Vector reuse passes on both cohorts, with inventory and replay checked together.}
Content-addressed reuse matches the rebuild on 27 of 30 held-out source
changes at a median 40.43$\times$. The two checks catch different things:
four held-out stores answer all 32 sampled top-10 queries identically yet
retain six extra documents, traced to ancestor-directory filtering that
differs between full and per-file chunking, while three original artifacts
reproduce identities and vectors yet fail ordered replay. Admission uses
both.

\paragraph{Reference stability.}
Repeat rebuilds on the held-out cohort match on whole-graph facts for 23/28
pairs and behavior for 23/24 evaluable pairs, so reference drift accompanies
a minority of symbol-repair mismatches. The follow-up also changes the repair
implementation, so cross-cohort differences do not isolate repository
generalization. \Detailsref{app:maintenance-holdout} gives the held-out
protocol, failure classifications, and cost distributions.

\ifvldbsubmission\else
\textbf{Answer to \qref{4}{sec:eval:incremental}.}
The offline comparison catches what execution status cannot. Vector reuse
matches an independent rebuild on 28/31 and 27/30 transitions across two
cohorts; symbol repair completes on every attempt and matches on 15/33 and
13/24, so the runtime admits it per repository and file replacement stays
close in admission while repair wins on speed. The comparison identifies
where a fast path preserved output; it is a measurement to repeat where the
views will run, not a runtime guarantee.
\fi

\subsection{Cross-Stage Lifecycle Accounting}\label{sec:eval:lifecycle}

\ifvldbsubmission
The controlled lifecycle trace verifies that the evaluated compiler and runtime
compose lexical, dense, and calibrated static definition operations over the
same repository commit while accounting for their stage costs.

Figure~\ref{fig:maintenance-lifecycle}(b) deliberately shows two granularities.
The upper group decomposes view construction into BM25, structural-graph, and
vector builders; these are internal contributions to the lower group's complete
materialization stage $B$, not query latency. Their medians are 0.81, 37.97,
and 59.48\,s, respectively; the vector builder materializes both granularities.
The lower group reports the complete lifecycle stages: materialization
$B=116.7$\,s (95\% CI 65.6--153.9), fresh-process runtime view loading
$L=7.40$\,s (6.99--8.07), and one warm service trace
$S=0.727$\,s (0.593--1.079). The six rows therefore must not be summed.
The median artifact occupies 160.3\,MiB, and the distributions retain the
observed MicroPython vector and SymPy graph long tails.

The controlled trace fixes $N=20$ source-query sessions, the reuse scale marked
in panel (c). Each session executes the same query once through dense retrieval
and once through BM25, producing 40 retrieval requests. Two additional,
calibrated definition probes include structural/static navigation, so the
recorded total is $20(1+1)+2=42$. Thus 42 is a derived request count, not a
chosen workload scale. The trace exercises all three view families through one
build--load--serve path; it is a reproducible integration workload, not a
distribution inferred from agent traces. Applying the accounting models from
Sec.~\ref{sec:framework:cost},
Figure~\ref{fig:maintenance-lifecycle}(c)
shows 13.20 versus 6.24\,s/session at the trace scale $q=20$, and 8.53 versus
1.27\,s/session at $q=100$, for process-isolated and shared-resident runtimes.
Each projection combines $1/q$ amortization with measured serve-only $S/N$,
not an end-to-end or hardware floor. Lines connect six accounting projections,
not measurements at six workload scales or a speedup/break-even comparison;
\Detailsref{app:lifecycle-boundary} gives intervals and the asymptotic
interpretation.
\else
Figure~\ref{fig:maintenance-lifecycle}(b) separates two granularities. The upper
group's BM25, structural-graph, and vector construction medians are 0.81, 37.97,
and 59.48\,s; they contribute to the lower group's complete materialization
stage and are not query latency. The lower group reports materialization
$B=116.7$\,s (95\% CI 65.6--153.9), fresh-process runtime loading
$L=7.40$\,s (6.99--8.07), and one warm trace $S=0.727$\,s
(0.593--1.079). These two granularities are non-additive. The median artifact
occupies 160.3\,MiB.

The trace fixes $N=20$ source-query sessions, the reuse scale in panel (c).
Each query runs once through dense retrieval and BM25; two calibrated definition
probes include static navigation. Hence $20(1+1)+2=42$ is a derived request
count, not a chosen workload scale. This reproducible build--load--serve trace
exercises all three view families but is not an observed agent distribution.
Panel (c) projects 13.20 versus 6.24\,s/session at $q=20$ and 8.53 versus
1.27\,s/session at $q=100$ for process-isolated and shared-resident runtimes.
The $1/q$ term approaches measured serve-only $S/N$, not an end-to-end or
hardware floor; connected points are accounting projections, not six executed
scales or a break-even claim.
\Detailsref{app:lifecycle-boundary} gives intervals and the asymptotic
interpretation.
\fi

\subsection{Q5: Agent Context Delivery}\label{sec:eval:context}

\ifvldbsubmission
\qref{5}{sec:eval:context} separates candidate delivery from a one-time history
transition. Grep/read performs per-issue discovery without $L_2$ candidates,
whereas Eager queries the materialized dense view once and injects
$C_{10}^{\mathrm{ctx}}$. Compact starts a new short prefix after the first
successful read; later turns append to that fixed prefix without rewriting it.
Eager and Eager+Compact use the same candidates, so their direct contrast
isolates this transition.
Because mean grep/read volume spans 25.9--159.7 thousand tokens per query,
Figure~\ref{fig:agent-runtime}(a--b) normalizes both series within model
against paired grep/read.
\else
\qref{5}{sec:eval:context} separates delivery from retention. Grep/read has no
injected $L_2$ candidates; Eager injects frozen $C_{10}^{\mathrm{ctx}}$; Compact
receives the same candidates and rewrites history once after the first
successful read, then appends. Thus Eager/Compact isolates retention.
Figure~\ref{fig:agent-runtime}(a--b) normalizes against paired grep/read because
baseline volume spans 25.9--159.7 thousand tokens per query.
\fi

Token axes sum provider-reported prompt/completion tokens rather than
cache-adjusted cost. Eager keeps its prefix; Compact resets once and reuses the
shorter prefix. Cache hits and prefill latency are unmeasured.

\ifvldbsubmission
Figure~\ref{fig:agent-runtime} uses ordinary tools, a 16-turn cap, and zero
recall for invalid answers. Among the core grep/eager/compact arms, the
lowest-token policy whose lower confidence bound clears the $-0.05$
localization margin selects Haiku eager and compact otherwise.

In panel order, selected token ratios are 49.9, 45.1,
44.8, 12.9, and 35.8\%, with respective $\Delta\mathrm{AR@5}$ of $-0.009$, $-0.006$,
$-0.003$, $+0.040$, and $+0.067$; every lower bound exceeds $-0.05$.
Qwen-27B eager alone misses the gate. The Compact/Eager token-usage ratio
ranges from 123.3\%
(Haiku) to 27.9\% (Gemma), and Gemma/Gemini lose 0.070/0.044 recall relative to
eager. Compaction is not uniformly dominant. Panel (c) is descriptive because
several subgroup intervals cross the margin; we make no routing claim.
\Detailsref{app:agent-protocol} gives protocol and validity audits.
\else
All models use ordinary tools, a 16-turn cap, and frozen top-10 candidates;
accounting includes full provider-reported trajectories and fixed answer-format
prompts, with zero recall for invalid answers. The rule in
Sec.~\ref{sec:framework:cost} selects Eager for Haiku and Compact otherwise.
In model order, these use 49.9, 45.1, 44.8, 12.9, and 35.8\% of paired
grep/read tokens; $\Delta\mathrm{AR@5}$ ranges from $-0.009$ to $+0.067$, with
all lower bounds above $-0.05$. Selection is unchanged at
$k\in\{1,3,5,10\}$; Qwen-27B Eager alone misses the margin
(Appendix Table~\ref{tab:agent-cutoff-sensitivity}).

Compaction is not uniformly dominant: Compact/Eager token usage ranges from
123.3\% (Haiku) to 27.9\% (Gemma). Gemma/Gemini trade 0.070/0.044
$\mathrm{AnswerRecall@5}$ against Eager while clearing the baseline gate.
Panel (c) is descriptive: several quality intervals cross the margin, so we
make no subgroup-routing claim. \Detailsref{app:agent-protocol} gives paired
intervals and validity audits.

\textbf{Answer to \qref{5}{sec:eval:context}.}
Candidate injection and one-time compaction occupy model-dependent
token--localization operating points; compaction is not uniformly dominant.
\fi

\providecommand{\ProviderExpandedTasks}{30}
\providecommand{\ProviderExpandedRepositories}{11}
\providecommand{\ProviderExpandedRepeats}{3}
\providecommand{\ProviderExpandedCells}{180}
\providecommand{\ProviderExpandedNativeRecall}{52.22}
\providecommand{\ProviderExpandedCodeNibRecall}{55.00}
\providecommand{\ProviderExpandedRecallDifference}{+2.78}
\providecommand{\ProviderExpandedRecallLower}{-19.44}
\providecommand{\ProviderExpandedRecallUpper}{+25.00}
\providecommand{\ProviderExpandedTokenTasks}{21}
\providecommand{\ProviderNormalizedTasks}{30}
\providecommand{\ProviderNormalizedRepositories}{11}
\providecommand{\ProviderNormalizedRepeats}{3}
\providecommand{\ProviderNormalizedCells}{180}
\providecommand{\ProviderNormalizedCellsPerProvider}{90}
\providecommand{\ProviderNormalizedNativeFinishedExecutions}{65}
\providecommand{\ProviderNormalizedNativeExecutionFailures}{25}
\providecommand{\ProviderNormalizedNativeUnavailableArtifacts}{0}
\providecommand{\ProviderNormalizedNativeUnknownOutcomes}{0}
\providecommand{\ProviderNormalizedCodeNibFinishedExecutions}{78}
\providecommand{\ProviderNormalizedCodeNibExecutionFailures}{9}
\providecommand{\ProviderNormalizedCodeNibUnavailableArtifacts}{3}
\providecommand{\ProviderNormalizedCodeNibUnknownOutcomes}{0}
\providecommand{\ProviderNormalizedRecallTasks}{30}
\providecommand{\ProviderNormalizedRecallNative}{51.67}
\providecommand{\ProviderNormalizedRecallCodeNib}{60.37}
\providecommand{\ProviderNormalizedRecallDifference}{+8.70}
\providecommand{\ProviderNormalizedRecallLower}{-12.41}
\providecommand{\ProviderNormalizedRecallUpper}{+29.50}
\providecommand{\ProviderNormalizedTimeSecondsTasks}{30}
\providecommand{\ProviderNormalizedTimeSecondsNative}{93.54}
\providecommand{\ProviderNormalizedTimeSecondsCodeNib}{74.92}
\providecommand{\ProviderNormalizedTimeSecondsDifference}{-18.62}
\providecommand{\ProviderNormalizedTimeSecondsLower}{-31.90}
\providecommand{\ProviderNormalizedTimeSecondsUpper}{-7.68}
\providecommand{\ProviderNormalizedTokensThousandsTasks}{23}
\providecommand{\ProviderNormalizedTokensThousandsNative}{222.79}
\providecommand{\ProviderNormalizedTokensThousandsCodeNib}{179.31}
\providecommand{\ProviderNormalizedTokensThousandsDifference}{-43.48}
\providecommand{\ProviderNormalizedTokensThousandsLower}{-82.01}
\providecommand{\ProviderNormalizedTokensThousandsUpper}{-13.80}
\def\ProviderNormalizedQualityRows{
File Recall@5 (primary) & 30/30 & 51.67 & 60.37 & +8.70 & $[-12.41,+29.50]$ \\
File Recall@1 & 30/30 & 48.61 & 52.96 & +4.35 & $[-15.13,+23.09]$ \\
File Recall@3 & 30/30 & 51.39 & 60.09 & +8.70 & $[-12.37,+29.37]$ \\
Symbol Recall@5 & 29/29 & 40.80 & 44.35 & +3.54 & $[-15.56,+22.58]$ \\}
\def\ProviderNormalizedCostRows{
Complete cell time (s) & 30/30 & 93.54 & 74.92 & -18.62 & $[-31.90,-7.68]$ \\
Model-request wall time (s) & 30/30 & 82.22 & 70.22 & -12.00 & $[-25.09,-1.59]$ \\
Tool-dispatch wall time (s) & 30/30 & 5.68 & 0.30 & -5.38 & $[-8.81,-2.43]$ \\
Input tokens (thousands) & 23/30 & 221.15 & 178.00 & -43.15 & $[-81.48,-13.55]$ \\
Output tokens (thousands) & 23/30 & 1.64 & 1.30 & -0.34 & $[-0.68,-0.06]$ \\
Total tokens (thousands) & 23/30 & 222.79 & 179.31 & -43.48 & $[-82.01,-13.80]$ \\}
\subsection{Q6: Manifest-Backed Provider Substitution}
\label{sec:eval:provider}

\qref{5}{sec:eval:context} fixed the harness we wrote. \qref{6}{sec:eval:provider}
and \qref{7}{sec:eval:repair} instead place \nickname behind agents we did
not write, on real SWE-bench Verified~\cite{swebench,swebench-verified}
tasks, and score what those agents produce. This study keeps LocAgent's
prompts, tool protocol~\cite{locagent}, model, and limits fixed and swaps
only the repository provider behind its tools: the
native arm loads LocAgent's own graph and BM25 indexes, and the \nickname arm
serves the same search and navigation tools from the manifest's BM25 and
symbol-graph views, so the comparison evaluates each provider as the agent
experiences it.

We freeze \ProviderNormalizedTasks{} SWE-bench Verified tasks from
\ProviderNormalizedRepositories{} repositories, at most three per repository
and disjoint from an earlier ten-task study, and run
\ProviderNormalizedRepeats{} paired repeats per provider, giving
\ProviderNormalizedCells{} \emph{cells}, where a cell is one run of one
provider on one task. Both arms share a local Qwen3.5-27B
checkpoint~\cite{qwen35}, server,
issue bytes, prompts, tool schemas, limits, and one experiment-level tool
boundary that decodes array arguments identically and cuts every observation
at 48,000 characters; an earlier execution without this shared boundary is
reported in the appendix and not pooled. The
analysis unit is the task with its repeats
nested inside it, and the primary estimand is the task-level paired
difference in File Recall@5 under 20,000 repository/task/repeat hierarchical
bootstrap draws. Relevance labels stay outside the agent input, and a failed
execution receives empty predictions and remains in the denominator.

\begin{table}[t]
  \caption{Paired provider substitution over \ProviderNormalizedTasks{} tasks
  and \ProviderNormalizedCells{} cells, \nickname minus native. Recall is a
  percentage; $n/N$ counts tasks with complete usage over eligible tasks.
  Cell time covers provider loading, agent execution, cleanup, and ledger
  sealing, and excludes index construction and model-server startup.}
  \label{tab:provider-expanded}
  \centering
  \scriptsize
  \setlength{\tabcolsep}{2.2pt}
  \renewcommand{\arraystretch}{1.0}
  \resizebox{\columnwidth}{!}{%
  \begin{tabular}{@{}lrrrrr@{}}
    \toprule
    \textbf{Measure} & $\boldsymbol{n/N}$ & \textbf{Native} & \textbf{\nickname}
    & $\boldsymbol{\Delta}$ & \textbf{95\% interval} \\
    \midrule
    \ProviderNormalizedQualityRows
    \midrule
    \ProviderNormalizedCostRows
    \bottomrule
  \end{tabular}}
\end{table}

\begin{figure*}[t]
  \centering
  \includegraphics[width=0.98\textwidth]{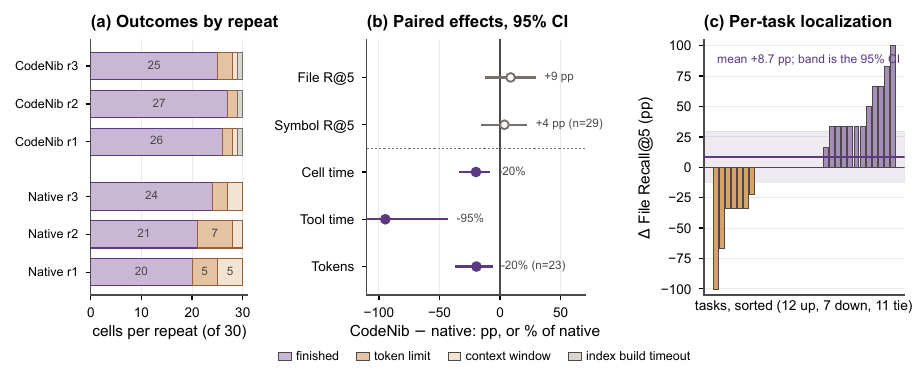}
  \Description{Panel a stacks execution outcomes for the two providers over
  90 cells each: native finishes 65 cells and fails 15 on the cumulative-token
  limit and ten on the context window, while CodeNib finishes 78 and fails
  seven on tokens, two on the context window, and three because one task's
  index build timed out. Panel b shows paired effects with 95 percent
  intervals: file and symbol recall differences in percentage points span
  zero, while cell time, tool-dispatch time, and tokens fall by 20, 95, and 20
  percent with intervals excluding zero. Panel c sorts the per-task paired
  difference in File Recall at 5, showing twelve tasks favouring CodeNib,
  seven favouring native, and eleven ties.}
  \caption{Provider substitution over \ProviderNormalizedTasks{} tasks.
  \textbf{(a)} Where each arm's 90 cells end. \textbf{(b)} Paired effects
  with 95\% intervals, in points for recall and as a share of native for
  cost. \textbf{(c)} Per-task File Recall@5 difference on the full
  denominator, where a failed cell scores zero.}
  \label{fig:provider-substitution}
\end{figure*}

\paragraph{Execution completion.}
The providers differ most in whether the agent finishes at all. \nickname
completes \ProviderNormalizedCodeNibFinishedExecutions{} of 90 cells against
\ProviderNormalizedNativeFinishedExecutions{} for native
(Figure~\ref{fig:provider-substitution}a). Native's 25 failures are 15
cumulative-token limits and ten context-window limits, whereas \nickname has
seven and two, plus all three repeats of one scikit-learn task whose SCIP
index reached the one-hour construction ceiling, leaving no manifest to
serve. Both arms issue about ten tool calls per cell without a tool error,
so the difference lies in what each call returns.

\paragraph{Serving cost.}
\nickname lowers mean cell time by 18.62\,s, total tokens by 43.48k, and
tool-dispatch time by 5.38\,s, all with intervals excluding zero
(Table~\ref{tab:provider-expanded}); on the 13 tasks where both arms
complete every repeat the time saving holds and the token saving no longer
excludes zero. The price is storage, 206.79\,MiB more per task for SCIP and
decoded artifacts, while build times are not distinguishable from native's
(\detailsref{app:provider-expanded}).

\paragraph{Why the agent finishes more often.}
The two providers answer the same calls with different amounts of text: on
the 60 cell pairs where both arms complete, \nickname returns a median 3,681
characters from \texttt{search\_code\_snippets} against native's 6,375, 9,535
from \texttt{get\_entity\_contents} against 12,593, and 1,364 from
\texttt{explore\_tree\_structure} against 1,894, a paired difference of
$-21.5$k characters per cell that resubmission compounds into the token gap.
That economy is a design property of the views, bounded candidate packs and
resolved occurrences rather than raw file text, and native LocAgent has no
mechanism to obtain it. Rerunning native with each observation cut at
\nickname's per-call median confirms length is the mechanism: the cut native
finishes all 90 cells and runs 15.2\,s faster, but its File Recall@5 falls
from 51.7 to 45.0 (\detailsref{app:provider-bytematched}). Native needs the
extra bytes to localize as well as it does; \nickname does not.

\paragraph{Localization.}
\nickname keeps localization while cutting cost. The primary File Recall@5
difference is \ProviderNormalizedRecallDifference{} points with interval
$[\ProviderNormalizedRecallLower,\ProviderNormalizedRecallUpper]$, and the
Recall@1 and symbol differences are also positive with intervals spanning
zero. Twelve tasks favor \nickname, seven favor native, and eleven tie
(Figure~\ref{fig:provider-substitution}c). Against the length-cut native,
which receives the same bytes per call, the estimate grows to $+15.4$ points
$[-5.6,+35.5]$.

\def\RepairUsageRows{On-demand & 158/200 & $\geq$ 714.61 & $\geq$ 11.09 & $\geq$ 0.847 \\
BM25 eager & 167/200 & $\geq$ 750.94 & $\geq$ 10.06 & $\geq$ 0.881 \\
Aider repo-map & 157/200 & $\geq$ 798.67 & $\geq$ 9.62 & $\geq$ 0.931 \\
CodeNib eager & 157/200 & $\geq$ 733.01 & $\geq$ 10.15 & $\geq$ 0.862 \\
CodeNib compact & 161/200 & $\geq$ 684.23 & $\geq$ 9.92 & $\geq$ 0.807 \\}
\providecommand{\HaikuCompletePairs}{199}
\providecommand{\HaikuCompleteDelta}{+6.53}
\providecommand{\HaikuCompleteTaskCI}{[+1.01,\,+12.06]}
\providecommand{\HaikuCompleteRepoCI}{[-3.92,\,+17.42]}
\providecommand{\HaikuCompleteExactP}{0.0294}
\subsection{Q7: Context Delivery and Retention on Real Repairs}
\label{sec:eval:repair}

The repair experiment runs Haiku 4.5~\cite{claude-haiku-45}, the smallest
tier of its model family, on 200 SWE-bench Verified tasks under the five
policies of Sec.~\ref{sec:system:context}; at its 33--40\% resolution rates a
context policy has room to move the outcome. The sample is a separately
frozen, repository-stratified draw of 200 of the 500 Verified tasks, sized to
the hosted budget, that excludes the 50 tasks used to build the harness and
contains the 30 tasks of
\qref{6}{sec:eval:provider}, with the model reached through a hosted alias
observed on 2026-09-05~\cite{claude-model-versioning} and the harness
mini-SWE-agent of the SWE-agent project~\cite{sweagent} with the official
evaluator; one run per task and policy
plus two extra runs on ten preselected tasks to measure variability. Every
cell has the same 100-step and 3,600-second limits and USD~1 stop rule, and
the independent evaluator scores each submitted patch with the official task
tests. Views are built per task on its base commit, so nothing is amortized
across tasks. The runner driving mini-SWE-agent's model calls inserts the
delivered context into the task message and, for compact, rewrites the
history after an attested read; the agent's only tool remains Bash and no
\nickname code runs inside the container. The predeclared primary contrast
is CodeNib compact versus On-demand; four secondary contrasts compare CodeNib
eager with On-demand, BM25 eager, and Aider eager, then CodeNib compact with
CodeNib eager, and only the last fixes candidate bytes and isolates retention.

\begin{figure*}[t]
  \centering
  \includegraphics[width=0.98\textwidth]{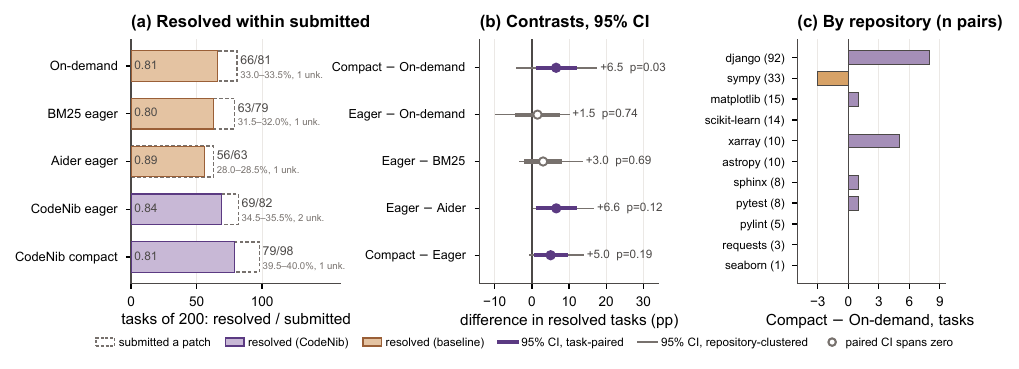}
  \Description{Panel a shows, for each of five context policies, the number of
  tasks with a submitted patch as a dashed outline and the number resolved as a
  filled bar inside it, with the resolved share printed on the bar and the
  identification range over unknown outcomes printed under the count. Panel b
  shows five paired contrasts as point estimates with task-paired and
  repository-clustered 95\% intervals. Panel c shows the compact minus
  on-demand difference in resolved tasks for each of eleven repositories.}
  \caption{Five context policies on 200 repairs. \textbf{(a)} Resolutions sit
  inside submissions and 0.80--0.89 of submissions resolve, so policies differ
  in reaching a patch, not in patch accuracy. \textbf{(b)} Compact versus
  on-demand is the one contrast whose task-paired interval excludes zero;
  repository-clustered intervals, over 11 repositories, all include zero.
  \textbf{(c)} Per-repository difference: django +8 of 92 pairs, xarray +5
  of 10, sympy \(-3\) of 33.}
  \label{fig:repair-policies}
\end{figure*}

\paragraph{Resolution and retention.}
Retiring the delivered candidates changes how often the agent reaches a patch
at all. \nickname compact records 79 confirmed resolutions against 66 for
On-demand, the largest of the five policy differences
(Figure~\ref{fig:repair-policies}a). The mechanism is submission rate rather
than accuracy: compact submits on 98 of 200 tasks against 81, and the two
resolve the same share of what they submit, 0.81 each. Under a fixed budget a
shorter prompt buys more turns, and more turns reach a patch.

The predeclared test looks only at tasks where the two policies disagree: on the
199 complete pairs, compact alone resolves 22 tasks and On-demand alone
resolves 9. If the two policies were equally good, those 31 tasks would
split about evenly, and a split as lopsided as 22 to 9 would arise by chance
with probability $p=\HaikuCompleteExactP$ (an exact McNemar test). The
complete-pair difference is $\HaikuCompleteDelta$ points with task-paired
interval $\HaikuCompleteTaskCI$. One pair has an unknown outcome on both
arms; every fill of it and of the five other unknown cells keeps the
difference between $+6.0$ and $+7.0$ points, with a worst-case $p$ of 0.0502
against the predeclared 0.05 (\detailsref{app:repair-missing-sensitivity}).
Clustering by repository, with 92 Django tasks, widens the interval to
$\HaikuCompleteRepoCI$; Django and xarray carry most of the gain and sympy
runs against it (Figure~\ref{fig:repair-policies}c,
\detailsref{app:repair-complete-pairs}).
Repeated runs flip outcomes on two of ten tasks for compact and four for
On-demand. Doubling the budget to USD~2 on the 108 non-Django tasks
resolves 50 against 46, $+3.7$ points $[-3.7,+11.2]$: both arms improve and
the gap narrows as the budget binds less, as a budget-mediated mechanism
predicts (\detailsref{app:repair-budget2}).

Candidate quality and retention each contribute. Among the eager policies,
\nickname's dense candidates resolve 69 tasks against 63 for BM25 and 56 for
Aider's map, although delivery alone lifts On-demand by only $+1.0$ to
$+2.5$ points. Retention supplies the rest: with candidate bytes held fixed,
compact adds $+4.0$ to $+5.5$ points over eager, with a Holm-adjusted exact
test above 0.05. Audited requests confirm that compact replaced the candidates on 193 of
200 tasks; the seven others stay in the denominator.

\paragraph{Outcome and usage observability.}
Every planned cell stays in its denominator. Figure~\ref{fig:repair-policies}a
counts all 1,000 primary cells, 994 with known outcomes; the six unknown
outcomes widen each policy's resolution share to the identification range
printed under its count, and the 200 cells with incomplete provider usage
make the appendix's usage figures lower bounds. Mean agent process time
ranges from 163.10 to 176.50\,s across policies. Building the context that a
policy delivers is charged to the harness and is comparable to the agent run
it precedes: a mean 149.21\,s for both \nickname views including model
loading, 35.21\,s for the Aider map, and zero for On-demand, which starts
from the issue alone (\detailsref{app:repair-preparation}). The budget
therefore covers agent execution, and preparation sits outside it
(\detailsref{app:repair-study}).

\section{Discussion}\label{sec:disc}

\subsection{Findings}\label{sec:findings}

The seven studies connect each serving choice to the outcome and cost it
changes for an agent, and together they tell a serving runtime which route
to take for which request.

\paragraph{F1: Retiring delivered candidates converts budget into completed
attempts.}
Under a common USD~1 stop rule, \nickname compact records the largest of the
five policy differences we measured, 79 confirmed resolutions against 66 for
on-demand exploration (Figure~\ref{fig:repair-policies}a). The difference
lies in reaching a submission rather than in per-submission accuracy: compact
submits on 98 tasks against 81, and both resolve 0.81 of what they submit.
Where the two policies disagree, retirement wins 22 tasks to 9; the gain is
concentrated in Django and xarray, and it narrows to $+3.7$ points when the
budget is doubled on the other 108 tasks, which is what a budget-mediated
mechanism predicts. Delivered eagerly, \nickname's candidates already beat
Aider's map, 69 to 56. Compaction of agent history
is itself studied~\cite{complexity-trap,agentdiet,swe-pruner}; the
eager/compact contrast isolates retiring one delivered block at a detected
event with candidate content held fixed.

\paragraph{F2: Substituting the repository provider lets the agent finish more
often and cost less, because the views answer in fewer bytes.}
With LocAgent's policy, prompts, tools, model, budgets, and tool boundary
fixed, the \nickname provider completes 78 of 90 cells against 65 for native
and lowers cell time by 18.6\,s and tokens by 43.5k, with intervals excluding
zero. The mechanism is response economy: bounded candidate packs and resolved
occurrences say in 3,681 characters what native's file text says in 6,375. Forcing native to \nickname's
lengths makes it finish every cell cheaper but drops its File Recall@5 from
51.7 to 45.0 against \nickname's 60.4, with intervals spanning zero on 30
tasks. The price is storage, 207\,MiB more per task.

\paragraph{F3: Every view has a cheap ready point, and static definitions can
be served from it.}
BM25 is queryable 0.19\,s after source-unit extraction where the dense view
needs 48.6\,s for 17 more points of file Recall@10, and the static graph
loads in 1.23\,s where a live server needs 6.35\,s, answering 403/470
definition requests with the server's locations. An agent can start on
lexical search and static definitions, then move to the dense view and the
live server as they become ready. Graph construction takes 592 to 613\,s per
repository, so the ready-point saving compounds each time a session reuses a
built graph, and F4 identifies the repositories where repair also carries it
across commits.

\paragraph{F4: The offline comparisons catch what execution status cannot,
and they are what locate the safe routes.}
Held-out symbol repair completes on every Go/Python attempt and reproduces a
rebuild on 13/24, with seven of the mismatches where repeat rebuilds agree;
static references match live output on 64\% of requests and C/C++ references
on 23/75. A runtime that admitted repairs by execution success or routed
every navigation request to the static graph would serve those outputs; ours
admits repair per repository and navigation per backend and capability
because the comparison says where each holds; symbol repair and file
replacement are admitted at similar rates held out, and repair is the faster
where admitted. Callers needing rebuild or live semantics route there.

\paragraph{F5: Content-addressed vector reuse is the dependable incremental
path, checked by inventory and replay together.}
Vector reuse passes on 27/30 held-out and 28/31 original source changes at
median ratios of 40.43$\times$ and 25.44$\times$. Three original artifacts
match identities and vectors yet fail top-10 replay, and four held-out stores
answer 32 sampled queries identically while retaining six extra documents.

\subsection{Threats to Validity}\label{sec:threats}

\paragraph{Construct validity.}
Retrieval targets are pre-patch files and callables intersected by developer
patches, so they omit context read without editing, and added-only units are
unretrievable. \qref{6}{sec:eval:provider} measures localization, and \qref{7}{sec:eval:repair}'s official tests cannot
establish correctness beyond their coverage. Navigation compares source
locations only, ready-state RSS is not peak memory, and maintenance compares
declared facts and sampled navigation outputs, so neither proves general
equivalence. Token counts exclude cache discounts and are lower bounds where
counters are missing, nested clocks do not give an end-to-end latency, and
\qref{1}{sec:eval:rag} leaves the route selector unmeasured.

\paragraph{Internal validity.}
Two analyses of the repair study were written after we saw its outcomes, the
complete-pair estimate and the missing-outcome sensitivity; the population,
ordering, and contrasts were frozen beforehand. Haiku is reached through a
hosted alias that can change between execution dates, and missing usage
counters rule out any token-saving claim. Because the USD~1 stop rule binds,
a non-submission is an outcome of this bounded deployment, the USD~2
follow-up covers only non-Django tasks, and view construction would enter an
end-to-end cost. Only the compact and eager arms hold candidate bytes fixed. \qref{6}{sec:eval:provider} and its length-matched control reuse one cohort
and its indexes after a first execution exposed two interface defects of
ours, so they correct the interface but are not independent replications;
an earlier ten-task study estimated a 17.5-point localization deficit with an
interval spanning zero.
The navigation audit reassigned 311 of 487 observed mismatches to our
decoder, and the residual case study is AI-assisted rather than
human-annotated; rebuilding the \qref{6}{sec:eval:provider} graphs with the
corrected decoder reproduces their facts exactly. Timing shares the host
with other work and is descriptive.

\paragraph{External validity.}
Retrieval and navigation cover 100 snapshots in five language groups, each
maintenance cohort eight repositories. The repair primary covers 200 Python
tasks from 11 repositories, 92 of them Django, with one model and unknown
benchmark exposure; provider substitution covers 30 tasks and one model. The
routes of Figure~\ref{fig:nav-agreement} hold for these repositories
and servers alone, and no experiment measures concurrent publication.

\subsection{Future Directions}

A concurrent repository database needs per-view versions, atomic
publication, recovery, and cost-based routing within the validity boundaries
measured here. Evaluated runtime traces could supervise routing, context
selection, and compaction in a post-training data flywheel, and a scheduler
should place lexical, graph, and navigation work on CPUs and embedding,
reranking, and inference on GPUs.
\FloatBarrier

\section{Conclusion}\label{sec:conclusion}

\nickname{} serves commit-linked lexical, dense, and structural views to
agents and admits each reused route only against a rebuild or live server.
Static definitions load in 1.23\,s against a live
server's 6.35\,s on the backends that pass the agreement rule, vector reuse is
exact on 28/31 and 27/30 source changes, and graph repair is admitted per
repository. Selected context policies use 50--87\% fewer tokens across five
models; behind LocAgent, the substituted provider finishes 78 of 90 runs
against 65; and on 200 real repairs, compact retention resolves 79 issues
against 66 for on-demand exploration and 56 for Aider's repository map.

\FloatBarrier
\clearpage

\bibliographystyle{ACM-Reference-Format}
\bibliography{sample}

\ifvldbsubmission
\else
  \clearpage
  \appendix
  \raggedbottom
  \setlength{\textfloatsep}{8pt plus 2pt minus 2pt}
  \setlength{\dbltextfloatsep}{8pt plus 2pt minus 2pt}
  \setlength{\floatsep}{8pt plus 2pt minus 2pt}
  \setlength{\dblfloatsep}{8pt plus 2pt minus 2pt}
  \setlength{\intextsep}{8pt plus 2pt minus 2pt}
  \section{Dataset Construction and Ground Truth}
\label{app:datasets}

\paragraph{Frozen dataset identities.}
CodeNib Base is the 100-row test split of
\texttt{fishmingyu/codenib-base-dataset} at Hub revision
\path{4eb84e2e8918474969ce68c5b06facf14d6be604}. The rows bind 100 unique
\texttt{(repository, base commit)} snapshots from 25 repositories. CodeNib
Synthesis is the five-config,
500-row test split of \texttt{sysevol-ai/codenib-synthesis} at revision
\path{5ac36c39ef69bbfe2e14dac58b6067b8c350c53e}. The artifact records the
parquet hashes and canonical row-identity hashes for both datasets.

\paragraph{Upstream provenance and task boundary.}
Neither split is a raw SWE-bench agent-resolution workload. CodeNib Base
resamples SWE-bench~\cite{swebench} issue--patch pairs:
Verified~\cite{swebench-verified} supplies Python, and
Multilingual~\cite{swebench-multilingual} supplies C/C++, Go, Rust, and
TypeScript/JavaScript. It retains the repository,
pre-solution \texttt{base\_commit}, issue body, and developer patch. CodeNib
does not run the upstream pass/fail tests.
\qref{1}{sec:eval:rag}--\qref{2}{sec:eval:index} use the patch only to derive
pre-patch localization labels; \qref{3}{sec:eval:lsp} reuses the snapshots but
samples independent graph-covered source positions. CodeNib Synthesis is a
second-stage derivative of Base, not another upstream sample. It reuses one selected Base
snapshot from each of the same 25 repositories and replaces the issue query
with source-grounded localization queries. Thus the two workloads permit paired
system analysis but are not independent evidence across repositories.

\begin{table}[H]
  \centering
  \scriptsize
  \setlength{\tabcolsep}{3.1pt}
  \begin{tabular}{@{}lrrrr@{}}
    \toprule
    Language & Rows & Repos & LOC p50 & $L_0$/$L_2$ p50 \\
    \midrule
    C/C++   & 20 & 5 & 206,900 & 279/7,483 \\
    Go      & 21 & 5 & 147,434 & 383/4,790 \\
    Python  & 20 & 5 & 158,938 & 379/6,690 \\
    Rust    & 20 & 5 & 104,917 & 404/4,110 \\
    TS/JS   & 19 & 5 &  57,942 & 251/1,751 \\
    \bottomrule
  \end{tabular}
  \caption{Frozen Base composition: per-snapshot medians, not sampling quotas.}
  \label{tab:base-composition}
\end{table}

\paragraph{Base repository and instance sampling.}
The collector first removes repositories with fewer than three candidate
issues. Within each language group, it counts non-hidden files in each
repository, sorts repositories by this count, and selects five approximately
at the minimum, 25th, 50th, 75th, and maximum ranks; rounded-rank collisions
use a deterministic center-out fallback. It then asks the recorded Claude Opus
alias to classify every issue as low, medium, or high from the issue body and
the first 6,000 patch characters. Classification runs in batches of ten, retries
in smaller batches, and uses medium only if structured parsing still fails.

Ground-truth extraction checks out the exact base commit, parses patch target
files, and tree-sitter-chunks supported files before and after applying the
developer patch. A pre-patch symbol is retained when its source range overlaps
a changed hunk and its content changes, or when it is deleted. The collector
rejects a row if extraction fails, the target set is empty, any new symbol is
required, or more than ten target blocks remain. It takes one available row
from each difficulty stratum per repository, fills the remaining per-repository
quota deterministically, and distributes shortfalls round-robin across language
groups until reaching 100. The frozen result contains 56 low-, 36 medium-, and
8 high-difficulty rows. Of its 151 target blocks, 74 rows have one block, 15
have two, and 11 have three to six. These are purposeful strata rather than a
population sample of repositories or issues.

\paragraph{Coordinates and evaluation targets.}
Tree-sitter chunk rows are stored as zero-based inclusive
\texttt{CodeChunk} ranges. Dataset export adds one to both endpoints and emits
one-based inclusive \texttt{CodeLocation} ranges. Retrieval runs only on the
base snapshot. Consequently, the scored targets are the pre-patch modified or
deleted source units and their enclosing files; added-only code has no
retrievable target. The issue body is the Base query. File and block recall use
the same frozen target records, without consulting the post-patch repository at
query time.

\paragraph{Synthesis construction.}
The synthesis planner takes all five Base repositories in each language group
and chooses one Base snapshot per repository: the snapshot whose prebuilt graph
contains the most symbol vertices. This yields 25 snapshots, each assigned 20
queries. For each snapshot the nominal budget is seven behavioral, three file
hint, three module hint, three traversal, two reasoning, and two symbol hint
queries. Behavioral anchors are non-test source symbols of at least 100
characters, sampled from at most 24 candidates using size, symbol type, and
graph connectivity; up to eight one-hop neighbors provide disambiguating
context. Traversal rows instead choose a two- or three-symbol reference chain,
expose only its anchor symbols, and reject verbatim leakage of hidden symbols.

Generation uses the recorded \texttt{opus} alias, a ten-turn cap, sampling seed
42, assignment seed 0, a 0.5 simple-query ratio, and three behavioral consensus
runs. A separate call to the same alias judges target discrimination; failed
rows enter at most three fix/regenerate attempts, while traversal generation
allows two judge retries. The provider alias has no recoverable immutable
revision. Deterministic validation rejects duplicate query IDs, empty targets,
and target files outside the declared language group. The frozen counts are
175 behavioral, 76 file hint, 75 module hint, 50 reasoning, 50 symbol hint, and
74 traversal rows; one Rust traversal replacement accounts for the 76/74
imbalance. The quality report has zero deterministic errors and five
non-\texttt{valid} judge warnings; excluding those five does not change which
\qref{5}{sec:eval:context} policy meets the localization margin for any model.

\section{Metric Definitions and Worked Examples}
\label{app:metrics}

The studies expose different output contracts, so their metrics are not
interchangeable. Recall and match indicators are computed per instance and then
macro-averaged; an instance with more targets therefore receives no extra
weight.

\paragraph{Patch-target coverage and all-file success.}
For instance $i$ at granularity $g\in\{\mathrm{file},\mathrm{symbol}\}$, let
$Y_i^g$ be the nonempty target set and $R_{i,k}^g$ the first $k$ distinct
returned units; symbol targets use the evaluated $L_2$ callable identities.
Over $M$ instances,
\[
\mathrm{TaskRecall}^g@k
=\frac{1}{M}\sum_{i=1}^{M}
\frac{|R_{i,k}^g\cap Y_i^g|}{|Y_i^g|}.
\]
At file granularity, repeated blocks from one path are collapsed before the
cutoff. The stricter
$\mathrm{FileSuccess}_i@k=\ind[Y_i^{\mathrm{file}}\subseteq
R_{i,k}^{\mathrm{file}}]$ requires every target file; its macro-average is the
reported success rate. For example, target files $\{A,B\}$ and ranked blocks
$[A{:}f_1,A{:}f_2,C{:}g,B{:}h]$ produce distinct-file order $[A,C,B]$.
At $k=2$, File Recall is $1/2$ and File Success is $0$; at $k=3$, both are $1$.

\paragraph{Physical-index fidelity.}
Let $E_{i,k}$ and $A_{i,k}$ be the $L_2$ identity sets returned by exact Flat
and an approximate index. Neighbor Recall is
\[
\mathrm{NeighborRecall}_i@k
=\frac{|A_{i,k}\cap E_{i,k}|}{|E_{i,k}|}.
\]
Thus eight shared identities at $k=10$ score $0.8$, while a permutation of all
ten scores $1.0$. This is set overlap with Flat, not patch relevance or an
ordered-rank comparison.

\paragraph{Navigation compatibility.}
For request $a_i$, the match indicator is
\[
I_i^{\mathrm{nav}}=\ind[
N(P_{\mathrm{static}}(a_i,M_c))=N(P_{\mathrm{live}}(a_i,c))].
\]
Its mean is the match rate. Providers returning the same normalized set
$\{(\texttt{a.py},10),(\texttt{b.py},30)\}$ match even if character columns,
end ranges, or metadata differ; omitting either pair or changing a start line
is a mismatch. Conditional latency summaries include only requests with
$I_i^{\mathrm{nav}}=1$.

\paragraph{Maintenance output checks.}
A graph transition matches its independent rebuild only when the whole-graph
and serving checks in Eqs.~\eqref{eq:graph-fact-equivalence} and
\eqref{eq:graph-serving-equivalence}, together with the changed-file fact
projection, all hold. A vector transition must reproduce document identities,
numerically equivalent vectors, and exact ordered Flat replay. One missing
typed edge, or a replay with ranks 9 and 10 swapped, fails the corresponding
check even if an auxiliary F1 or set-overlap score is near one. Its raw measured
ratio remains visible, but it is excluded from the matching-transition speedup
summary.

\paragraph{Committed-answer localization.}
Let $Y_i$ be the target source spans and $A_{i,k}$ the first $k$ final-answer
spans after parsing structured locations, resolving usable symbol entries, and
removing overlapping duplicate predictions. A target is covered when an answer
span has the same path and an overlapping inclusive line range:
\[
\mathrm{AnswerRecall}_i@k
=\frac{1}{|Y_i|}\sum_{y\in Y_i}
\ind[\exists a\in A_{i,k}:\operatorname{overlap}(a,y)].
\]
The reported score macro-averages this quantity over instances.
The main result fixes $k=5$ for reporting and does not limit the number of
locations admitted by the answer schema; retrieval and ANN use their separate
top-10 contracts.
For targets
$\{A{:}10\text{-}20,B{:}30\text{-}40,C{:}50\text{-}60\}$, an answer committing
to $A{:}15\text{-}18$, $C{:}55\text{-}65$, and $D{:}1\text{-}8$ covers two
targets and scores $2/3$. An answer with no usable in-scope committed span
scores zero.

\paragraph{Trajectory-token accounting.}
For $m$ recorded model invocations, let $U_t^{\mathrm{in}}$ and
$U_t^{\mathrm{out}}$ be the provider-reported prompt and completion tokens.
The reported trajectory token usage is
\begin{equation}
T_{\mathrm{traj}}=\sum_{t=1}^{m}
\left(U_t^{\mathrm{in}}+U_t^{\mathrm{out}}\right).
\label{eq:trajectory-token-accounting}
\end{equation}
The count $m$ includes any triggered answer-format invocation; it is neither
the number of tool calls nor necessarily the 16-turn main-loop count. For an
observation $o_j$ produced after invocation $j$, let $I_{j,t}=1$ when it is
retained in the prompt for invocation $t$, and let $b_{j,t}$ be its marginal
serialized token count there. Its repeated prompt-token contribution is
\[
V(o_j)=\sum_{t=j+1}^{m} b_{j,t}I_{j,t}.
\]
Only if the observation remains unchanged in every later prompt with marginal
size $b$ does this reduce to $b(m-j)$. For $m=6$, $j=2$, and $b=100$, it appears
in prompts 3--6 and contributes $400$ input tokens. Eviction or a
history rewrite makes the corresponding indicators zero; neither can erase
usage already recorded through invocation $j$. The experiment reports the
direct provider sum in Eq.~\eqref{eq:trajectory-token-accounting}, not this
decomposition, and does not estimate cache hits, prefill work, KV memory,
latency, or billing.

\section{Multilingual Source Units}
\label{app:source-units}

\paragraph{Levels.}
The levels are semantic adapter outputs, not fixed AST depths shared by every
grammar. $L_0$ emits at most one signature-only skeleton per processed file;
it contains top-level declarations and, where available, member signatures.
Files with no recognized declaration produce no $L_0$ document. $L_1$ denotes
the adapter's named top-level constructs. The evaluated $L_2$ configuration sets
\texttt{l2\_level\_exclusive=true}: it retains top-level leaf declarations and
nested members but omits their class, struct, trait, interface, or impl
containers. Table~\ref{tab:multilang-units} lists the concrete five-language
mapping. $L_1$ defines the hierarchy but is not separately materialized in
\qref{1}{sec:eval:rag}--\qref{2}{sec:eval:index}.

\begin{table*}[t]
  \centering
  \scriptsize
  \setlength{\tabcolsep}{4pt}
  \begin{tabularx}{\textwidth}{@{}p{1.2cm}p{2.4cm}XX@{}}
    \toprule
    Language & Extensions/parser & $L_1$ recognized constructs & Evaluated exclusive $L_2$ output \\
    \midrule
    Python & \texttt{.py/.pyi/.pyx}; Python
      & Module functions and classes, including async/decorated forms
      & Module functions and class methods; class containers are omitted. \\
    Go & \texttt{.go}; Go
      & Functions, structs, named types, interfaces, \texttt{var}, and \texttt{const}
      & Functions, receiver methods, \texttt{var}, and \texttt{const}; type containers are omitted. \\
    Rust & \texttt{.rs}; Rust
      & Functions, structs, enums, traits, impls, constants, statics, and type aliases
      & Free functions, methods defined in impl blocks, constants, statics, and type aliases; struct/enum/trait/impl containers are omitted. \\
    C/C++ & \texttt{.c/.h/.cc/}\newline\texttt{.cpp/.cxx/.hpp}; C++
      & Function definitions, classes/structs, non-prototype declarations, and macros
      & Function definitions, in-class method definitions/declarations, non-prototype declarations, and macros; class containers are omitted. \\
    TS/JS & \texttt{.ts/.tsx/.js/.jsx}; JavaScript/TypeScript
      & Functions, classes, initialized variables, object literals, and exported/assigned function forms
      & Functions and assignments, class methods, initialized variables, and objects; class containers are omitted. \\
    \bottomrule
  \end{tabularx}
  \caption{Five-language source-unit definitions; JavaScript and TypeScript share
  one adapter, and \texttt{.c} uses the C++ adapter.}
  \label{tab:multilang-units}
\end{table*}

\paragraph{Repository and document filters.}
Repository traversal uses only the declared language extensions. It excludes
the standard VCS/cache/environment/build directories, files larger than 10
MiB, hidden/binary/backup/minified/bundle basename patterns, and files
containing a line longer than 10,000 characters. A test path has a
\texttt{test}, \texttt{tests}, \texttt{\_\_tests\_\_}, \texttt{spec}, or
\texttt{specs} directory; a basename starting with \texttt{test}; or a
\texttt{\_test}, \texttt{\_spec}, \texttt{.test.}, or \texttt{.spec.} marker.
Headers, imports, module docstrings, and trailing free text are not emitted as
separate $L_2$ chunks.
\qref{1}{sec:eval:rag}--\qref{2}{sec:eval:index} set no line-count split,
preserving each recognized logical unit; model token caps may still truncate
its encoded text. The production compiler used by lifecycle and incremental-vector experiments
instead caps a logical $L_2$ piece at 300 lines. Split pieces retain the same
symbol identity and contiguous ranges.

An $L_2$ document prepends its repository-relative \texttt{path:symbol}
identity to the source span and, for class methods, adds a compact enclosing
class line. An $L_0$ document embeds the signature skeleton while retaining
path and full-file range in metadata. Source metadata stores path, symbol type,
name, node ID, and range alongside each vector.

\section{Parsing and Semantic-Navigation Backends}
\label{app:semantic-backends}

\paragraph{Separation of responsibilities.}
Tree-sitter supplies deterministic syntax trees and source-unit ranges; it
does not resolve imports, types, or cross-file symbols~\cite{treesitter}.
SCIP is an offline interchange format from which CodeNib builds persistent
semantic artifacts~\cite{scip}. LSP is the live workspace protocol used for
the \qref{3}{sec:eval:lsp} reference provider and for
\qref{4}{sec:eval:incremental} repair requests~\cite{lsp}. A file can therefore
remain available to lexical and dense retrieval when its semantic backend is
unavailable; the manifest reports the missing graph or exact-position
capability instead of treating a syntax-only chunk as resolved semantic data.

\begin{table*}[t]
  \centering
  \scriptsize
  \setlength{\tabcolsep}{3.6pt}
  \begin{tabularx}{\textwidth}{@{}p{1.15cm}p{2.55cm}XXp{2.65cm}@{}}
    \toprule
    Group & Tree-sitter source-unit adapter & Cold structural artifact & Live JSON-RPC provider & \qref{3}{sec:eval:lsp} static position path \\
    \midrule
    Python & Python grammar
      & \texttt{scip-python} $\rightarrow$ SCIP
      & \texttt{basedpyright-langserver}
      & SCIP occurrence index, then graph fallback \\
    Go & Go grammar
      & \texttt{scip-go} $\rightarrow$ SCIP
      & \texttt{gopls}
      & SCIP occurrence index, then graph fallback \\
    Rust & Rust grammar
      & \texttt{rust-analyzer scip} $\rightarrow$ SCIP
      & \texttt{rust-analyzer}
      & SCIP occurrence index, then graph fallback \\
    C/C++ & C++ grammar, including \texttt{.c}
      & \texttt{clangd} background \texttt{.idx} plus compilation database
      & \texttt{clangd}
      & graph position/range index \\
    TS/JS & Shared adapter over TypeScript and JavaScript grammars
      & \texttt{scip-typescript} $\rightarrow$ SCIP
      & \texttt{typescript-language-server}
      & SCIP occurrence index, then graph fallback \\
    \bottomrule
  \end{tabularx}
  \caption{Evaluated five-language backend routes. Chunking, cold semantic
  construction, and live serving are selected independently by the language
  registry; C/C++ is the non-SCIP cold-build route.}
  \label{tab:semantic-backends}
\end{table*}

\paragraph{SCIP lowering and persistence.}
For the four SCIP-backed groups, the indexer emits documents keyed by
repository-relative path and occurrences carrying a declared position encoding,
half-open line/character range, resolved symbol identifier, and role bitfield.
CodeNib lowers definition occurrences to source-linked graph vertices and
resolved uses to source-anchored relationships, then writes a versioned
\texttt{graph.pkl}. A separate \texttt{lsp\_index.pkl} retains every occurrence
for native position requests. Its lookup first selects the smallest occurrence
containing the zero-based query position, keys global symbols by SCIP identity
and local symbols by file plus identity, and returns deduplicated, sorted source
locations. Definition roles are selected by the SCIP definition bit; reference
queries optionally retain the declaration. The evaluated limits are eight
definition and 40 reference locations. If an exact occurrence does not yield a
location, the static provider maps the position to the enclosing graph symbol;
provider metadata identifies this fallback behavior. C/C++ instead decodes
clangd background-index records into the graph/range representation and does
not claim a SCIP occurrence path.

\paragraph{Project preparation and capability checks.}
Cold semantic construction runs against the exact detached checkout. Go uses
module metadata, Rust the repository Cargo workspace/toolchain, and TS/JS the
detected npm/yarn/pnpm/bun workspace plus a generated or patched
\texttt{tsconfig}/\texttt{jsconfig} with JavaScript enabled. C/C++ requires a
nonempty \texttt{compile\_commands.json}; the builder reuses one when present or
attempts CMake and Bear-based generation. Its artifact report records compilation
database entries, resolved-path ratio, source coverage, graph coverage, and
range-bearing files. Missing tools, failed preparation, or failed quality gates
produce an unavailable capability rather than a smaller artifact advertised
under the same semantic profile.

\paragraph{Live-LSP replay boundary.}
\qref{3}{sec:eval:lsp} gives the static and live providers the same
repository-relative file, zero-based line/character, options, and limit. Each
of the 100 snapshots supplies five definition and five reference positions.
After JSON-RPC initialization,
the runner requires at least two and at most eight warmup rounds until normalized
live outputs stabilize; it then measures each request ten times. The idle grace
is ten seconds for clangd background indexing and one second for the other
servers. Both definition and reference comparisons use deduplicated,
path/start-line-sorted location sets. Graph loading,
static-provider initialization, live process startup, idle waiting, and warmup
are recorded separately from the reported marginal request latency. Tree-sitter
chunk ranges and SCIP/LSP positions remain zero-based internally; only exported
\texttt{CodeLocation} lines cross the one-based user-facing boundary.

\section{Retrieval Models and Frozen Parameters}
\label{app:retrieval-config}

\paragraph{Embedding stack.}
All five embedders run through SentenceTransformers with their shipped pooling
modules and a Flat inner-product FAISS index. We apply no extra vector
normalization beyond modules contained in each model snapshot. Table
\ref{tab:embedding-config} reports parameter counts obtained by summing the
stored safetensor shapes and architectural fields from the recovered cache
revisions. Those revisions are post-hoc provenance for historical
\qref{1}{sec:eval:rag}--\qref{2}{sec:eval:index} runs, not execution-time
locks; the lifecycle run separately pins Qwen3-Embedding-0.6B at execution
time.

\begin{table*}[t]
  \centering
  \scriptsize
  \setlength{\tabcolsep}{3.4pt}
  \begin{tabularx}{\linewidth}{@{}l l r r l r l X@{}}
    \toprule
    Model & Backbone (layers/heads) & Params & $d$ & Pool/norm & Token cap & Batch & Query/document prefix \\
    \midrule
    SR-Small & NomicBERT (12/12) & 137M & 768 & CLS/none & 8,192 & $8\!\rightarrow\!4$ & shipped \texttt{query}/none \\
    Qwen3-Embed-0.6B & Qwen3 (28/16) & 596M & 1,024 & last/$\ell_2$ & 8,192 & $8\!\rightarrow\!4$ & code-task/empty \\
    Jina-Code-1.5B & Qwen2 (28/12) & 1.54B & 1,536 & last/$\ell_2$ & 8,192 & $4\!\rightarrow\!2$ & \texttt{nl2code\_query}/\texttt{nl2code\_document} \\
    Qwen3-Embed-4B & Qwen3 (36/32) & 4.02B & 2,560 & last/$\ell_2$ & 8,192 & $2\!\rightarrow\!1$ & code-task/empty \\
    SR-Large & Qwen2 (28/28) & 7.07B & 3,584 & last/none & 32,768 & $2\!\rightarrow\!1$ & shipped \texttt{query}/none \\
    \bottomrule
  \end{tabularx}
  \caption{Dense-index configuration; batch arrows denote failure-only retry.}
  \label{tab:embedding-config}
\end{table*}

The SR aliases resolve to \path{Salesforce/SweRankEmbed-Small} and
\path{fishmingyu/SweRankEmbed-Large}. The artifact ledger records all five Hub
IDs plus their recovered commit and tensor hashes.
SR-Small's query prefix is ``Represent this query for searching relevant
code.'' SR-Large and both Qwen embedders use ``Given a github issue, identify
the code that needs to be changed to fix the issue'' in an instruction/query
template; this overrides Qwen's shipped web-search instruction. Jina uses its
shipped natural-language-to-code query and document prompts. The document side
is unprefixed for both SweRank models, explicitly empty for Qwen, and prefixed
with ``Candidate code snippet'' for Jina.

\begin{figure}[t]
  \centering
  \includegraphics[width=\columnwidth,trim=0 5pt 0 5pt,clip]
    {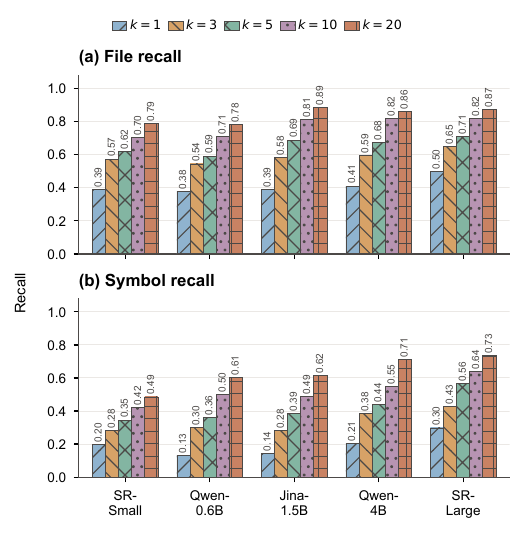}
  \Description{Two grouped bar charts compare five pure embedding models at
  retrieval cutoffs 1, 3, 5, 10, and 20. File recall rises from roughly
  0.38--0.50 at cutoff 1 to 0.78--0.89 at cutoff 20; symbol recall rises from
  roughly 0.13--0.30 to 0.49--0.73.}
  \caption{Pure-embedding Recall@$k$ on the 100 CodeNib Base snapshots used by
  \qref{1}{sec:eval:rag}, without graph expansion or reranking.
  Bar labels are macro means. \textbf{(a)} File results use path deduplication;
  \textbf{(b)} symbol results use $L_2$ identities.
  Figure~\ref{fig:rag-pareto} shows the $k=10$ slice.}
  \label{fig:embedding-cutoff-sensitivity}
\end{figure}

\paragraph{Pointwise reranking.}
Qwen3-Reranker-0.6B, 4B, and 8B use Qwen3 causal backbones with respectively
28/16, 36/32, and 36/32 layers/attention heads. The wrapper formats one
instruction--query--document prompt, reserves an 8,192-token left-padded input,
and scores relevance as
$\exp z_{\mathrm{yes}}/(\exp z_{\mathrm{yes}}+\exp z_{\mathrm{no}})$ at the
last position. Batch size starts at eight and halves on CUDA out-of-memory; a
single still-failing document receives score zero and remains in the recorded
run. The instruction is the same GitHub-issue localization task used for the
Qwen embeddings. Each curve retrieves $k'\in\{30,50,100\}$ dense $L_2$
candidates, reranks them pointwise, and evaluates the top ten. SR-Small is
paired with all three rerankers; Qwen3-Embedding-0.6B, Jina-Code-1.5B, and
Qwen3-Embedding-4B are paired with the 4B and 8B rerankers. SR-Large is a dense
baseline only. The corresponding Hub IDs are
\path{Qwen/Qwen3-Reranker-0.6B}, with analogously suffixed 4B and 8B variants.

\begin{table}[t]
  \centering
  \scriptsize
  \setlength{\tabcolsep}{3.4pt}
  \begin{tabularx}{\columnwidth}{@{}lX@{}}
    \toprule
    Operator & Frozen parameters \\
    \midrule
    Dense & Flat IP; $L_2$; output $k=10$; model-native query/document prompts. \\
    Graph expansion & Dense pool 300; first 10 chunks seed one-hop incoming and outgoing reference expansion; at most 10 unique neighbors/seed; semantic weight 1; graph weight 0.5; RRF constant 60. \\
    IVF-Flat & $n_{\mathrm{list}}=\min(\operatorname{round}(4\sqrt{N}),\lfloor N/39\rfloor)$; probe 1, 2, 5, 10, 25, 50, or 100\% of lists. \\
    HNSW-Flat & $M=32$; $ef_{\mathrm{construction}}=200$; $ef_{\mathrm{search}}\in\{16,32,64,128\}$. \\
    ANN timing & Qwen3-Embedding-0.6B frozen float32 vectors/query; one FAISS CPU thread; search $k=10$; 10 warmups and 100 timed repetitions per method--snapshot cell. \\
    BM25 lifecycle & Production $L_2$ chunks; 300-line chunk cap; maximum requested $k=128$. \\
    \bottomrule
  \end{tabularx}
  \caption{Retrieval/index parameters; graph weight is development-selected
  and ANN recall is against exact Flat top 10.}
  \label{tab:retrieval-parameters}
\end{table}

\section{Incremental-Maintenance Protocol}
\label{app:incremental}

\paragraph{Update sequences and execution.}
The workload selects five first-parent commit transitions from each of eight
repositories, with two repositories per language; hence
$n_{\mathrm{share}}=5$.
File/symbol arms independently advance the same base graph with one long-lived
LSP; fresh targets are independent. In Eq.~\eqref{eq:update-speedups},
$T_u^{G,a}$ covers detection, repair, and persistence, while arm-specific
$T_s^{G,a}$ covers server start and workspace warmup. Each update is therefore
charged $T_s^{G,a}/n_{\mathrm{share}}$. A scheduled transition with no source
change does not alter $n_{\mathrm{share}}$, because the setup was incurred for
the full sequence. Checkout and base loading are excluded. One resumed,
nonmatching Ruff point uses $n_{\mathrm{share}}=1$ and is outside conditional
summaries. Vector fresh/delta arms share a preloaded model; model loading, base
construction, checkout, and post-timing comparison are excluded, and their
targets remain independent. Graph/vector selectors yield 33/31 source-changing
transitions; no-source rows remain outside ratios. Reported protocols are 21/4.

\paragraph{Repair mechanics.}
Before each batch, the patcher synchronizes Git-modified files into the
long-lived server with versioned \texttt{didOpen}/\texttt{didChange}
notifications. It obtains incoming edges with \texttt{references}; method
locations are retained only when a definition query resolves back to that
concrete method, preventing interface--implementation expansion from creating
false call edges. Outgoing repair filters semantic tokens to changed ranges
and resolves definitions by source position. Repeated semantic tokens at the
same \texttt{(line, character)} are deduplicated, but equal token text at
different positions is not because shadowing and imports can resolve it
differently. Latency therefore depends on affected-symbol and reference
fan-out, not only changed-line count.

\paragraph{Offline output comparison.}
After timing, the evaluation compares each update with an independent rebuild.
Graph comparison requires exact vertex and typed anchored-edge multisets, both
globally and on changed-file facts, plus exact normalized definition/reference
replay between the reloaded artifacts. F1 and serving agreement remain
diagnostics, not tolerances. Vector comparison requires equal document
identities, numerically equal reconstructed vectors, and exact ordered Flat
top-10 replay; model revision, dimension, token cap, metric, and chunking levels
are compared explicitly. Fresh-target construction and comparison are excluded
from maintenance latency. Mismatches retain latency but are excluded from the
matching-transition speedup summary.

\begin{table}[H]
  \centering
  \scriptsize
  \setlength{\tabcolsep}{2.8pt}
  \begin{tabular}{@{}llrrrl@{}}
    \toprule
    View & Lang. & $n$ & Exact & Speedup p50 [IQR] & Aux. fidelity \\
    \midrule
    Graph & Go     & 7 & 7 & 8.89 [6.99, 10.01]  & E/B 100/100\% \\
    Graph & Python & 8 & 8 & 6.95 [6.22, 13.00]  & E/B 100/100\% \\
    Graph & Rust   & 9 & 0 & 17.00 [8.97, 40.42]$^*$ & E/B 99.12/97.40\% \\
    Graph & TS/JS  & 9 & 0 & 1.97 [1.32, 7.59]$^*$ & E/B 97.61/99.53\% \\
    \addlinespace[1.5pt]
    Vector & Go     & 7 & 7 & 27.61 [22.18, 31.77] & A/R 7/7 \\
    Vector & Python & 6 & 6 & 38.18 [37.42, 38.75] & A/R 6/6 \\
    Vector & Rust   & 9 & 6 & 19.43 [14.26, 24.79] & A/R 9/6 \\
    Vector & TS/JS  & 9 & 9 & 13.63 [8.92, 25.53]  & A/R 9/9 \\
    \bottomrule
  \end{tabular}
  \caption{Source-changing transitions. Exact counts pass the checks; E/B is
  median edge F1/serving agreement and A/R artifact/replay exact count.
  Unstarred speedups use passing rows; starred values are raw ratios when none
  pass and are excluded from conditional aggregates.}
  \label{tab:incremental-full}
\end{table}

\paragraph{Mismatch audit.}
Across all graph rows, symbol and file maintenance match 15/33 and 14/33
independent rebuilds, with median speedups of 8.67$\times$ and 1.95$\times$ on
those transitions. The 14 rows where both paths match yield a 4.25$\times$
median file/symbol time ratio.
Four independent Rust/TS rebuilds reach 99.75\% median edge F1 and 99.35\%
serving agreement, but one is not exact on the changed scope; strict checks thus
expose maintenance or live-provider repeatability rather than tune a tolerance.
All 31 vector targets match artifact identities and vectors, while three Rust
rows fail exact replay. Protocol 22's later provider-hierarchy fallback is not
pooled with the reported protocol-21 campaign.

\section{Held-Out Maintenance Cohort}
\label{app:maintenance-holdout-section}
\def\MaintenanceHoldoutRows{
Hugo & 5 & 5 & 5 & 5 & 5 \\
Terraform & 3 & 1 & 0 & 3 & 3 \\
Prometheus & 3 & 1 & 1 & 3 & 3 \\
Astropy & 3 & 3 & 3 & 3 & 3 \\
Matplotlib & 5 & 3 & 5 & 4 & 4 \\
SymPy & 5 & 0 & 0 & 5 & 5 \\
\midrule
Nushell & 4 & 0 & 0 & 4 & 1 \\
Babel & 3 & 0 & 0 & 3 & 3 \\}
\def\MaintenanceSetupRows{
Admitted & 13 & 1 & 4.30 & [1.22, 9.58] \\
Admitted & 13 & 5 & 11.62 & [1.23, 16.11] \\
Not admitted & 11 & 1 & 1.74 & [1.71, 2.91] \\
Not admitted & 11 & 5 & 5.53 & [1.76, 5.95] \\}
\def\MaintenanceStabilityRows{
Whole-graph facts & 23 & 5 & 3 \\
Changed-scope facts & 28 & 0 & 3 \\
Navigation replay & 23 & 1 & 7 \\}
\paragraph{Repository-disjoint follow-up.}\label{app:maintenance-holdout}
The follow-up freezes five first-parent transitions in each of eight new
repositories before observing maintenance outcomes. It selects all remaining
Go/Python repositories in the Base universe after excluding the original
cohort, plus one Rust and one TypeScript boundary repository.
Table~\ref{tab:maintenance-holdout} retains every repository. The graph
source-changing denominator is 24 for Go/Python and seven for Rust/TS;
nine no-source positions complete the 40-position workload. Vector filtering
excludes tests and yields 30 source-changing positions.

This cohort uses graph protocol 23 and vector protocol 4. Relative to the
original graph study, symbol repair falls back when matching source
coordinates and names conceal a changed declaration hierarchy. Go's explicit
unsupported-builtin-reference diagnostic maps to an empty repository-reference
set in both fresh construction and repair. Other errors retain their failure
semantics. Both the implementation and repository population change, so we
report this cohort separately from protocol 21.

\begin{table}[htbp]
  \centering
  \small
  \setlength{\tabcolsep}{5pt}
  \begin{tabular}{@{}lrrrrr@{}}
    \toprule
    Repository & Graph $n$ & Symbol & File & Vector $n$ & Vector \\
    \midrule
    \MaintenanceHoldoutRows
    \bottomrule
  \end{tabular}
  \caption{Held-out source-changing denominators and admitted updates.
  Every repository contributes five planned positions. Symbol and File count
  complete graph-comparison passes out of Graph $n$; the final column counts
  complete vector-comparison passes out of Vector $n$. The last two rows are
  the predeclared language-boundary population.}
  \label{tab:maintenance-holdout}
\end{table}

\paragraph{Completion and unavailable comparisons.}
Graph execution yields 26 successful observations, six no-source positions,
and eight failure positions. Five failures share one Babel base-build failure,
whose observed elapsed cost is 4,938.38\,s; this cost is recorded once.
Prometheus file replacement fails on its third transition after an LSP
definition error for Go's \texttt{unsafe} package. Its remaining two positions
are aborted because that file-replacement state is unusable. The symbol arm
and independent rebuild remain available at all three positions and stay in
their respective analyses. Successful execution alone does not establish
output equivalence.

Symbol repair passes 13/24 primary source changes and file replacement passes
14/24. File replacement passes two Matplotlib positions where symbol repair
fails, while symbol repair passes one Terraform position where file
replacement fails. The Rust boundary admits no source-changing position:
four positions have no eligible navigation probes. Babel's three planned
source changes have unavailable graph prerequisites.

\paragraph{Setup sharing and complete-workload costs.}
For each arm, the analysis recomputes update cost as
$t_{\mathrm{update}}+(t_{\mathrm{setup}}+t_{\mathrm{recovery}})/n_{\mathrm{share}}$
from the recorded update, provider-setup, and recovery times.
Table~\ref{tab:maintenance-setup} gives both the isolated-update setting
($n_{\mathrm{share}}=1$) and the executed five-transition sharing setting.
The intervals use 20,000 repository-cluster bootstrap draws and are pointwise.
All positions from a sampled repository move together in each draw.

\begin{table}[htbp]
  \centering
  \small
  \begin{tabular}{@{}lrrrr@{}}
    \toprule
    Primary symbol subset & $n$ & $n_{\mathrm{share}}$ & Median ratio & 95\% interval \\
    \midrule
    \MaintenanceSetupRows
    \bottomrule
  \end{tabular}
  \caption{Per-transition rebuild/update ratios for all 24 Go/Python source
  changes, separated by the complete offline comparison. Non-admitted ratios
  describe time spent producing outputs that fail the comparison and do not
  imply a usable speedup. The statistic is a median of paired ratios.}
  \label{tab:maintenance-setup}
\end{table}

The frozen language policy attempts symbol repair for Go/Python and uses
rebuilds for Rust/TS. Across all 30 primary positions, including six with no
source change, mean update cost is 302.46\,s with isolated setup and
243.97\,s with shared setup; mean reference cost is 686.33\,s.
These trace-derived costs include failed equivalence checks. Every primary
symbol attempt completes, so execution-failure fallback cannot detect its
eleven source-changing mismatches. The five missing Babel costs prevent a
complete 40-position estimate. For no-source graph positions, the reference
worker records zero after reusing its preceding graph; this is a skipped-build
sentinel, not a measured zero-cost request. The analysis reconstructs costs
from separately executed arms and does not measure an end-to-end router.

\paragraph{Independent rebuilds and mismatch attribution.}
The stability phase retains 40 positions: 28 source-changing observations,
seven no-source observations, and five unavailable Babel prerequisites.
Table~\ref{tab:maintenance-stability} separates fact and behavior comparisons
on the 31 planned source changes. Four Rust positions yield zero eligible
behavior requests. Their behavior checks are unmeasured, rather than observed
disagreements; graph-fact checks remain available.

\begin{table}[htbp]
  \centering
  \small
  \begin{tabular}{@{}lrrr@{}}
    \toprule
    Independent rebuild comparison & Exact & Different & Unmeasured \\
    \midrule
    \MaintenanceStabilityRows
    \bottomrule
  \end{tabular}
  \caption{Repeat-rebuild checks over all 31 graph-source changes. The
  unmeasured column includes unavailable prerequisites and, for navigation,
  empty probe populations. Component denominators therefore differ.}
  \label{tab:maintenance-stability}
\end{table}

Among the 24 primary source changes, twelve have matching symbol repair and
matching repeat rebuilds. Seven symbol mismatches occur where the independent
rebuild pair agrees; four accompany a differing rebuild pair. One additional
position has matching symbol repair despite a differing rebuild pair.
These are observable classifications: one repeat cannot establish backend
determinism, and reference drift does not exclude a simultaneous repair error.
Neither classification changes admission or removes a workload position.

\section{Lifecycle Execution Boundary}
\label{app:lifecycle-boundary}

Each materialization uses an isolated, quiescent checkout and fresh output and
runtime processes, but does not flush host caches or reload remote model
weights; reported construction is therefore warm-host, not machine cold-start.
The evaluated vector builder materializes both $L_0$ and $L_2$ even though the
trace queries only $L_2$, and lifecycle cost includes both. After fresh-process
runtime view loading, replay runs one warmup and three measured repetitions;
$S$ is one warm service trace over $N=20$ sessions. Panel (c) evaluates
$B/q+L+S/N$ and $(B+L)/q+S/N$ at $q\in\{1,5,10,20,50,100\}$; only $q=20$ is
the controlled trace scale, while the others project the same measured terms.
At $q=20$, process-isolated and shared-resident projections are
13.20\,s/session (95\% CI 10.43--15.96) and 6.24\,s/session (3.66--8.11);
at $q=100$, medians are 8.53 and 1.27\,s/session. Runtime view loading floors
the former model; the latter approaches measured serve-only $S/N$, not a
hardware or cross-system floor.

\section{Agent-Context Protocol and Validity Details}
\label{app:agent-protocol}

\paragraph{Compaction transition.}
After the complete tool batch containing the first successful \texttt{read},
and before the next model invocation, the runner deterministically restores the
clean issue, deduplicated paths, newest successful read, and a direction cue
copied from the first 600 characters of the latest nonempty assistant message.
Only this cue is capped; the issue, retained read, and final answer are not. It
removes the candidate dump and prior assistant/tool messages; later messages
append normally. Accounting retains all pre-rewrite usage.

\paragraph{Local-model execution.}
Both Qwen models use vLLM 0.23.0 with a 65,536-token server cap. Gemma 4-12B
uses vLLM 0.25.1 with a pinned Hub revision, the versioned Gemma tool template,
and a 131,072-token cap. Local runs use 48,000-token history, 4,096-token
completion windows, and prefix caching. Gemini 2.5 Flash uses Vertex AI with
\texttt{thinkingBudget=0}; manifests retain these controls and access date.
Compact starts one new prefix at its transition; subsequent calls append to and
can reuse that prefix's KV cache. The transition is the policy's single cache
discontinuity: we do not claim that the pre-transition cache survives it.
\qref{5}{sec:eval:context} reports provider-reported trajectory tokens, not
transition latency or cache-adjusted cost.

\paragraph{Answer-format accounting.}
When a run terminates without the required answer schema, the runner may issue
fixed answer-format invocations after early termination or around the 16-turn
cap; their usage is included. Baseline/eager/compact trigger rates are
24.4/16.6/13.2\% (9B), 10.8/7.6/3.6\% (27B), and zero (Haiku).
The corresponding rates are 12.0/1.8/0.6\% for Gemma and 3.2/0/0\% for
Gemini. Invalid answers remain in the quality metric with zero recall.

\paragraph{Direct history-policy contrast.}
For Haiku/9B/27B, compact/eager token ratios are 123.3\% ($[113.4,134.1]$),
87.6\% ($[82.3,93.0]$), and 78.7\% ($[72.9,85.4]$); paired
$\mathrm{AnswerRecall@5}$ changes are
$+0.018$ ($[-0.013,+0.051]$), $-0.023$ ($[-0.052,+0.005]$), and $+0.008$
($[-0.014,+0.032]$). Gemma/Gemini ratios are 27.9\% ($[23.4,32.5]$) and
76.3\% ($[63.9,92.3]$), with recall changes of $-0.070$
($[-0.097,-0.045]$) and $-0.044$ ($[-0.074,-0.018]$); all intervals are 95\%
CIs.

\begin{table}[H]
  \centering
  \scriptsize
  \setlength{\tabcolsep}{2.7pt}
  \renewcommand{\arraystretch}{1.04}
  \begin{tabular}{@{}l*{5}{r}@{}}
    \toprule
    \textbf{Metric} &
    \multicolumn{1}{c}{\textbf{Haiku}} &
    \multicolumn{1}{c}{\textbf{Qwen-9B}} &
    \multicolumn{1}{c}{\textbf{Qwen-27B}} &
    \multicolumn{1}{c}{\textbf{Gemma}} &
    \multicolumn{1}{c}{\textbf{Gemini}} \\
    \textit{Selected policy} &
    \multicolumn{1}{c}{Eager} &
    \multicolumn{1}{c}{Compact} &
    \multicolumn{1}{c}{Compact} &
    \multicolumn{1}{c}{Compact} &
    \multicolumn{1}{c}{Compact} \\
    \midrule
    $k=1$  & $+0.5$ & $+3.0$ & $+0.2$ & $+7.6$ & $+10.0$ \\
    $k=3$  & $-0.5$ & $+0.2$ & $\phantom{+}0.0$ & $+4.5$ & $+7.2$ \\
    \rowcolor{black!7}
    $\mathbf{k=5}$ \textbf{ (main)} &
      $-0.9$ & $-0.6$ & $-0.3$ & $+4.0$ & $+6.7$ \\
    $k=10$ & $-0.7$ & $-0.6$ & $+0.1$ & $+4.0$ & $+6.6$ \\
    \midrule
    \shortstack[l]{Worst 95\% CI\\lower bound} &
      $-4.3$ & $-4.9$ & $-3.7$ & $-1.2$ & $+2.1$ \\
    \bottomrule
  \end{tabular}
  \caption{\qref{5}{sec:eval:context} final-answer cutoff sensitivity.
  Cutoff rows are selected-policy-minus-grep/read changes in
  $\mathrm{AnswerRecall}$, in percentage points; $k=5$ is the shaded main-text
  setting. Reapplying the $-5.0$-point lower-bound gate and minimum-token rule
  leaves each selected policy unchanged. The final row is the worst lower
  endpoint among the four snapshot-clustered 95\% CIs.}
  \label{tab:agent-cutoff-sensitivity}
\end{table}

\paragraph{Validity.}\label{app:validity} Removing the five judge-warning queries
does not change which policy meets the localization margin. Haiku backfill
retains the observed model ID and harness, but provider-time drift remains
unresolved.

\section{Author Contributions}
\label{app:author-contributions}

{
\noindent\textbf{Zhongming Yu:} Conceptualization and Methodology (lead);
Software (lead: LSP, retrieval, incremental indexes, agent/web infrastructure,
CI/CD, and deployment); Data curation; Investigation; Formal analysis;
Validation; Visualization; Writing -- original draft; Writing -- review and
editing; Project administration; Supervision.
\textbf{Hengjia Yu:} Software (LSP, incremental graphs, and agent
infrastructure); Investigation; Validation; Visualization.
\textbf{Boqin Yuan:} Data curation; Software (dataset, web, and deployment).
\textbf{Shuting Zhao:} Methodology and Software (retrieval).
\textbf{Yizhao Chen:} Software (agent infrastructure).
\textbf{Aryan Dokania:} Software (graph retrieval); Investigation; Formal
analysis. \textbf{Mihir Jagtap:} Software (incremental vector indexing).
\textbf{Jiayu Chang:} Software (agent infrastructure).
\textbf{Yitong Ma:} Visualization; Writing -- review.
\textbf{Yash Jayswal:} Software (web); Writing -- review.
\textbf{Wentao Ni:} Methodology (vector indexes); Writing -- review.
\textbf{Hejia Zhang:} Software (agent infrastructure); Writing -- review.
\textbf{Zhaoling Chen:} Software (agent infrastructure); Writing -- review.
\textbf{Gangda Deng:} Conceptualization; Methodology (graph retrieval, agent
methods, benchmarks, and experiments); Writing -- review.
\textbf{Jishen Zhao:} Supervision (PI); Project
administration; Writing -- review.
}

\section{Additional Distributional Views}
\label{app:distributional-views}

Figure~\ref{fig:agent-cutoff-sensitivity} visualizes
Table~\ref{tab:agent-cutoff-sensitivity}, and
Figure~\ref{fig:agent-runtime-breakdown} expands
Figure~\ref{fig:agent-runtime}(c).
Figures~\ref{fig:lsp-replay-groups} and~\ref{fig:lsp-replay-ecdf} retain the
superseded navigation replay of the first version of this paper: a
capped-output run against occurrence indexes decoded by the text decoder whose
range-parsing defect Sec.~\ref{sec:eval:lsp} reports. That run matched on 632
of 1,000 requests (437/500 definitions, 195/500 references) and measured a
median live/static ratio of 4.72$\times$ on matching requests; the corrected
uncapped replay in Sec.~\ref{sec:eval:lsp} and Figure~\ref{fig:nav-agreement}
supersedes both numbers.

\begin{figure}[H]
  \centering
  \includegraphics[width=\columnwidth]
    {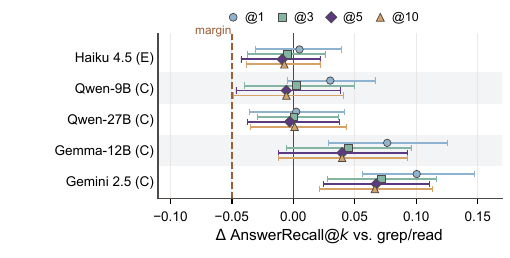}
  \Description{A forest plot shows the paired AnswerRecall changes and
  snapshot-clustered confidence intervals of the selected Eager or Compact arm
  at cutoffs 1, 3, 5, and 10 for five agent models. A dashed vertical line marks
  the negative 0.05 reporting margin and a solid line marks zero.}
  \caption{Complete interval view behind
  Table~\ref{tab:agent-cutoff-sensitivity}. Points are paired mean
  $\Delta\mathrm{AnswerRecall@}k$ from grep/read; whiskers are
  snapshot-clustered 95\% CIs. The Eager/Compact arm named beside each model is
  invariant across the four cutoffs.}
  \label{fig:agent-cutoff-sensitivity}
\end{figure}

  \newpage

\begin{figure}[H]
  \centering
  \includegraphics[width=0.95\columnwidth]{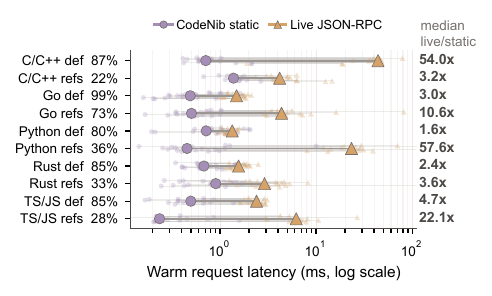}
  \Description{Static-index and live language-server replay latency by
  language and capability. Each row shows faint per-snapshot median pairs,
  pooled static and live request medians, its normalized-location match rate,
  and the median per-request live-to-static latency ratio.}
  \caption{Superseded capped-provider replay by language and capability.
  Labels give normalized path/start-line match rates; faint pairs are
  per-snapshot medians of request-level repetition medians. Large points are
  pooled request medians, and right-hand values are medians of per-request
  live/static ratios. Nonmatches are excluded from all latency summaries.}
  \label{fig:lsp-replay-groups}
\end{figure}

\begin{figure}[H]
  \centering
  \includegraphics[width=\columnwidth]{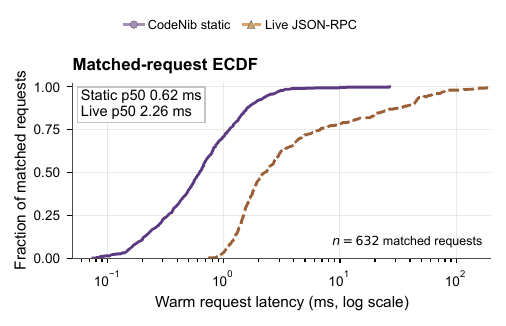}
  \Description{Empirical cumulative distributions compare warm static-index
  and live JSON-RPC latency over the 632 navigation requests whose normalized
  path and start-line sets match. The static distribution lies to the left of
  the live distribution, with medians of 0.62 and 2.26 milliseconds.}
  \caption{Superseded capped-provider matched-request latency distribution.
  Each observation is one request's median over ten repetitions. The ECDF
  excludes 368 nonmatching requests, startup, loading, and warmup; it
  characterizes the conditional compatible subset of that run, not an
  achieved workload speedup.}
  \label{fig:lsp-replay-ecdf}
\end{figure}

\begin{figure}[H]
  \centering
  \includegraphics[width=\columnwidth]{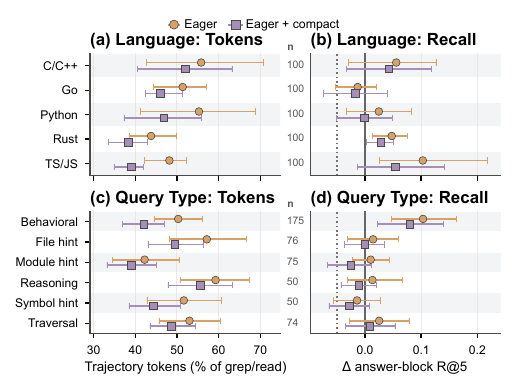}
  \Description{Four forest plots expand token and AnswerRecall-at-5 effects by
  programming language and query type for Eager and Eager plus Compact against
  paired grep and read. Whiskers show snapshot-clustered confidence intervals.}
  \caption{Expanded workload-slice view underlying
  Figure~\ref{fig:agent-runtime}(c).
  \textbf{(a--b)} Language and \textbf{(c--d)} query-type effects on
  provider-reported tokens and $\mathrm{AnswerRecall@5}$.
  Whiskers are snapshot-clustered 95\% CIs. These descriptive intervals do not
  define a subgroup-routing policy.}
  \label{fig:agent-runtime-breakdown}
\end{figure}

\def\ProviderConstructionRows{
Astropy 14182 & 97.10 & 2706.07 & 54.46 & 251.25 & 0.57 & 10.03 \\
Django 10973 & 147.39 & 206.55 & 58.89 & 308.17 & 0.62 & 3.12 \\
Django 13121 & 181.90 & 219.52 & 63.02 & 330.30 & 0.64 & 3.35 \\
Django 13551 & 181.84 & 218.43 & 64.15 & 336.15 & 0.66 & 3.45 \\
Django 14725 & 192.52 & 249.12 & 68.22 & 358.31 & 0.67 & 3.73 \\
Django 15814 & 176.09 & 234.55 & 71.70 & 365.94 & 0.70 & 3.82 \\
Matplotlib 21568 & 80.58 & 451.82 & 37.30 & 171.79 & 0.70 & 1.97 \\
scikit-learn 13124 & 68.10 & 1602.67 & 33.87 & 171.69 & 0.70 & 3.63 \\
SymPy 13615 & 452.45 & 971.12 & 97.16 & 802.67 & 0.88 & 9.18 \\
SymPy 20438 & 3502.50 & 1504.05 & 115.12 & 1307.57 & 1.64 & 11.42 \\
\midrule
Mean & 508.05 & 836.39 & 66.39 & 440.38 & 0.78 & 5.37 \\
Median & 178.97 & 350.47 & 63.59 & 333.22 & 0.68 & 3.68 \\}
\def\ProviderCutoffRows{
File recall & Native & 0.525 & 0.675 & 0.675 \\
 & \nickname & 0.400 & 0.500 & 0.500 \\
\rowcolor{black!5}
Symbol recall & Native & 0.400 & 0.450 & 0.450 \\
\rowcolor{black!5}
 & \nickname & 0.300 & 0.300 & 0.300 \\
File all-target success & Native & 0.500 & 0.600 & 0.600 \\
 & \nickname & 0.400 & 0.500 & 0.500 \\
\rowcolor{black!5}
Symbol all-target success & Native & 0.400 & 0.400 & 0.400 \\
\rowcolor{black!5}
 & \nickname & 0.200 & 0.200 & 0.200 \\}
\def\ProviderCostIntervalRows{
Cell elapsed time (s) & 10 & $-30.78$ & $[-58.01,-3.56]$ \\
\rowcolor{black!5}
Tokens (thousands) & 9 & $-72.56$ & $[-134.19,-10.94]$ \\
Provider load (s) & 7 & $+6.00$ & $[+1.98,+10.01]$ \\
\rowcolor{black!5}
Tool time (s) & 7 & $-7.19$ & $[-14.10,-0.29]$ \\}
\section{Provider-Substitution Protocol and Full Results}
\label{app:provider-study}

Two provider studies share the loop, the checkpoint, and the limits described
below. The main paper reports the later cohort, 30 tasks selected by a fixed
hash ranking over the 200-task Verified pool with at most three per
repository, disjoint from the earlier sample, each run with three paired
repeats for 180 cells, in its rerun under a shared tool boundary
(Appendix~\ref{app:provider-normalized}). This appendix retains the shared
configuration and the earlier ten-task study in full;
Appendix~\ref{app:provider-expanded} reports the first execution of the
30-task cohort.

That earlier sample contains five Django tasks, two SymPy tasks, and one each
from Astropy, Matplotlib, and scikit-learn, with one run per provider and an
alternating seeded starting arm. Both inventories have ten ready artifacts,
and earlier construction diagnostics and interrupted attempts remain separate
from its twenty evaluated cells. Its file and symbol Recall@5 point estimates
favor the native provider by 17.5 and 15.0 points with intervals spanning
large gains and losses; the later study does not reproduce that direction,
and its own primary interval also spans zero.

\paragraph{Execution configuration.}
The LocAgent protocol revision begins \texttt{ef170542a5cc}; the Qwen3.5-27B
checkpoint revision begins \texttt{fc05daec18b0}. The artifact retains both
full commit identifiers. Both arms use that checkpoint, served locally by vLLM
0.23.0 on one H100 80-GB GPU in bfloat16, with a 65,536-token window and
thinking disabled. The request leaves sampling parameters unset and uses the
checkpoint defaults: temperature 0.6, top-$p$ 0.95, and top-$k$ 20.
The loop allows ten exploration iterations followed by a final-answer turn,
with a 4,096-token response ceiling. Limits are 600 seconds per cell,
60 seconds per tool, and cumulative 300k input/20k output tokens. The token
limit is checked after responses and can overshoot by one response.
At most two explicit transport retries are permitted for the frozen transient
error classes; every started attempt remains in the ledger. Semantic failures
and model-window errors do not trigger new scientific cells.
The \nickname adapter loads only BM25 and symbol-graph views and caps each
rendered response at 48,000 characters. Native rendering follows the frozen
upstream implementation. Response selection, content, and truncation are
therefore part of the provider intervention; the study does not isolate the
manifest or one view in that provider.

\paragraph{Index construction and retained storage.}
Table~\ref{tab:provider-construction} reports the successful construction
attempt that produced each evaluated artifact. Both providers build their
graph and BM25 indexes from the same source commit. The supervised clock
includes worker startup, imports, construction, and exit; checkout
materialization, environment installation, and model-server startup lie
outside it. Worker and stage timers are nested components of that clock.
These shared-host measurements are descriptive results from one retained
successful attempt per provider and task. They are separate from the
predeclared four-metric serving-cost inference below.

\begin{table*}[t]
  \caption{Retained construction costs on the ten evaluated tasks. N denotes
  native and C denotes \nickname. Time measures the complete successful worker
  process. Storage counts logical bytes in the retained view directories,
  plus the \nickname manifest. RSS columns record different scopes: N uses
  the Python worker's own high-water mark; C uses the maximum of its own and
  waited-for children's high-water marks. Neither is simultaneous process-tree
  memory, and the two columns do not support a paired memory-saving claim.}
  \label{tab:provider-construction}
  \centering
  \small
  \setlength{\tabcolsep}{5pt}
  \begin{tabular}{@{}lrrrrrr@{}}
    \toprule
    & \multicolumn{2}{c}{Build (s)} & \multicolumn{2}{c}{Storage (MiB)}
      & \multicolumn{2}{c}{Recorded RSS (GiB)} \\
    \cmidrule(lr){2-3}\cmidrule(lr){4-5}\cmidrule(l){6-7}
    \textbf{Task} & N & C & N & C & N & C \\
    \midrule
    \ProviderConstructionRows
    \bottomrule
  \end{tabular}
\end{table*}

The storage audit counts retained files after the provider runs. Source
checkouts, installed environments, model weights, logs, and earlier failed
artifacts are excluded. \nickname's directories include raw SCIP, decoded,
and cached navigation artifacts, so these sizes describe retained storage
with the executed configuration, not a minimum serving footprint. Native
view-tree hashes match their pre-execution bindings. The \nickname manifest
also matches its pre-execution hash; the audit records its view-file
inventory after collection. Different graph builders produce different
artifacts, and these differences cannot be attributed to manifest overhead.

One earlier \nickname Matplotlib attempt retains a failed worker receipt
with \texttt{SystemExit} after 318.82\,s, but no supervisor terminal receipt.
That worker duration contributes an additional cost lower bound. Across
the supplied construction batches, successful supervised times sum to
5,080.48\,s for native and 8,363.90\,s for \nickname; adding the earlier
worker raises the latter lower bound to 8,682.73\,s. These are sums of observed
attempt durations, not end-to-end elapsed times. Earlier diagnostic batches
and setup costs remain outside this accounting.

\paragraph{Targets and denominators.}
All ten tasks have file and existing-symbol labels, derived from the developer
patch and held outside the agent inputs. File predictions are deduplicated
repository-relative paths. Symbol predictions are qualified source identities;
the retained metric field is named \texttt{function}, but includes class
identities as well as functions and methods. Recall is the fraction of gold
targets recovered in the first $k$ predictions. All-target success requires
every corresponding gold target in those predictions. Execution failures
contribute empty predictions. Table~\ref{tab:provider-cutoffs} retains the same
ten tasks at every cutoff.

\begin{table*}[t]
  \caption{Provider localization across all ten tasks. All-target success
  differs from recall when a task has multiple targets. The two native
  token-limit failures and one native context-window failure remain included.}
  \label{tab:provider-cutoffs}
  \centering
  \small
  \setlength{\tabcolsep}{6pt}
  \begin{tabular}{@{}llrrr@{}}
    \toprule
    \textbf{Measure} & \textbf{Provider} & $k=1$ & $k=3$ & $k=5$ \\
    \midrule
    \ProviderCutoffRows
    \bottomrule
  \end{tabular}
\end{table*}

\paragraph{Uncertainty and missing measurements.}
The frozen analysis uses 20,000 paired repository-clustered bootstrap draws,
seed 20260904, and task-weighted means over resampled repository clusters.
Simultaneous intervals use maximum standardized deviation within the
predeclared localization/availability and complete-case cost families.
The file and symbol Recall@5 differences are $-0.175$ and $-0.150$, with
simultaneous 95\% intervals $[-0.711,0.361]$ and $[-0.743,0.443]$.
All localization intervals include zero. The five repository clusters and
single run per cell limit inference about other tasks or model-run variation.

Token accounting is complete for nine paired tasks. The context-window
failure has incomplete total usage, which remains unknown. Load and tool
timers survive in only seven native results, so these components use seven
identical task pairs on both sides. Provider-specific means over different
task subsets must not be subtracted to estimate a paired effect.
The native load counter covers issue/index setup after its subprocess starts
and imports; the \nickname counter wraps manifest loading in the controller.
These component scopes differ. Complete cell time includes both paths from
provider initialization and supplies the common operational comparison.

\paragraph{Uniform cell-time accounting.}
The original result field \texttt{elapsed\_seconds} starts at the agent loop
for successful cells but at the outer execution deadline for failed cells.
We retain that historical analysis and correct this mixed scope using the
independently recorded \texttt{observed\_cell\_elapsed\_ns} field for every
cell. This outer clock includes provider initialization, the agent loop,
cleanup, and attempt-ledger sealing, while excluding prior index construction
and model-server startup. Load and tool timers are components of this clock.
The correction changes no task, response, prediction, failure, or token count;
the original resampling configuration applies to the corrected time values.
Mean complete cell time becomes 97.789\,s for native and 67.006\,s for
\nickname over all ten tasks. Table~\ref{tab:provider-cost-intervals} uses this
uniform clock and recomputes the same simultaneous cost family. The original
mixed-scope means, 94.493 and 59.502\,s, are retained only in the audit history.

\begin{table}[t]
  \caption{Corrected cost effects, \nickname minus native. Each row uses
  identical complete task pairs on both sides and a simultaneous 95\% interval
  from the four-metric cost family. Components must not be added to cell time.}
  \label{tab:provider-cost-intervals}
  \centering
  \small
  \setlength{\tabcolsep}{6pt}
  \begin{tabular}{@{}lrrr@{}}
    \toprule
    \textbf{Measure} & $n$ & \textbf{Mean difference} & \textbf{95\% interval} \\
    \midrule
    \ProviderCostIntervalRows
    \bottomrule
  \end{tabular}
\end{table}

\FloatBarrier
\providecommand{\ProviderExpandedTasks}{30}
\providecommand{\ProviderExpandedRepositories}{11}
\providecommand{\ProviderExpandedRepeats}{3}
\providecommand{\ProviderExpandedCells}{180}
\providecommand{\ProviderExpandedNativeRecall}{52.22}
\providecommand{\ProviderExpandedCodeNibRecall}{55.00}
\providecommand{\ProviderExpandedRecallDifference}{+2.78}
\providecommand{\ProviderExpandedRecallLower}{-19.44}
\providecommand{\ProviderExpandedRecallUpper}{+25.00}
\providecommand{\ProviderExpandedTokenTasks}{21}
\def\ProviderExpandedExecution{
Finished execution & 61 & 82 \\
Execution failure & 29 & 5 \\
Unavailable artifact & 0 & 3 \\
Cumulative-token limit & 16 & 4 \\
Request context-window limit & 13 & 1 \\
Construction unavailable & 0 & 3 \\}
\def\ProviderExpandedQuality{
File Recall@5 (primary) & 30 & 52.22 & 55.00 & $+2.78$ & $[-19.44,+25.00]$ \\
File Recall@1 & 30 & 46.39 & 49.44 & $+3.06$ & $[-18.33,+25.00]$ \\
File Recall@3 & 30 & 52.22 & 53.89 & $+1.67$ & $[-20.16,+23.48]$ \\
Symbol Recall@5 & 29 & 39.66 & 44.92 & $+5.27$ & $[-16.05,+25.57]$ \\}
\def\ProviderExpandedCost{
Complete cell time (s) & 30 & 90.73 & 70.52 & $-20.21$ & $[-40.46,-4.09]$ \\
Model-request wall time (s) & 30 & 82.40 & 66.55 & $-15.85$ & $[-35.73,-0.30]$ \\
Tool-dispatch wall time (s) & 30 & 4.55 & 0.20 & $-4.35$ & $[-8.23,-1.72]$ \\
Input tokens (thousands) & 21 & 215.35 & 157.77 & $-57.58$ & $[-86.75,-28.03]$ \\
Output tokens (thousands) & 21 & 1.46 & 1.31 & $-0.14$ & $[-0.41,+0.13]$ \\
Total tokens (thousands) & 21 & 216.81 & 159.08 & $-57.72$ & $[-86.95,-28.16]$ \\}
\def\ProviderExpandedConditionalCost{
Complete cell time (s) & 14 & 78.51 & 60.50 & $-18.01$ & $[-29.81,-4.91]$ \\
Total tokens (thousands) & 14 & 194.94 & 137.57 & $-57.37$ & $[-91.76,-28.17]$ \\
Reported provider load (s) & 14 & 0.16 & 2.39 & $+2.24$ & $[+0.91,+3.87]$ \\
Historical tool timer (s) & 14 & 2.21 & 0.17 & $-2.04$ & $[-3.39,-0.94]$ \\}
\def\ProviderExpandedConstruction{
Supervised build (s) & 29 & 209.89 & 591.71 & $+381.82$ & $[-107.28,+997.64]$ \\
Worker build (s) & 30 & 286.29 & 612.80 & $+326.52$ & $[-189.91,+953.40]$ \\
Retained files (MiB) & 30 & 35.10 & 241.89 & $+206.79$ & $[+60.99,+437.15]$ \\}
\section{Expanded Provider Study with Repeated Runs}
\label{app:provider-expanded}

This is the first execution of the 30-task cohort, run before the shared
tool boundary of Appendix~\ref{app:provider-normalized}: its \nickname arm
rejected JSON-encoded array arguments that native accepted, failing 147 of
240 tree-traversal calls, and capped only its own observations at 48,000
characters. The main text reports the rerun; this appendix retains the first
execution in full, and the two are never pooled. It contains
\ProviderExpandedTasks{} tasks from \ProviderExpandedRepositories{}
repositories, with two providers and \ProviderExpandedRepeats{} runs per
task and provider. All \ProviderExpandedCells{} scheduled cells have
recorded outcomes.
The original ten-task study remains separately reported in
Appendix~\ref{app:provider-study}; its results are not pooled with this extension.

\paragraph{Selection and execution.}
A frozen hash ranking with repository-balanced round-robin selection caps
each repository at three tasks. Selection excludes the original ten Q4 tasks
and precedes model execution. The tasks come from the existing 200-task
Verified pool, so they overlap the Q5 sampling pool.
Each native/\nickname pair runs adjacently; each provider runs first in 45
of the 90 pairs. The model checkpoint, LocAgent-compatible policy, prompts,
tool schemas, and per-cell limits follow Appendix~\ref{app:provider-study}.
Three repetitions share one continuous local Qwen3.5-27B serving session,
using checkpoint sampling defaults and engine seed zero. They are repeated
trajectories, without separately assigned request seeds.

\paragraph{Primary estimator.}
For each task, we average File Recall@5 over its three runs per provider,
then average the paired differences across all 30 tasks. Observed execution
failures and unavailable artifacts contribute empty predictions.
The analysis was frozen before execution: 20,000 paired hierarchical
bootstrap draws resample repositories, tasks within repositories, and paired
repetition IDs, with PCG64 seed 20260909. The analysis has 30 task units and
11 repository clusters. Table~\ref{tab:provider-expanded-quality} reports
the primary result and separately labeled exploratory cutoffs.

\begin{table*}[t]
  \caption{Expanded-study localization. Provider means are percentages;
  differences and intervals are percentage points, \nickname minus native.
  File Recall@5 is the frozen primary outcome. Other rows use exploratory
  pointwise 95\% intervals without multiplicity adjustment. All observed
  failures remain included. Symbol recall uses the 29 tasks with existing-symbol
  labels; as in the original study, these include classes and functions.}
  \label{tab:provider-expanded-quality}
  \centering
  \small
  \setlength{\tabcolsep}{4pt}
  \begin{tabularx}{\textwidth}{@{}Xrrrrr@{}}
    \toprule
    \textbf{Measure} & $n$ & \textbf{Native} & \textbf{\nickname}
      & $\boldsymbol{\Delta}$ & \textbf{95\% interval} \\
    \midrule
    \ProviderExpandedQuality
    \bottomrule
  \end{tabularx}
\end{table*}

The primary difference is \ProviderExpandedRecallDifference{} percentage
points, with a 95\% interval from \ProviderExpandedRecallLower{} to
\ProviderExpandedRecallUpper{}. This interval supports neither superiority
nor equivalence or non-inferiority. The corrected-range reconstruction
reproduces the original ten graphs exactly, so this extension does not
identify an effect of that decoder correction.

\begin{table}[t]
  \caption{Terminal outcomes for 90 scheduled cells per provider.
  The first three rows partition all cells. The last three rows decompose
  execution failures and unavailable artifacts; they must not be added to
  the first three. Finished execution does not imply correct localization.
  These terminal reasons are descriptive, without graph-defect attribution.}
  \label{tab:provider-expanded-execution}
  \centering
  \small
  \begin{tabular}{@{}lrr@{}}
    \toprule
    \textbf{Outcome or terminal reason} & \textbf{Native} & \textbf{\nickname} \\
    \midrule
    \ProviderExpandedExecution
    \bottomrule
  \end{tabular}
\end{table}

\paragraph{Tool-call outcomes.}
The saved ledgers contain 147 errors among 883 \nickname tool calls and two
among 825 native calls. These differ from trajectory terminal failures in
Table~\ref{tab:provider-expanded-execution}: completed trajectories can contain
tool errors. All 147 \nickname errors occur in tree traversal when the model
supplies a JSON-encoded string for \texttt{dependency\_type\_filter}, whose
schema declares an array or null. The adapter rejects the resulting character
sequence. Native returns without a tool exception on 118 such string-filter
calls, without establishing correct traversal semantics. Thus the provider
comparison includes argument-handling differences; completion and cost
contrasts do not isolate graph-representation effects. Original arguments and
observations remain unchanged.

\paragraph{Incurred costs and accounting coverage.}
Table~\ref{tab:provider-expanded-cost} includes failed executions.
Complete cell time starts at provider initialization and includes loading,
the policy loop, cleanup, and ledger sealing. Prior index construction,
model-server startup, and pre-cell artifact hash checks lie outside this clock.
Model-request wall time includes failed calls and permitted transport retries;
it is elapsed SDK-request time, not GPU computation time. Tool-dispatch wall
time includes retained failed dispatch attempts. These components overlap
the complete-cell clock and must not be added to it.

\begin{table*}[t]
  \caption{Expanded-study incurred costs. Each row averages three repetitions
  per provider and task over identical task pairs. Time rows cover all 30
  tasks, including observed failures and proven no-call zeros. Token rows
  cover only 21 tasks with complete usage in all six cells. Intervals are
  exploratory, pointwise paired hierarchical 95\% intervals; $\Delta$ is
  \nickname minus native.}
  \label{tab:provider-expanded-cost}
  \centering
  \small
  \setlength{\tabcolsep}{4pt}
  \begin{tabularx}{\textwidth}{@{}Xrrrrr@{}}
    \toprule
    \textbf{Measure} & $n$ & \textbf{Native} & \textbf{\nickname}
      & $\boldsymbol{\Delta}$ & \textbf{95\% interval} \\
    \midrule
    \ProviderExpandedCost
    \bottomrule
  \end{tabularx}
\end{table*}

Fifteen cells retain incomplete usage: fourteen native and one \nickname.
Requiring all six usage records leaves \ProviderExpandedTokenTasks{} paired
tasks. These records support a conditional token contrast; full-cohort token
savings remain unknown. The three unavailable \nickname cells reuse one
failed scikit-learn construction. They issue no model calls, incur zero
execution cost, and retain zero localization credit. Construction cost is
charged once for that task and provider.

Table~\ref{tab:provider-expanded-conditional} conditions further on both
providers finishing all three runs, leaving 14 tasks. This subset describes
completed trajectories and does not replace the all-task comparison.
The reported loading timers have different scopes: native excludes worker
startup and imports, whereas \nickname times manifest loading. Complete
cell time includes both startup paths.

\begin{table*}[t]
  \caption{Costs conditional on both providers completing all three runs
  (14 tasks). Intervals are exploratory and pointwise; $\Delta$ is \nickname
  minus native. Reported load and historical tool timers are nested,
  provider-specific components, not full cold-readiness measurements.}
  \label{tab:provider-expanded-conditional}
  \centering
  \small
  \setlength{\tabcolsep}{4pt}
  \begin{tabularx}{\textwidth}{@{}Xrrrrr@{}}
    \toprule
    \textbf{Measure} & $n$ & \textbf{Native} & \textbf{\nickname}
      & $\boldsymbol{\Delta}$ & \textbf{95\% interval} \\
    \midrule
    \ProviderExpandedConditionalCost
    \bottomrule
  \end{tabularx}
\end{table*}

\paragraph{Construction and retained artifacts.}
All 60 build attempts remain included: native has 30 ready artifacts,
and \nickname has 29 plus one SCIP timeout after 3,603.30\,s.
One successful native SymPy build lacks the supervisor's terminal receipt.
Its worker record and retained artifact were verified without rebuilding;
the outer duration remains unknown. Table~\ref{tab:provider-expanded-build}
keeps the 29-task paired outer clock separate from the 30-task worker clock.
Worker time is never substituted for the missing outer duration.

\begin{table*}[t]
  \caption{Expanded-study construction costs, one build per provider and
  task. Failed attempts remain included. Each row uses identical task pairs;
  $\Delta$ is \nickname minus native. Intervals are exploratory, pointwise
  repository/task bootstrap intervals. Retained bytes include SCIP
  intermediates and failed-build files, excluding source checkouts,
  environments, logs, and model weights. This is retained storage rather
  than minimum serving storage.}
  \label{tab:provider-expanded-build}
  \centering
  \small
  \setlength{\tabcolsep}{4pt}
  \begin{tabularx}{\textwidth}{@{}Xrrrrr@{}}
    \toprule
    \textbf{Measure} & $n$ & \textbf{Native} & \textbf{\nickname}
      & $\boldsymbol{\Delta}$ & \textbf{95\% interval} \\
    \midrule
    \ProviderExpandedConstruction
    \bottomrule
  \end{tabularx}
\end{table*}

Recorded RSS counters differ in scope across providers and do not support
a paired process-tree memory comparison. The lower incurred execution-cost
estimates also do not establish quality-preserving acceleration.

\paragraph{Verification.}
The original frozen scorer reproduces all 180 scores. Independent executions
of the archived primary analyzer, secondary summarizer, and accounting auditor
reproduce their reports byte for byte. The accounting audit verifies 177
sealed ledgers, 1,880 outbound request digests, and metadata from 1,864
responses. Capture-time response checks cover identifiers, model, fingerprint,
and usage; full response text is inventoried at audit time.
All failures, incomplete usage, original records, and frozen configurations
remain retained. The table export verifies report hashes, matching replay
files, and the retained input-file hashes without issuing model calls.
\FloatBarrier

\def\ProviderNormalizedQuality{
File Recall@5 (primary) & 30/30 & 51.67 & 60.37 & +8.70 & $[-12.41,+29.50]$ \\
File Recall@1 & 30/30 & 48.61 & 52.96 & +4.35 & $[-15.13,+23.09]$ \\
File Recall@3 & 30/30 & 51.39 & 60.09 & +8.70 & $[-12.37,+29.37]$ \\
Symbol Recall@5 & 29/29 & 40.80 & 44.35 & +3.54 & $[-15.56,+22.58]$ \\}
\def\ProviderNormalizedExecution{
Finished execution & 65 & 78 \\
Execution failure & 25 & 9 \\
Unavailable artifact & 0 & 3 \\
Unknown outcome & 0 & 0 \\}
\def\ProviderNormalizedCost{
Complete cell time (s) & 30/30 & 93.54 & 74.92 & -18.62 & $[-31.90,-7.68]$ \\
Model-request wall time (s) & 30/30 & 82.22 & 70.22 & -12.00 & $[-25.09,-1.59]$ \\
Tool-dispatch wall time (s) & 30/30 & 5.68 & 0.30 & -5.38 & $[-8.81,-2.43]$ \\
Input tokens (thousands) & 23/30 & 221.15 & 178.00 & -43.15 & $[-81.48,-13.55]$ \\
Output tokens (thousands) & 23/30 & 1.64 & 1.30 & -0.34 & $[-0.68,-0.06]$ \\
Total tokens (thousands) & 23/30 & 222.79 & 179.31 & -43.48 & $[-82.01,-13.80]$ \\}
\def\ProviderNormalizedConditionalCost{
Complete cell time (s) & 13/30 & 81.95 & 69.33 & -12.62 & $[-23.39,-3.08]$ \\
Total tokens (thousands) & 13/30 & 184.40 & 156.54 & -27.87 & $[-62.84,+3.57]$ \\
Reported provider load (s) & 13/30 & 0.26 & 3.56 & +3.30 & $[+1.13,+6.03]$ \\
Historical tool timer (s) & 13/30 & 4.10 & 0.48 & -3.62 & $[-7.13,-1.61]$ \\}
\section{Provider Comparison with Shared Argument and Output Handling}
\label{app:provider-normalized}

This correction reuses the 30 tasks, 11 repositories, indexes, and paired
three-repeat schedule of Appendix~\ref{app:provider-expanded}. We specified
it after diagnosing that study's argument-handling and output-budget
differences. It is a same-cohort follow-up; the historical results remain
separately reported. The model checkpoint, prompts, tool schemas, execution
limits, and primary estimator are unchanged.

\paragraph{Shared interface.}
Both providers accept arrays or one JSON serialization layer for the six
array-valued tool arguments. The transform preserves values, order, and
duplicates and checks element types. Successful provider outputs share a
48,000-character limit, measured in Unicode code points and including any
truncation marker. The experiment bypasses \nickname's previous final cap
before applying this common limit. Provider-internal selection limits remain
unchanged. Error messages retain the policy's formatting; the sole observed
tool error has a 56-character body. All 288 tree-traversal calls complete
without a tool error. These interface changes act jointly, so the follow-up
does not separate their individual effects or isolate graph representation.

\paragraph{Localization and execution.}
All 180 scheduled cells have known outcomes. We apply the same 30-task paired
estimator and 20,000-draw repository/task/repeat bootstrap as in the historical
extension, including empty predictions for observed failures and unavailable
artifacts. Table~\ref{tab:provider-normalized-quality} reports localization;
its primary interval spans both signs. Table~\ref{tab:provider-normalized-execution}
reports terminal outcomes. Execution failures comprise seven cumulative-token
and two context-window failures for \nickname, versus 15 and ten for native.
The three unavailable-artifact cells reuse the original failed construction.

\begin{table*}[t]
  \caption{Localization with shared argument and output handling. $n/N$ gives
  paired/eligible tasks; three repeats are nested within each task. Provider
  means are percentages, and $\Delta$ and its interval are percentage points,
  \nickname minus native. File Recall@5 is primary. Other rows use exploratory
  pointwise 95\% intervals. Symbol recall uses existing-class/function labels.}
  \label{tab:provider-normalized-quality}
  \centering
  \small
  \setlength{\tabcolsep}{4pt}
  \begin{tabularx}{\textwidth}{@{}Xrrrrr@{}}
    \toprule
    \textbf{Measure} & $n/N$ & \textbf{Native} & \textbf{\nickname}
      & $\boldsymbol{\Delta}$ & \textbf{95\% interval} \\
    \midrule
    \ProviderNormalizedQuality
    \bottomrule
  \end{tabularx}
\end{table*}

\begin{table}[t]
  \caption{Disjoint terminal outcomes for 90 scheduled cells per provider
  in the shared-interface correction. Finished execution alone does not
  establish correct localization. All failures remain in the analysis.}
  \label{tab:provider-normalized-execution}
  \centering
  \small
  \begin{tabular}{@{}lrr@{}}
    \toprule
    \textbf{Outcome} & \textbf{Native} & \textbf{\nickname} \\
    \midrule
    \ProviderNormalizedExecution
    \bottomrule
  \end{tabular}
\end{table}

\paragraph{Cost coverage.}
Table~\ref{tab:provider-normalized-cost} retains the historical clock scopes:
complete cell time includes provider startup, loading, policy execution, and
cleanup, while prior construction, server startup, and pre-cell artifact
checks lie outside that clock. Its components overlap and must not be added.
Twelve cells have incomplete reported token usage, ten native and two
\nickname; requiring complete usage in all six cells leaves 23 paired tasks.
Thus the token contrast describes that subset. The all-task time interval is
below zero, while the localization interval does not establish superiority,
equivalence, or non-inferiority.

\begin{table*}[t]
  \caption{Incurred costs with shared argument and output handling, including
  failed executions wherever a metric is known. $n/N$ gives paired/eligible
  tasks. Each row averages three repeats per provider and task. $\Delta$ is
  \nickname minus native; intervals are exploratory and pointwise. Token
  contrasts use complete-usage pairs and do not estimate full-cohort savings.}
  \label{tab:provider-normalized-cost}
  \centering
  \small
  \setlength{\tabcolsep}{4pt}
  \begin{tabularx}{\textwidth}{@{}Xrrrrr@{}}
    \toprule
    \textbf{Measure} & $n/N$ & \textbf{Native} & \textbf{\nickname}
      & $\boldsymbol{\Delta}$ & \textbf{95\% interval} \\
    \midrule
    \ProviderNormalizedCost
    \bottomrule
  \end{tabularx}
\end{table*}

Table~\ref{tab:provider-normalized-conditional} further conditions on both
providers completing all three runs, leaving 13 tasks. These conditional
results supplement the all-task comparison. Construction was reused;
Appendix~\ref{app:provider-expanded} retains its costs and missing-clock scope.

\begin{table*}[t]
  \caption{Costs conditional on both providers finishing all three runs in
  the shared-interface correction. $n/N$ is paired/eligible tasks; $\Delta$ is
  \nickname minus native. Intervals are exploratory and pointwise. Reported
  load and historical tool timers retain their provider-specific scopes.}
  \label{tab:provider-normalized-conditional}
  \centering
  \small
  \setlength{\tabcolsep}{4pt}
  \begin{tabularx}{\textwidth}{@{}Xrrrrr@{}}
    \toprule
    \textbf{Measure} & $n/N$ & \textbf{Native} & \textbf{\nickname}
      & $\boldsymbol{\Delta}$ & \textbf{95\% interval} \\
    \midrule
    \ProviderNormalizedConditionalCost
    \bottomrule
  \end{tabularx}
\end{table*}

\paragraph{Recomputation.}
The full audit checks 1,886 model requests, 1,874 responses, and 177 execution
ledgers. A separate CPU invocation reproduces the audit, all 180 saved-answer
scores, primary input rows, and both statistical summaries byte for byte.
This check reuses recorded answers and outputs; re-executing the corrected
providers is a separate check. That CPU-only replay reproduces all 1,717 saved
calls across the 180-cell schedule, including raw and delivered success outputs,
normalized arguments, boundary metadata, and the saved exception. Independent
readback verifies each stored response against its original execution ledger.
An additional archive-only check reproduces the numerical outputs and all saved
provider calls from packaged inputs in isolated CPU containers. These checks
do not rerun model decisions or measure new latency. Final anonymity and
distribution review remain separate from recorded-result reproduction.

\section{Response-Length Control for Provider Substitution}
\label{app:provider-bytematched}

This control reruns only the native arm of the shared-boundary study
(Appendix~\ref{app:provider-normalized}) and compares it with that study's
retained \nickname cells; the historical native arm is never pooled. It keeps
the 30 tasks, three repeats, frozen pair order, checkpoint, prompts, schemas,
limits, and shared array-argument boundary. The one change is that, after the
shared 48,000-code-point cap, each native observation is cut at the
\nickname per-call median for that tool in the shared-boundary run:
\texttt{search\_code\_snippets} 3,681, \texttt{get\_entity\_contents} 9,535,
and \texttt{explore\_tree\_structure} 1,364 code points, with the same
provider-neutral marker. These are per-tool ceilings on Unicode code points,
not exact matches of response distributions or trajectories; the harness
prefix of 13 code points sits outside the cap. Raw, shared-capped, and
delivered lengths are retained for every call.

\begin{table*}[t]
  \caption{Capped native against the retained \nickname arm. Recall rows use
  the frozen paired estimator over 30 tasks (29 for symbols); cost rows
  condition on the 24 tasks whose six cells all complete and use pointwise
  intervals. Construction is reused, not remeasured.}
  \label{tab:provider-bytematched}
  \centering
  \small
  \setlength{\tabcolsep}{5pt}
  \begin{tabular}{@{}lrrrrr@{}}
    \toprule
    \textbf{Measure} & $n$ & \textbf{Capped native} & \textbf{\nickname}
      & $\boldsymbol{\Delta}$ & \textbf{95\% interval} \\
    \midrule
    Finished cells (of 90) & 30 & 90 & 78 & $-12$ & \\
    File Recall@5 (\%) & 30 & 45.00 & 60.37 & $+15.37$ & $[-5.56,\,+35.47]$ \\
    Symbol Recall@5 (\%) & 29 & 33.72 & 44.35 & $+10.63$ & $[-9.44,\,+29.76]$ \\
    Complete cell time (s) & 24 & 58.25 & 74.77 & $+16.52$ & $[+9.96,\,+23.58]$ \\
    Model-request wall time (s) & 24 & 49.71 & 70.19 & $+20.49$ & $[+13.87,\,+27.64]$ \\
    Tool-dispatch wall time (s) & 24 & 3.78 & 0.29 & $-3.50$ & $[-6.27,\,-1.60]$ \\
    Input tokens (thousands) & 24 & 104.51 & 181.00 & $+76.49$ & $[+55.90,\,+101.14]$ \\
    \bottomrule
  \end{tabular}
\end{table*}

All 90 capped native cells complete; \nickname's non-completions are the
three construction-unavailable cells, seven cumulative-token limits, and two
context-window limits already reported. Table~\ref{tab:provider-bytematched}
lists the paired results. The control therefore removes the completion and
cost advantages of Sec.~\ref{sec:eval:provider} and reverses their sign,
while \nickname keeps faster tool dispatch and higher localization point
estimates whose intervals include zero. Cutting native's observations
lowers its File Recall@5 from 51.67 in the shared-boundary run to 45.00 here,
which is the reading we give in the main text: the two providers differ in
what a delivered byte carries. The rerun is a native-only, median-derived
ceiling study on one cohort and does not identify how much of the
shared-boundary difference each tool contributes.

\section{Residual Navigation Mismatches: A Case Study}
\label{app:navigation-residual}

From the 235 path/start-line requests on which the corrected static provider
and the live server disagree, a seeded, language-stratified sample of 50
\texttt{references} requests was drawn: all three Go mismatches, and 12 each
for C/C++, Python, and Rust and 11 for TS/JS. For each case the request, both
answer sets, the static occurrence or graph records at the position (SCIP for
four languages, the clangd-derived graph for C/C++), and the saved server
logs were assembled. Two independent AI-assisted passes, locked before
unblinding, assigned one primary cause from four: indexer coverage (dynamic
dispatch, re-exports, macros, files outside the compile database), decoder or
normalization, a scope or identity difference between the two answers, and
live drift. Agreement before adjudication was 35/50 ($\kappa=0.56$); each of
the 15 disagreements has a written adjudication. Fifteen cases keep an
explicit low-confidence flag within their category.

\begin{table}[t]
  \caption{Primary cause per language for 50 sampled reference mismatches.
  Counts describe mechanisms in the sample and are not prevalence estimates
  over all residual traffic. A scope or identity difference records that the
  two answers resolve different sets, not that either is correct.}
  \label{tab:navigation-residual}
  \centering
  \small
  \setlength{\tabcolsep}{4pt}
  \begin{tabular}{@{}lrrrrr@{}}
    \toprule
    \textbf{Lang.} & $n$ & \textbf{Cover.} & \textbf{Decode}
      & \textbf{Scope} & \textbf{Drift} \\
    \midrule
    C/C++ & 12 & 7 & 4 & 1 & 0 \\
    Go & 3 & 1 & 0 & 2 & 0 \\
    Python & 12 & 7 & 0 & 5 & 0 \\
    Rust & 12 & 1 & 1 & 10 & 0 \\
    TS/JS & 11 & 0 & 6 & 3 & 2 \\
    \midrule
    Total & 50 & 16 & 11 & 21 & 2 \\
    \bottomrule
  \end{tabular}
\end{table}

The sample points at three distinct remedies. C/C++ and Python mismatches
are mostly coverage, where the static index has no occurrence for the
symbol; Rust mismatches are mostly scope, where both providers answer but
resolve different identities; and TS/JS mismatches include six decoding or
normalization cases, three of which coincide with the replay client sending
\texttt{.cjs}/\texttt{.mjs} files as plain text, a candidate cause that this
study records but does not test. The passes were AI-assisted rather than
human, share one evidence collection, and lack per-request server transcripts,
so the attributions are provisional.

\section{Second Budget Level for the Repair Study}
\label{app:repair-budget2}

To test whether the compact-versus-On-demand gap depends on the USD~1 stop
rule and on Django, both policies were rerun on all 108 non-Django tasks of
the repair sample with the per-cell budget doubled to USD~2 and every other
limit unchanged. All 216 cells have terminal records; one task
(\texttt{psf\_\_requests-1142}) fails before any model call under both
policies because its pinned image carries untracked build files, exactly as
in the original study, and its outcome stays unknown. The frozen estimator of
Sec.~\ref{sec:eval:repair} is reused without change.

Compact resolves 50 tasks against 46 for On-demand, with identification
bounds of 46.3--47.2\% and 42.6--43.5\% over all 108 tasks. On the 107
complete pairs, compact alone resolves 11 and On-demand alone 7, an exact
McNemar $p$ of 0.48; the paired difference is $+3.7$ points with task-paired
interval $[-3.7,+11.2]$ and repository-clustered interval $[-6.7,+13.4]$.
Both policies still hit the token ceiling on about half of the tasks (49 and
52 cells), so the doubled dollar budget did not remove the binding limit; it
raised both resolution rates and narrowed the gap. Observed cost was USD~95
for compact and USD~99 for On-demand. This follow-up is a same-sample
sensitivity analysis under one model, not an independent replication, and its
single unknown pair keeps token inference ineligible.

\def\RepairUsageRows{On-demand & 158/200 & $\geq$ 714.61 & $\geq$ 11.09 & $\geq$ 0.847 \\
BM25 eager & 167/200 & $\geq$ 750.94 & $\geq$ 10.06 & $\geq$ 0.881 \\
Aider repo-map & 157/200 & $\geq$ 798.67 & $\geq$ 9.62 & $\geq$ 0.931 \\
CodeNib eager & 157/200 & $\geq$ 733.01 & $\geq$ 10.15 & $\geq$ 0.862 \\
CodeNib compact & 161/200 & $\geq$ 684.23 & $\geq$ 9.92 & $\geq$ 0.807 \\}
\def\RepairPolicyRows{On-demand & 66/200 & 1/200 & $[33.0,\, 33.5]$ \\
BM25 eager & 63/200 & 1/200 & $[31.5,\, 32.0]$ \\
Aider repo-map & 56/200 & 1/200 & $[28.0,\, 28.5]$ \\
CodeNib eager & 69/200 & 2/200 & $[34.5,\, 35.5]$ \\
CodeNib compact & 79/200 & 1/200 & $[39.5,\, 40.0]$ \\}
\section{Real-Repair Protocol and Outcome Observability}
\label{app:repair-study}
\providecommand{\HaikuCompletePairs}{199}
\providecommand{\HaikuCompleteDelta}{+6.53}
\providecommand{\HaikuCompleteTaskCI}{[+1.01,\,+12.06]}
\providecommand{\HaikuCompleteRepoCI}{[-3.92,\,+17.42]}
\providecommand{\HaikuCompleteExactP}{0.0294}

\paragraph{Population and frozen execution.}
The repair study samples 200 tasks from SWE-bench Verified after excluding all
50 IDs in the engineering-pilot pool. Repository-stratified largest-remainder
allocation and SHA-256 ordering determine the sample before treatment
outcomes. The frozen population contains 92 Django, 33 SymPy, 15 Matplotlib,
14 scikit-learn, ten Astropy, ten xarray, eight pytest, eight Sphinx, five
Pylint, four Requests, and one Seaborn task. Ten preselected tasks receive two
additional runs with every policy. The 100 extra cells measure variability;
the 1,000 repetition-zero cells supply the primary estimates.

The experiment pins mini-SWE-agent~\cite{sweagent} at commit
\texttt{bc85a456} and the official SWE-bench evaluator at
\texttt{87ab1f6c}. Full commit identities and source hashes are retained in
the execution records. Each runner case contains only its instance ID,
repository, base commit, and issue body. The independent evaluator receives
the golden patch, test patch, and test partitions. Runner and evaluator use
the same immutable per-task container image.

\paragraph{Model and limits.}
The hosted primary requests \texttt{vertex\_ai/}\allowbreak\texttt{claude-haiku-4-5}
in \texttt{us-east5}, as observed on 2026-09-05, with temperature zero and no
provider decoding seed. This alias does not identify immutable weights.
Each cell has limits of 100 model steps, 3,600 wall-seconds, 60 seconds per
shell command, one million cumulative input tokens, 30,000 cumulative output
tokens, and 4,096 tokens per completion. The agent stops after a response
reaches its USD~1 usage-based cost limit; this is distinct from a hard
per-request billing cap. Four task workers run concurrently, with policies
serial within each task. The phase freezes policy order by a hash permutation
inside each task/repetition block. Repetition identities label runs rather
than provider random seeds.

\paragraph{Context interventions.}
All five policies use the same prompt, Bash tool, base checkout, submission
format, and limits. On-demand receives no initial repository context. BM25
and dense eager receive source-addressed candidate packs; Aider eager receives
the raw upstream RepoMap~\cite{aider}. Dense retrieval uses the frozen
Qwen3-Embedding-0.6B exact index~\cite{qwen3-embed}. Candidate packs retain at
most ten nonoverlapping $L_2$ blocks under a 16-KiB byte ceiling. The same
byte ceiling applies to the RepoMap, whose tokenizer and representation remain
upstream-specific. These arms therefore compare complete context policies,
not equal-byte or equal-candidate interventions.

Dense eager and CodeNib compact begin with byte-identical context. Compact
changes only the initial candidate block in later model queries, replacing it
with a content-hash receipt after the first attested source read. The detector
accepts one literal \texttt{cat} or bounded \texttt{sed} action over a safe
repository-relative path, optionally prefixed with \texttt{/testbed/} or the
literal \texttt{cd /testbed \&\&} command. Before the action, the runner
attests a regular UTF-8 file in the same container. Its nonempty successful
output must match those bytes and occur in the agent-visible observation.
Pipes, redirects, unsupported readers, arbitrary chaining, expansion, and
missing or mismatched attestations preserve the full history. Detector I/O
and time remain part of execution. The issue, observed source, and editing
and testing history are preserved; the original raw trajectory retains the
initial candidate bytes.

\paragraph{Observed compaction use.}
\label{app:repair-compaction-use}
Across all 200 primary Compact tasks, 193 send at least one request with the
candidate block replaced, six retain the original block, and one lacks an
auditable projection. These counts include unsuccessful repairs and incomplete
usage. Among triggered tasks, the first compacted request has median position
five (range three to 37), counting started provider attempts including retries.
The 199 available projections verify 1,360 full and 7,181 compacted request
hashes, with the remaining history preserved. This use does not recover
missing token counters.

\paragraph{Failures and reporting correction.}
Continuation stages execute only never-started identities, preserving every
earlier attempt. They make no whole-cell retries or outcome-based task
replacement. A narrowly checked readiness-parser amendment accepts a known
Python startup warning before otherwise valid JSON; the scientific worker,
model, prompt, candidates, and policy settings remain frozen.

The original auditor jointly gated outcome reporting and complete billing.
The separate post-execution reporting correction retains known test results
when an earlier provider failure omitted usage. It also derives non-resolution
for completed no-patch attempts only from source-bound terminal, process, and
trajectory evidence. Original rows remain unchanged. A submitted patch with
missing grading, an incomplete runtime, or a pre-call failure keeps an unknown
repair outcome. Unknown usage stays unknown even when the outcome is known.
The correction first replays every original raw cell and then records a proof
for each derived label. It changes reporting rather than execution.

\paragraph{Outcome ranges and inference.}
For a policy with $s$ known resolutions and $u$ unknown outcomes among 200
tasks, its full-population resolution range is
$[s/200,(s+u)/200]$. Paired-difference ranges assign both possible outcomes
to each unknown member of the same fixed pair. These are identification
ranges, not confidence intervals; a narrow range does not imply statistical
significance. These full-population ranges retain every task and leave unknown
outcomes unimputed.

The predeclared primary contrast is Compact minus On-demand. The four
secondary contrasts are Dense eager minus On-demand, Dense eager minus BM25
eager, Dense eager minus Aider eager, and Compact minus Dense eager.
Only a contrast with all 200 paired outcomes known receives its original
20,000-draw task-paired percentile-bootstrap interval and exact two-sided
McNemar test. Repository-then-task resampling supplies a separate sensitivity
interval. The four secondary tests retain their complete Holm family; if any
member is unavailable, all four adjusted values remain unavailable. There is
no noninferiority margin or post-hoc policy selection.

\def\RepairContrastRows{CodeNib compact vs. On-demand (primary) & 1 & $[6.0,\, 7.0]$ & -- & -- \\
CodeNib eager vs. On-demand & 2 & $[1.0,\, 2.5]$ & -- & -- \\
CodeNib eager vs. BM25 eager & 2 & $[2.5,\, 4.0]$ & -- & -- \\
CodeNib eager vs. Aider repo-map & 2 & $[6.0,\, 7.5]$ & -- & -- \\
CodeNib compact vs. CodeNib eager & 2 & $[4.0,\, 5.5]$ & -- & -- \\}
\begin{table*}[htbp]
  \centering
  \small
  \setlength{\tabcolsep}{3pt}
  \begin{tabularx}{\textwidth}{@{}Xrrrr@{}}
    \toprule
    Contrast & Unknown pairs & Range (pp) & Estimate [95\% CI] & $p$ \\
    \midrule
    \RepairContrastRows
    \bottomrule
  \end{tabularx}
  \caption{All five frozen Haiku resolution contrasts. Ranges retain unknown
  outcomes in the 200-pair denominator. Dashes denote unavailable inference,
  not a zero effect. The primary test is separate from the four-member
  secondary Holm family.}
  \label{tab:repair-contrasts}
\end{table*}

The same missing-outcome rule leaves all five original
repository-hierarchical intervals unavailable. The reproducibility records
retain these entries alongside the task-paired results.

\paragraph{Observed Complete Pairs.}
\label{app:repair-complete-pairs}
A post-execution supplement estimates each of the five original contrasts
on tasks with both repair outcomes observed. It retains known outcomes with
incomplete billing and uses the original 20,000-draw methods and seed.
These conditional estimates describe the observed subsets; the original
200-task ranges and missing-outcome sensitivity remain separate.

Compact and On-demand have \HaikuCompletePairs{} complete pairs, with 79 and
66 resolutions. Compact alone resolves 22 tasks and On-demand alone nine.
The paired difference is $\HaikuCompleteDelta$ percentage points, with
task-paired 95\% interval $\HaikuCompleteTaskCI$ and exact McNemar
$p=\HaikuCompleteExactP$. Repository-then-task resampling gives
$\HaikuCompleteRepoCI$, which includes zero. Every contrast involving Eager
has 198 complete pairs; all four secondary Holm-adjusted tests exceed 0.05.
Conditioning on observed outcomes does not establish robustness to missing
outcomes or to sampling different repositories.

\def\RepairCompleteRows{Compact vs. On-demand & 199 & $+6.53$ & $[+1.01,\,+12.06]$ & $[-3.92,\,+17.42]$ & 0.0294 \\
Eager vs. On-demand & 198 & $+1.52$ & $[-4.55,\,+7.58]$ & $[-9.80,\,+9.92]$ & 0.7428 \\
Eager vs. BM25 & 198 & $+3.03$ & $[-2.02,\,+8.08]$ & $[-3.33,\,+13.54]$ & 0.6899 \\
Eager vs. Aider & 198 & $+6.57$ & $[+1.01,\,+12.12]$ & $[+0.50,\,+16.54]$ & 0.1178 \\
Compact vs. Eager & 198 & $+5.05$ & $[+0.51,\,+9.60]$ & $[-0.48,\,+13.79]$ & 0.1917 \\}
\begin{table*}[htbp]
  \centering
  \small
  \setlength{\tabcolsep}{3pt}
  \begin{tabularx}{\textwidth}{@{}Xrrrrr@{}}
    \toprule
    Contrast & Pairs & $\Delta$ (pp) & Task 95\% CI (pp)
      & Repo. 95\% CI (pp) & $p$ \\
    \midrule
    \RepairCompleteRows
    \bottomrule
  \end{tabularx}
  \caption{Post-execution Haiku estimates on observed outcome pairs.
  Eager and Compact denote the CodeNib policies; BM25 and Aider denote their
  eager context arms. Intervals are pointwise. The primary exact test is
  unadjusted; the four secondary exact tests receive Holm correction.
  Pair counts condition only on outcome availability, independently of billing.
  Table~\ref{tab:repair-contrasts} retains the original 200-pair analysis.}
  \label{tab:repair-complete-pairs}
\end{table*}

\begin{table}[htbp]
  \caption{Primary Haiku repair outcomes over all 200 tasks per policy
  (Aider repo-map is the Aider eager policy). The resolution range assigns
  each unknown outcome both possible values; it is not a confidence interval.
  Known resolutions require official test evidence. Additional repetitions do
  not enter these counts.}
  \label{tab:repair-outcomes}
  \centering
  \small
  \setlength{\tabcolsep}{4pt}
  \begin{tabular}{@{}lrrr@{}}
    \toprule
    \textbf{Policy} & \textbf{Resolved} & \textbf{Unknown}
      & \textbf{Range (\%)} \\
    \midrule
    \RepairPolicyRows
    \bottomrule
  \end{tabular}
\end{table}

\begin{table*}[htbp]
  \caption{Observed Haiku usage per planned task. Input and output are in
  thousands of provider-reported tokens; amounts use the frozen regional list rates of USD~1.10 and
  5.50 per million input and output tokens. A $\geq$ marker denotes incomplete usage and a downward-rounded lower
  bound. These are neither complete token means nor invoices and cannot
  establish a token-saving ratio.}
  \label{tab:repair-usage}
  \centering
  \small
  \setlength{\tabcolsep}{5pt}
  \begin{tabular}{@{}lrrrr@{}}
    \toprule
    \textbf{Policy} & \textbf{Usage-Complete} & \textbf{Input (k)}
      & \textbf{Output (k)} & \textbf{USD/Task} \\
    \midrule
    \RepairUsageRows
    \bottomrule
  \end{tabular}
\end{table*}

\paragraph{Exhaustive missing-outcome sensitivity.}
\label{app:repair-missing-sensitivity}
A separate post-execution analysis enumerates all $2^6=64$ binary assignments
to the six unknown primary outcomes. Each completion retains all 200 task
pairs, the frozen resampling settings, and the same four-test Holm family.
Shared task/policy outcomes take the same hypothetical value in every contrast.
For each resampling scheme, Table~\ref{tab:repair-missing-sensitivity} gives the
smallest lower and largest upper endpoint across all completion-specific
95\% intervals. These interval hulls inherit the percentile bootstrap's
approximate coverage and sampling assumptions; they are not simultaneous
intervals or finite-sample coverage guarantees.

The primary $p_{\max}$ is the largest exact McNemar $p$-value over completions.
Each secondary value maximizes its Holm-adjusted $p$-value over the same joint
completions. The true outcome vector, if observed, must be among them, so
maximization preserves the underlying complete-data test's error control.
It does not remove task dependence or sampling limits. For Compact versus
On-demand, the task-paired interval hull lies above zero, whereas the
repository-hierarchical hull spans $-4.9$ to $+18.5$ percentage points.
The worst-case exact test also exceeds 0.05. The approximate percentile
interval and discrete exact test use different procedures and need not
yield identical threshold decisions. Original outcomes and inference
eligibility remain unchanged; this sensitivity supplies no missing usage.

\def\RepairMissingRows{CodeNib compact vs. On-demand & $[+0.5,\, +12.5]$ & $[-4.9,\, +18.5]$ & 0.0502 \\
CodeNib eager vs. On-demand & $[-5.0,\, +8.5]$ & $[-10.7,\, +12.0]$ & 0.9166 \\
CodeNib eager vs. BM25 eager & $[-3.0,\, +9.5]$ & $[-4.3,\, +16.1]$ & 0.9166 \\
CodeNib eager vs. Aider repo-map & $[+0.5,\, +13.0]$ & $[-0.7,\, +18.7]$ & 0.2005 \\
CodeNib compact vs. CodeNib eager & $[-1.0,\, +10.5]$ & $[-3.0,\, +14.9]$ & 0.5060 \\}
\begin{table*}[htbp]
  \centering
  \small
  \setlength{\tabcolsep}{4pt}
  \begin{tabularx}{\textwidth}{@{}Xrrr@{}}
    \toprule
    Contrast & Task-paired hull (pp) & Repository hull (pp) & $p_{\max}$ \\
    \midrule
    \RepairMissingRows
    \bottomrule
  \end{tabularx}
  \caption{Post-execution sensitivity over all missing-outcome completions.
  Hulls cover all completion-specific two-sided 95\% bootstrap intervals.
  The primary test is unadjusted; the four secondary tests retain Holm
  adjustment before maximization. Printed $p$-values round upward to four
  decimal places.}
  \label{tab:repair-missing-sensitivity}
\end{table*}

\paragraph{Usage and process time.}
Input/output counts sum observed provider responses, including responses
before later retries. Missing response usage makes the total a lower bound.
Regional-list-rate amounts use the frozen USD~1.10/5.50 per million
input/output rates; they are observed usage-based amounts rather than
invoices. All per-task usage amounts divide by 200, including tasks with
incomplete counters. Lower bounds cannot establish a token-saving ratio.

Outer agent-process time includes the process's setup, model interaction,
and cleanup. Each mean prints its observed denominator. Context-build,
image-preparation, nested runtime, and official-grader clocks have separate
scopes and are not summed into an end-to-end or cold-policy latency.
Model-server startup and local GPU allocation also require separate accounting.

\def\RepairTimeRows{On-demand & 176.50 & 200/200 & 0/200 \\
BM25 eager & 163.76 & 200/200 & 0/200 \\
Aider repo-map & 163.10 & 200/200 & 0/200 \\
CodeNib eager & 167.53 & 200/200 & 0/200 \\
CodeNib compact & 168.48 & 200/200 & 0/200 \\}
\begin{table}[htbp]
  \centering
  \small
  \begin{tabular}{@{}lrrr@{}}
    \toprule
    Policy & Mean agent-process seconds & Observed & Missing \\
    \midrule
    \RepairTimeRows
    \bottomrule
  \end{tabular}
  \caption{Observed outer agent-process clocks for the primary Haiku cohort.
  Missing times are neither zero nor inferred from token usage.}
  \label{tab:repair-time}
\end{table}

\def\RepairRepeatRows{On-demand & 6/10 & 4/10 & 0/10 \\
BM25 eager & 7/10 & 3/10 & 0/10 \\
Aider repo-map & 9/10 & 1/10 & 0/10 \\
CodeNib eager & 10/10 & 0/10 & 0/10 \\
CodeNib compact & 8/10 & 2/10 & 0/10 \\}
\begin{table}[htbp]
  \centering
  \small
  \begin{tabular}{@{}lrrr@{}}
    \toprule
    Policy & Equal outcomes & Different outcomes & Unknown \\
    \midrule
    \RepairRepeatRows
    \bottomrule
  \end{tabular}
  \caption{Three-run outcome agreement on ten preselected repeated tasks.
  Each row retains all ten tasks; additional runs never enter the 200-task
  primary estimates. Equal outcomes may be successes or failures.}
  \label{tab:repair-repeat}
\end{table}

\FloatBarrier
\subsection{Preparation Components and Retained Storage}
\label{app:repair-preparation}

All 200 tasks have successful BM25/dense preparation and ready Aider maps.
Table~\ref{tab:repair-preparation-time} separates the worker that builds both
views and candidate packs from its nested timers. Mean worker time is
149.21\,s (median 83.73\,s); Aider's mean preparation-process time is 35.21\,s.
These measurements describe different preparation scopes.

\def\RepairPrepTimeRows{Context worker (both views) & 149.206 & 83.727 & 200/200 \\
BM25 build & 4.480 & 5.378 & 200/200 \\
Dense build & 134.486 & 67.583 & 200/200 \\
BM25 load & 0.293 & 0.191 & 200/200 \\
BM25 retrieval & 0.361 & 0.276 & 200/200 \\
Dense document load & 0.036 & 0.024 & 200/200 \\
FAISS load & 0.020 & 0.017 & 200/200 \\
Dense retrieval & 0.134 & 0.100 & 200/200 \\
BM25 pack construction & 0.002 & 0.002 & 200/200 \\
Dense pack construction & 0.002 & 0.002 & 200/200 \\
Aider preparation process & 35.208 & 39.450 & 200/200 \\
Aider map build & 33.155 & 37.381 & 200/200 \\}
\begin{table}[htbp]
  \centering
  \small
  \setlength{\tabcolsep}{8pt}
  \begin{tabular}{@{}lrrr@{}}
    \toprule
    Component & Mean (s) & Median (s) & Observed \\
    \midrule
    \RepairPrepTimeRows
    \bottomrule
  \end{tabular}
  \caption{Retained preparation measurements for the 200 repair tasks.
  Build, load, retrieval, and pack timers are nested in the context worker;
  map build is nested in the Aider process. Rows must not be added.
  These are preparation components, separate from agent and grader clocks.}
  \label{tab:repair-preparation-time}
\end{table}

Successful context workers accumulate 29,841.19 process-seconds.
Four prior failures add 106.83\,s; one interrupted attempt lacks a clock.
The resulting lower bound, 29,948.02\,s, is summed work rather than cohort
elapsed time. Across all phases, 213 recorded image-preparation events cover
200 tasks and total 4,301.26\,s for pin verification, optional digest pulls,
and inspection. Bound audit lineage excludes copied timing records from
these totals.

\def\RepairPrepStorageRows{BM25 retained files & 8.502 & 7.374 & 200/200 \\
Dense retained files & 40.546 & 36.677 & 200/200 \\
Aider raw map & 0.013 & 0.013 & 200/200 \\}
\begin{table}[htbp]
  \centering
  \small
  \setlength{\tabcolsep}{8pt}
  \begin{tabular}{@{}lrrr@{}}
    \toprule
    Retained representation & Mean (MiB) & Median (MiB) & Observed \\
    \midrule
    \RepairPrepStorageRows
    \bottomrule
  \end{tabular}
  \caption{Retained index and map bytes, with $1$ MiB $=2^{20}$ bytes.
  BM25/dense include the files named by their build receipts; Aider includes
  only its raw UTF-8 map. These different persistence scopes exclude
  checkouts, model weights, and preparation logs.}
  \label{tab:repair-preparation-storage}
\end{table}

Verification rehashes all 1,200 index files and each raw map.
Aider maps have median size 13,888 bytes and 3,881.5 upstream-tokenizer tokens.
Candidate workers have median peak RSS 4.21 GiB, excluding external GPU
allocation. Initial source setup, earlier source-attestation failures,
controller/publication work, and resident GPU costs remain incompletely
measured; component clocks cannot establish complete cold or amortized
repair latency.

\fi

\end{document}